\documentclass[superscriptaddress, twocolumn,showkeys,
amssymb,amsmath,nobibnotes,aps,prd,
nofootinbib]{revtex4-2}
\pdfoutput=1
\usepackage{graphicx,subfigure,bm,color,psfrag,hyperref}
\usepackage[export]{adjustbox}
\usepackage{amsfonts}
\usepackage{lipsum}
\usepackage{mathtools}
\usepackage{verbatim}
\usepackage[normalem]{ulem}
\usepackage[dvipsnames]{xcolor}
\hypersetup{colorlinks,linkcolor={blue},citecolor={red},urlcolor={greenW}}  
\definecolor{greenW}{rgb}{0.0, 0.55, 0.1} 
\begin{document}

\title{Matter Creation Cosmologies and Accelerated Expansion}

\author{Sudip Halder}
\email{sudip.rs@presiuniv.ac.in}
\affiliation{Department of Mathematics, Presidency University, 86/1 College Street, Kolkata 700073, India}

\author{Jaume  de Haro}
\email{jaime.haro@upc.edu}
\affiliation{Departament de Matem\`atiques, Universitat Polit\`ecnica de Catalunya, Diagonal 647, 08028 Barcelona, Spain}

\author{Supriya Pan}
\email{supriya.maths@presiuniv.ac.in}
\affiliation{Department of Mathematics, Presidency University, 86/1 College Street, Kolkata 700073, India}
\affiliation{Department of Mathematics, Faculty of Applied Sciences, \& Institute of Systems Science, Durban University of Technology, Durban 4000, Republic of South Africa}

\author{Tapan Saha}
\email{tapan.maths@presiuniv.ac.in}
\affiliation{Department of Mathematics, Presidency University, 86/1 College Street, Kolkata 700073, India}

\author{Subenoy Chakraborty}
\email{schakraborty.math@gmail.com}
\affiliation{Department of Mathematics, Brainware University, Barasat, West Bengal 700125, India}
\affiliation{Shinawatra University, 99 Moo 10, Bangtoey, Samkhok, Pathum Thani 12160 Thailand}
\affiliation{INTI International University, Persiaran Perdana BBN, Putra Nilai, 71800 Nilai, Malaysia}

%%%%%%%%%%%%%%%%%%%%%%%%%%%%%%%%%%%%%%%%%%%%%%%%%%%%%%%%%%%%%%%%%%%
\begin{abstract}
Non-conservation of dark matter can lead to late-time cosmic acceleration. This mechanism is known as the matter creation theory and this replaces the need of dark energy and modified gravity theories. We consider a two-fluid system consisting of a cold dark matter and a second fluid with constant barotropic equation of state.  
We performed detailed investigations of such cosmologies using the powerful techniques of qualitative analysis of dynamical systems. Considering a wide variety of the creation rates, we examine the phase space analysis of the individual scenario. According to our analyses, these scenarios predict decelerating unstable dark matter (or second fluid) dominated critical points, accelerating attractors dominated either by dark matter or the second fluid, accelerating scaling attractors  in which dark matter and the second fluid co-exist.  The regime of late-time accelerating expansion can be classified as either quintessence, phantom or driven by a cosmological constant. This huge variety of critical points makes these scenarios phenomenologically rich, and naturally   
suggests that such scenarios can be viewed as viable and potential alternatives to the mainstream cosmological models.

\end{abstract}

%%%%%%%%%%%%%%%%%%%%%%%%%%%%%%%%%%%%%%%%%%%%%%%%%%%%%%%%%%%%%%%%%%
\maketitle
%%%%%%%%%%%%%%%%%%%%%%%%%%%%%%%%%%%%%%%%%%%%%%%%%%%%%%%%%%%%%%%%%%

%--------------------------------------------------------------
\section{Introduction}
\label{sec-introduction}

Accelerating expansion of the universe \cite{SupernovaSearchTeam:1998fmf,SupernovaCosmologyProject:1998vns} is a major discovery of astrophysics and cosmology but the intrinsic mechanism behind this phenomenon is not clearly understood. Usually two well known approaches are considered in this regard: one is the introduction of some hypothetical dark energy (DE) fluid with sufficient negative pressure~\cite{Caldwell:1997ii,Caldwell:1999ew,Amendola:1999er,Brax:1999yv,Astier:2000as,Boyle:2001du,Kamenshchik:2001cp,Linder:2002et,Padmanabhan:2002cp,Debnath:2004cd,Li:2004rb,Feng:2004ff,Guo:2004fq,Vikman:2004dc,Jassal:2004ej,Gong:2004fq,Babichev:2004qp,Chiba:2005tj,Cataldo:2005qh,Zhang:2004gc,Zhang:2005eg,Zimdahl:2005bk,Scherrer:2005je,Li:2006ci,Guo:2006ab,Cai:2007us,Feng:2009jr,Linder:2008ya,Li:2009mf,Dutta:2009yb,dePutter:2010vy,Feng:2011zzo,Ma:2011nc,Sola:2015rra,Das:2017gjj,Pan:2017ios,Pan:2017zoh,Yang:2018qmz,Almaraz:2019zxy,Yang:2020zuk,Saridakis:2020zol,Hernandez-Almada:2021aiw,Yang:2023qqz,Kumar:2023bqj,Rezaei:2023xkj,Giare:2024gpk} (also see~\cite{Peebles:2002gy,Copeland:2006wr,Frieman:2008sn,Li:2011sd,Li:2012dt,Bamba:2012cp}) assuming that the Einstein's General Relativity (GR)
is the correct theory of gravity in the cosmological scales, and secondly, the modifications of the Einstein's GR ~\cite{Nojiri:2003ft,Nojiri:2005vv,Nojiri:2005jg,Nojiri:2006gh,Amendola:2006kh,Li:2006ag,Brookfield:2006mq,Nojiri:2007as,Cognola:2007zu,Amendola:2006we,Li:2007jm,Tsujikawa:2007xu,
Fay:2007uy,Li:2007xw,delaCruz-Dombriz:2008ium,Brax:2008hh,DeFelice:2008wz,Thongkool:2009js,Zhou:2009cy,Saridakis:2009bv,Miranda:2009rs,Li:2010cg,Nojiri:2010wj,Li:2011wu,Paliathanasis:2011jq,He:2011qn,Geng:2011aj,Li:2011wu,Harko:2011kv,Li:2012by,Chakraborty:2012kj,Bamba:2012vg,Odintsov:2013iba,He:2013vwa,Nojiri:2014zqa,Thomas:2015dfa,Liu:2016xes,Nunes:2016qyp,Paliathanasis:2016vsw,Nunes:2016plz,Kennedy:2017sof,Nunes:2016drj,Hernandez-Aguayo:2018oxg,Arnold:2019vpg,Atayde:2021pgb,Lobato:2021uff,Geng:2021hqc,Pan:2021tpk,dosSantos:2021owt,Santos:2022atq,Davies:2024nlc} (known as Modified gravity, hereafter MG in short) can also explain this accelerating expansion~(see the review articles  in this direction~\cite{Nojiri:2006ri,DeFelice:2010aj,Capozziello:2011et,Clifton:2011jh,Koyama:2015vza,Cai:2015emx, Nojiri:2017ncd,CANTATA:2021asi,Bahamonde:2021gfp}). Despite many efforts in constructing a variety of DE and MG models, an ultimate cosmological scenario consistent with all the observational datasets is still under the sea. Although the $\Lambda$-Cold Dark Matter ($\Lambda$CDM) model (constructed within the framework of GR) in which $\Lambda$ acts as DE, has been found to be quite successful with many astronomical probes, but the assumption of independent conservation of the cold dark matter (DM) and DE within this framework has no well founded explanation. It is well known that  $\Lambda$CDM already faces many challenges, such as the cosmological constant problem \cite{Weinberg:1988cp}, cosmic coincidence problem~\cite{Zlatev:1998yg}, and cosmological tensions in recent times~\cite{DiValentino:2021izs, Perivolaropoulos:2021jda, Abdalla:2022yfr}. 
Thus, in principle, there is no reason to prefer any cosmological proposal over the other and new models are welcome provided they are found to be consistent with the observational data. 
An alternative scenario to both DE and MG approaches, namely, the theory of gravitationally induced adiabatic particle creation or matter creation was proposed in the literature \cite{Alcaniz:1999hu,Freaza:2002ic,Steigman:2008bc,Lima:2009ic,Basilakos:2010yp,Lima:2015xpa} that can explain the late-time accelerating expansion of the universe. In fact, matter creation models can also explain the early inflationary era as well~\cite{Abramo:1996ip,Gunzig:1997tk} which is quite promising for a new cosmological scenario aiming to compete with existing cosmological proposals.  The development of this theory started with a pioneering work by Prigogine, Geheniau, Gunzig and  Nardone~\cite{Prigogine:1988jax} who established the link between the matter creation and the Einstein's gravitational equations.\footnote{We refer to \cite{Calvao:1991wg,Lima:1992np,Zimdahl:1993cu,Gariel:1995kh,Lima:1995xz,Lima:1996mp} discussing the connection between thermodynamics and matter creation and some of its important aspects, such as the equivalence of matter creation and bulk viscosity~\cite{Zimdahl:1996ka,Fabris:2005ts,Colistete:2007xi,Li:2009mf}.  }
In matter creation theory, the creation pressure caused by the produced particles plays the central role in driving the 
accelerating expansion of the universe. The creation pressure is directly linked with the rate of particle creation and the properties of the matter component from which the particles are created. As a result of which, without adding any hypothetical fluid or modifying the underlying gravitational theory, one can explain the accelerating phase of the universe.  While looking at the large amount of works in the direction of DE (Refs. \cite{Caldwell:1997ii,Caldwell:1999ew,Amendola:1999er,Brax:1999yv,Astier:2000as,Boyle:2001du,Kamenshchik:2001cp,Linder:2002et,Padmanabhan:2002cp,Debnath:2004cd,Li:2004rb,Feng:2004ff,Guo:2004fq,Vikman:2004dc,Jassal:2004ej,Gong:2004fq,Babichev:2004qp,Chiba:2005tj,Cataldo:2005qh,Zhang:2004gc,Zhang:2005eg,Zimdahl:2005bk,Scherrer:2005je,Li:2006ci,Guo:2006ab,Cai:2007us,Feng:2009jr,Linder:2008ya,Li:2009mf,Dutta:2009yb,dePutter:2010vy,Feng:2011zzo,Ma:2011nc,Sola:2015rra,Das:2017gjj,Pan:2017ios,Pan:2017zoh,Yang:2018qmz,Almaraz:2019zxy,Yang:2020zuk,Saridakis:2020zol,Hernandez-Almada:2021aiw,Yang:2023qqz,Kumar:2023bqj,Rezaei:2023xkj,Giare:2024gpk}), and MG (Refs. \cite{Nojiri:2003ft,Nojiri:2005vv,Nojiri:2005jg,Nojiri:2006gh,Amendola:2006kh,Li:2006ag,Brookfield:2006mq,Nojiri:2007as,Cognola:2007zu,Amendola:2006we,Li:2007jm,Tsujikawa:2007xu,
Fay:2007uy,Li:2007xw,delaCruz-Dombriz:2008ium,Brax:2008hh,DeFelice:2008wz,Thongkool:2009js,Zhou:2009cy,Saridakis:2009bv,Miranda:2009rs,Li:2010cg,Nojiri:2010wj,Li:2011wu,Paliathanasis:2011jq,He:2011qn,Geng:2011aj,Li:2011wu,Harko:2011kv,Li:2012by,Chakraborty:2012kj,Bamba:2012vg,Odintsov:2013iba,He:2013vwa,Nojiri:2014zqa,Thomas:2015dfa,Liu:2016xes,Nunes:2016qyp,Paliathanasis:2016vsw,Nunes:2016plz,Kennedy:2017sof,Nunes:2016drj,Hernandez-Aguayo:2018oxg,Arnold:2019vpg,Atayde:2021pgb,Lobato:2021uff,Geng:2021hqc,Pan:2021tpk,dosSantos:2021owt,Santos:2022atq,Davies:2024nlc}),  it is fairly understood that even though the DE and MG models got massive popularity in the community compared to the models arising from the theory of particle creation, however, the matter creation models are equally compelling due to their simplicity and elegant nature, see also \cite{deRoany:2010jq,Lima:2011hq,Jesus:2011ek,Lima:2012cm,Lima:2014qpa,Lima:2014hda,Ramos:2014dba,Fabris:2014fda,Chakraborty:2014fia,Baranov:2015eha,deHaro:2015hdp,Pan:2016jli,Paliathanasis:2016dhu,Biswas:2016idx,Bhattacharya:2017lvr,Pan:2018ibu,Ivanov:2019vkl,Ivanov:2019van,Singh:2019uwv,Cardenas:2020grl,Cardenas:2020exv,Gohar:2020bod,Kaur:2021dix,
Cardenas:2021wxj,Cardenas:2022vuw,Pinto:2023tof,Trevisani:2023wpw,Banerjee:2023teh,Montani:2024xys,Elizalde:2024rvg,Mandal:2024tng,Cardenas:2025sqf}.

The heart of this theory is the matter creation rate $-$ the rate at which matter particles are created. For a given particle creation rate, one can in principle determine the dynamics of the universe. However, one of the limitations of matter creation theory is the unavailability of any fundamental theory which can evaluate the particle creation rate, hence, different choices for the particle creation rate are considered and the resulting cosmological scenario is tested with the observational data. This treatment is almost identical with the DE and MG gravity theories where a hypothetical fluid (in the context of GR) or an unknown modified gravity is considered at the beginning, and no fundamental principle is available so far that can correctly describe everything.  Even though,  
for some specific phenomenological choices of $\Gamma$, see for instance \cite{Lima:2009ic,Lima:2015xpa}, one can mimic the $\Lambda$CDM-like cosmology, however, based on the ongoing debates in the scientific community \cite{DiValentino:2021izs, Perivolaropoulos:2021jda, Schoneberg:2021qvd,Abdalla:2022yfr,Kamionkowski:2022pkx}, $\Lambda$CDM is not the final destination of the universe, and most probably, new physics beyond $\Lambda$CDM is needed. 
Therefore, examining different matter creation rates is  equally compelling and worth exploring given the fact that currently, the dynamics of the universe is still elusive.

Now under the assumption of any arbitrary matter creation rate, depending on its complexity, the gravitational field equations become complicated and finding the analytical solutions of the cosmological variables is not always possible. Even though one can always adopt the numerical techniques, however, in the present article, we have considered one of the potential tools in cosmology, namely, the qualitative analysis of dynamical systems which has been extensively used in the context of cosmological models. As far as we are concerned with the literature,  the dynamical analysis of matter creating cosmologies  did not get considerable attention to the community without any specific reason. In the present article, we have therefore performed a detailed phase space analysis of various matter creation scenarios using the techniques of dynamical analysis. According to our analysis, we found that the cosmological models endowed with matter creation are phenomenologically very rich and attractive.

The paper has been organized as follows. In section~\ref{sec-basic-eqns}, we describe the gravitational equations of a two-fluid matter creation scenario and present the matter creation models that we wish to study. In section~\ref{sec-dyn-systems} we present the autonomous systems for the proposed matter creation models and examine the phase space analysis of the individual scenario. 
Finally, in section~\ref{sec-summary} we conclude the present article with the main findings.

\section{Matter Creation Cosmology}
\label{sec-basic-eqns}

We assume that the gravitational sector of the universe is described by the Einstein's General Relativity and our universe in its large scale is described  by the Friedmann-Lema\^{i}tre-Robertson-Walker (FLRW) line element 
\begin{eqnarray}
ds^2=-{dt}^2+ a^2(t) \left[\frac{dr^2}{1-kr^2} + r^2 \left(d\theta^2 + \sin^2 \theta d\phi^2 \right) \right],
\end{eqnarray}
where $a(t)$ is the expansion scale factor of the universe and $(t, r, \theta, \phi)$ are the co-moving coordinates. Here $k$ is the spatial curvature and its null value corresponds to the spatially flat universe while  $k =-1$ indicates that our spatial geometry of the universe is open while $k =+1$ corresponds to the closed spatial geometry of the universe. 
We consider a system having a volume $V$ which attain $N$ number of particles. In an open thermodynamical system, $N$ should be a function of time, i.e. it should change with time. Thus, the conservation equation will transform to the form
\begin{eqnarray}
  N^{\mu}_{;\mu}\equiv \dot{n}+\Theta n=n\Gamma \Longleftrightarrow N^{\mu}_{,\mu}u^{\mu}=\Gamma N, \label{particle-cons} 
\end{eqnarray}
where $n=N/V$ is the particle density, $N^{\mu}=n u^{\mu}$ corresponds to the particle flow vector ($u^{\mu}$ represents the four velocity of the particle); `dot' stands for the derivative with respect to the cosmic time $t$;  $\Theta=u^{\mu}_{;\mu}$ is the expansion scalar of the fluid which in the FLRW universe becomes $3H$ ~($H=\dot{a}(t)/a(t)$ is the Hubble rate of the FLRW universe), and $\Gamma$ is the rate of change of particle number. Here $\Gamma > 0$ indicates the production of particles while $\Gamma <0$ indicates particle annihilation. 
Using eqn. (\ref{particle-cons}) in the Gibb's eqn.~\cite{Zimdahl:1996ka}
\begin{eqnarray}
    T ds=d\left(\frac{\rho}{n}\right)+p d\left(\frac{1}{n}\right), \label{gibbs-eqn}
\end{eqnarray}
where $T$ corresponds to the fluid temperature, $s$ indicates the specific entropy (or, the entropy per particle), $\rho$ denotes the total energy density and $p$ represents the total thermodynamic pressure, one appears with the following conservation equation 
\begin{eqnarray}
    n T \dot{s}=\dot{\rho}+3H\left(1-\frac{\Gamma}{3H}\right)(\rho+p).\label{gibbs-cons}
\end{eqnarray}
Under the adiabatic (also known as  isentropic) thermodynamic process where the rate of change of specific entropy vanishes (i.e. $\dot{s}=0$), eqn. (\ref{gibbs-cons}) reduces to 
\begin{eqnarray}\label{new-Gamma-equation}
\dot{\rho}+3H(\rho+p)=\Gamma(\rho+p),
\end{eqnarray}
which can alternatively be put as 
\begin{eqnarray}
    \dot{\rho}+3H(\rho+p+p_c)=0,
\end{eqnarray}
where the new term $p_c$ is termed as the creation pressure due to the particle production and it takes the form $p_c = -(\Gamma/3H) \times (p+\rho)$. This creation pressure has some interesting implications: if the fluid under consideration is a normal fluid, then for $\Gamma >0$, one can induce a negative creation pressure. As we shall explain in the following, this creation pressure can drive the accelerating phase of the universe and hence particle production scenario can be treated as a mirage of a DE. 
Now, we assume that the total energy density of the universe is comprised of a pressure-less DM sector endowed with gravitationally induced adiabatic matter creation, and a second fluid which does not take part in the matter creation hypothesis. Additionally we consider that the fluids under consideration do not take part in an energy exchange mechanism, that means they are not interacting with each other. 
The gravitational equations in the context of the FLRW universe now explicitly be written as 
\begin{eqnarray}
H^2 + \frac{k}{a^2}  = \frac{\kappa^2}{3}(\rho_{\rm dm}+\rho_f),\label{friedmann-1A}\\
2\dot{H}+3H^2 + \frac{k}{a^2} =-{\kappa}^2(p_c+p_f),\label{friedmann-2A}
\end{eqnarray}
where $\kappa^2 = 8 \pi G$  is the Einstein's gravitational constant ($G$ is the Newton's gravitational constant); $\rho_f$, $p_f$ are respectively the energy density and pressure of the second fluid;  $\rho_{\rm dm}$ is the energy density of the pressure-less DM, $p_c$ is the creation pressure which is related to the rate of particle production $\Gamma$ as \cite{Steigman:2008bc,Lima:2009ic,Basilakos:2010yp,deHaro:2015hdp}
\begin{eqnarray}\label{creation-pressure}
p_c = - \frac{\Gamma}{3H}\rho_{\rm dm}. 
\end{eqnarray}
Here $\Gamma > 0$ indicates the creation of particles and $\Gamma < 0$ indicates the particle annihilation.\footnote{Since the DM sector is the only responsible fluid for the creation, it is natural to depict that the created particles are the DM particles.} The conservation equation of the second fluid is, $\dot{\rho_f} + 3 H (1+w) \rho_f =0$, where $w = p_f/\rho_f$ is the barotropic equation-of-state (EoS) of the second fluid and here we assume it to be a constant. This EoS is indefinite in sign. Although this fluid does not take part in the matter creation but it influences the expansion history of the universe, therefore, it is evident that the EoS, $w$, plays a crucial role in this context. In particular, the dynamics of the universe in presence of a normal fluid (characterized by $w \geq 0$) and hypothetical fluid (characterized by $w < 0$) should be different. Considering this,
in this article we explore all kind of possibilities through its EoS as follows: $0 \leq w \leq 1$ (this corresponds to a normal fluid where $w =1$, $w =1/3$, $w=0$ respectively corresponds to a stiff fluid, radiation and dust), $-1/3 \leq w < 0$ (this is a hypothetical fluid in the sense that the pressure is negative), $-1 < w \leq -1/3$ (quintessence-like fluid), $w =-1$ (it corresponds to the cosmological constant), $w< -1$ (phantom fluid). While our main intention is to focus on the matter creation scenarios in presence of a normal fluid, however, we consider the remaining cases for completeness. 
On the other hand, it is interesting to note that the matter creating DM fluid is equivalent to a non-interacting DM fluid with variable EoS. This can be quickly understood by recasting  the conservation equation of the DM fluid as  
\begin{eqnarray}\label{new-eqn-cons-DM}
\dot{\rho}_{\rm dm}^{\rm eff}+ 3H (1+w_{\rm dm}^{\rm eff}) \rho_{\rm dm}^{\rm eff} = 0,
\end{eqnarray}
where $w_{\rm dm}^{\rm eff} = - \Gamma/3H$ represents the effective EoS for DM and considering this effective description of DM, we use $\rho_{\rm dm}^{\rm eff}$ to represent $\rho_{\rm dm}$ and from now on we shall use $\rho_{\rm dm}^{\rm eff}$. For $\Gamma >0$, $w_{\rm dm}^{\rm eff}$ becomes negative in an expanding universe. This is not surprising since the possibility of negative EoS of DM has recently been reported in several articles~\cite{Naidoo:2022rda,Yao:2023ybs,Wang:2025zri,Abedin:2025dis,Li:2025eqh,Wang:2025hlh}. This establishes the equivalency of matter creation cosmologies and DM system with dynamical EoS. 
Now, concerning the second fluid, one can find the evolution of $\rho_f$ as, $\rho_f = \rho_{f0} (a/a_0)^{-3 (1+w)}$ where $\rho_{f0}$ is the present value of $\rho_f$ and $a_0$  is the scale factor at present time which without any loss of generality, we set $a_0 =1$. If, on the contrary, $w$ is dynamical, then depending on the nature of $w$, one can either solve $\rho_f$ analytically or numerically. However, the overall evolution of this matter creating cosmology depends on the matter creation rate $\Gamma$. In this work we focus on the constant $w$ in order to start with a simple scenario of matter creation cosmology. 
Finally, the accelerating equation in this case becomes, 

\begin{eqnarray}\label{acc-eqn-two-fluid}
    \frac{\ddot{a}}{{a}} = -\frac{\kappa^2}{6} \left[ \left(1 - \frac{\Gamma}{H}\right) \rho_{\rm dm}^{\rm eff}  + (1+3w) \rho_f \right]. 
\end{eqnarray}
From eqn. (\ref{acc-eqn-two-fluid}) one can clearly see that  if we consider a normal fluid characterized by $w \geq 0$, then for $\Gamma <0$, we will never realize an accelerating expansion of the universe because in this case $\ddot{a} <0$; however, for $\Gamma >0$, it is possible to get $\ddot{a} >0$ provided the sum of the terms inside the third brace of the right hand side of eqn. (\ref{acc-eqn-two-fluid}) becomes negative. This can be obtained for some suitable choices of $\Gamma$. However, we further note that 
for $\Gamma <0$, it is not impossible to obtain the accelerating expansion of the universe, because  in this case we need to allow $w< 0$, that means particle annihilation together with some hypothetical fluid with negative pressure can drive the cosmic acceleration. This further strengthens the presence of a hypothetical fluid with negative pressure in the context of particle annihilation. However, in the present article we restrict ourselves to $\Gamma >0$ and explore the cosmological scenarios for both $w \geq 0$ and $w <0$.

Now, in order to examine the cosmological scenarios, $\Gamma$ should be prescribed. As already argued, no fundamental theory is available so far which allows us to determine the actual form of the matter creation rate, and hence this motivates us to consider some phenomenological choices of $\Gamma$. 
Over the last couple of years, a variety of matter creation rates have been proposed in the literature \cite{Lima:2011hq,Jesus:2011ek,Lima:2012cm,Lima:2014qpa,Lima:2014hda,Chakraborty:2014fia,Baranov:2015eha,deHaro:2015hdp,Pan:2016jli,Paliathanasis:2016dhu,Pan:2018ibu}. In most of the cases, the choice of the matter creation rate $\Gamma$ is assumed to be dependent only on the Hubble rate. As matter creation rate is directly linked with the expansion rate of the universe, therefore, $\Gamma \propto H^n$ ($n$ is any real number) is a natural assumption to model the expansion history of the universe.  One can notice that for $\Gamma = 3 \epsilon H$ where $\epsilon$ is a constant, solving (\ref{new-eqn-cons-DM}), DM sector evolves as $\rho_{\rm dm}^{\rm eff} \propto a^{-3 (1- \epsilon)}$ and hence, it differs from its standard evolution $a^{-3}$. More interestingly, as described in \cite{Lima:2009ic}, if the creation rate is assumed to be $\Gamma = 3 \psi H (\rho_{c0}/\rho_{\rm dm})$, where $\psi >0$ is a constant and $\rho_{c0}$ is the present day value of the critical energy density of the universe, then solving (\ref{new-eqn-cons-DM}), one gets $\rho_{\rm dm}^{\rm eff} = (\rho_{\rm dm,0}^{\rm eff} - \psi \rho_{c0})a^{-3} + \psi \rho_{c0}$ ($\rho_{\rm dm,0}^{\rm eff}$ is the present value of $\rho_{\rm dm}^{\rm eff}$). Now if the matter sector consists of the DM fluid, then in a spatially flat FLRW universe, the Hubble equation becomes, $H^2/H_0^2 = (\Omega_{\rm dm,0}^{\rm eff} - \psi)a^{-3} + \psi$ ($\Omega_{\rm dm,0}^{\rm eff}$ is the present day value of the effective DM density parameter) which resembles the $\Lambda$CDM model with $\psi$ representing the cosmological constant. This clearly shows that $\Lambda$CDM like evolution can be realized in this framework. However, other choices for $\Gamma$ are equally allowed. In fact, looking at the left hand side of eqn. (\ref{new-Gamma-equation}) representing the conservation equation of pressure-less DM fluid endowed with matter creation, one can understand that $\Gamma$ may involve $H$, $\dot{H}$ and the scale factor of the FLRW universe.  On the other hand, considering the facts that, matter creation scenarios are equivalent to the bulk viscous models~\cite{Gariel:1995kh,Zimdahl:1996ka,Fabris:2005ts,Colistete:2007xi,Li:2009mf,Yang:2019qza,Cardenas:2020exv,Yang:2023qqz} (although both the prescriptions differ thermodynamically~\cite{Lima:1992np}) and DM EoS could be time-dependent~\cite{Naidoo:2022rda,Yao:2023ybs,Wang:2025zri,Abedin:2025dis,Li:2025eqh,Wang:2025hlh}, one can expect $\Gamma$ to be dependent on the higher order derivatives of $H$.    
Considering these altogether, a general form of the matter creation rate therefore could be the following

\begin{eqnarray}
    \Gamma=\Gamma\left(a, H,\dot{H},\ddot{H},\dddot{H},...\right).
\end{eqnarray}
Usually one can consider a number of choices for $\Gamma$ and investigate the viability of the resulting scenario. In the present work we consider the following two choices of $\Gamma$: 
\begin{eqnarray}
    \Gamma  = \Gamma (H), \quad \quad \quad  \Gamma= \gamma_{H} \dot{H},
\end{eqnarray}
where $\gamma$ is a function of the Hubble rate and the subscript denotes the derivative with respect to the Hubble rate.\footnote{At this point we would like to remark that there are no such restrictions in considering models of $\Gamma$ involving higher order derivatives of $H$ beyond $\dot{H}$, since as mentioned earlier, there are no such fundamental principles available in the literature that can derive the matter creation rate $\Gamma$. Although the inclusion of higher order derivatives of $H$ (e.g. $\ddot{H}$, $\dddot{H}$, etc) may increase the mathematical complexities, but the possible complications due to the presence of $\ddot{H}$, $\dddot{H}$, in $\Gamma$ do not restrict us in considering such scenarios, rather in this article we restrict our analyses to the models involving only up to $\dot{H}$. } The second model has some interesting features since under some certain choices for $\gamma$ and imposing some restrictions, one can retrieve the $\Lambda$CDM cosmology.  Using the conservation equation for DM, its energy density for the first choice of the matter creation rate can be solved as 
\begin{equation}\label{solution-1}
    \rho_{\rm dm}^{\rm eff}=\rho_{\rm dm,0}^{\rm eff} a^{-3} \exp\left(\int_{1}^{a}\frac{\Gamma}{aH} \, da \right)
\end{equation}
where $\rho_{\rm dm,0}^{\rm eff}$ is the present-day value of the DM density. While for the remaining matter creation rate, one can solve the energy density for DM as 

\begin{eqnarray}\label{solution-2}
    \rho_{\rm dm}^{\rm eff} = \rho_{\rm dm,0}^{\rm eff} a^{-3} \exp\Bigl[\gamma (H) - \gamma (H_0) \Bigr]
\end{eqnarray}
It is interesting to mention that with the proper choices of $\Gamma$, known cosmological scenarios can be recovered.  Concerning the variable $\gamma (H)$ we consider two choices~\cite{Cardenas:2020grl}: 

\begin{eqnarray}
\gamma(H)=\left(1-\frac{1}{\xi}\right) \ln\left(\frac{3H^2}{H_0^2}\right),\label{choice-1}
\end{eqnarray}
where $\xi \neq 0$ and 
\begin{eqnarray}\label{choice-2}
 \gamma(H)=\left\{\begin{array}{ccc}
             & \ln \left(\frac{H^2}{H_0^2}\right)-\frac{9}{4} \ln\left(\frac{H^2}{H_0^2}-1\right)\mbox{for}& H>H_0\\
             0&\mbox{for}& H = H_0.\\
            \end{array}\right.
\end{eqnarray}
As noticed in Ref. \cite{Cardenas:2020grl}, the above two choices of $\gamma (H)$ can lead to some interesting consequences.
In the following we perform the phase space analysis of the individual model assuming the spatially flat FLRW universe (i.e. $k =0$).

\section{Phase space Analysis}
\label{sec-dyn-systems}

In this section we present a detailed phase space analysis of the two-fluid matter creation scenarios considering various matter creation rates. {\it As far as we are concerned with the literature, no systematic analyses of such matter creation models are present, and hence, such analyses and results are completely new in this direction. }
In order to proceed with the dynamical analysis, we introduce the following dimensionless variables
\begin{equation}
	x=\frac{\kappa^2 \rho_{\rm dm}^{\rm eff}}{3 H^2}, \quad
 \mbox{and} \quad
	z=\frac{H_0}{H_0+H},
	\label{dimensionless-variables}
\end{equation}
where $H_0$ represents the present value of the Hubble rate.\footnote{Note that without any loss of generality, one can use $H_*$ (the value of the Hubble parameter at $t=t_*$) instead of $H_0$ and this will not change the physics.} Therefore, the variables $x$ and $z$, taking values in the intervals $0\leq x\leq 1$ and $0\leq z\leq 1$, define the phase space as a unit square ${\bf R}=[0,1]^2$. 
Now, using the dimensionless variables in (\ref{dimensionless-variables}), the gravitational equations describing the dynamics of the fluids can be converted into the autonomous systems.  For any matter creation rate $\Gamma$, the decelerating parameter, $q \equiv -1 -\dot{H}/H^2$, in terms of the dimensionless variables in (\ref{dimensionless-variables}), takes the form
\begin{equation}
    q=\frac{1}{2}\left[1+3w(1-x)-\frac{\Gamma xz}{(1-z)H_0}\right].\label{two-f-dec}
\end{equation}
Therefore, given a particular model of $\Gamma$, one can determine the evolution of the deceleration parameter.
In what follows, we present the autonomous systems for the proposed matter creation models and perform the stability of the critical points. 

%%%%%%%   Table 1  %%%%%%%%
\begin{table*}[t]
\centering
%\resizebox{0.9\textwidth}{!}{%
	\begin{tabular}{|c c c c c c c c|}\hline\hline
{\bf Critical point} & {\bf Existence} & {\bf Eigenvalue} & {\bf Stability} & $\mathbf{\Omega_f}$ &  $\mathbf{\Omega_{\rm dm}^{\rm eff}}$ & $\mathbf{q}$ & {\bf Acceleration} \\ \hline
%   &&&&&     \\

$A_{0}(0,0)$  & Always & $\left(3w,\frac{3}{2}(1+w)\right)$  & {\bf Stable} if $w<-1$;  &  1  &  0  &  $\frac{1}{2}(1+3w)$  &  $w<-\frac{1}{3}$  \\ 
 &&& Saddle if $-1<w<0$;  &&&&  \\
  &&& Unstable if $w> 0$  &&&&  \\ \hline
 
$A_{1}(1,0)$  & Always & $\left(-3w,\frac{3}{2}\right)$   &  Saddle if $w>0$;   & 0   &  1  &  $\frac{1}{2}$  & No \\
 &&& Unstable if $w< 0$ &&&&  \\  \hline
%   &&&&&  \\
$A_{2}(0,1)$  & Always  & $(\alpha,0)$   &  Non-hyperbolic Saddle if $w>-1$   & 1   &  0  & Undefined   &  Undetermined \\
  &&& Unstable if $w<-1$ &&&&  \\ \hline
%   &&&&&  \\
$A_{3}(1,1)$  & Always  & $\left(-\alpha,\frac{\alpha}{2}\right)$  & Always Saddle  &  0  & 1   &  $-\infty$  &  Yes \\  \hline
%   &&&&&  \\ 
$A_{4}\left(1,\frac{3}{3+\alpha}\right)$  &  $\alpha>0$ & $\left(-\frac{3\alpha(1+w)}{3+\alpha},-\frac{3\alpha}{2(3+\alpha)}\right)$  &  {\bf Stable} if $w>-1$    &  0  &  1  &  $-1$  &  Yes     \\ 
  &&&  Saddle if $w<-1$ &&& &    \\ \hline
%   &&&&& \\ 
 $A_5\left(0,z_c\right)$  &  $w=-1$ & $\left((3+\alpha)z_c-3,0\right)$  & {\bf Stable} if $z_c<\frac{3}{3+\alpha}$;  &  $1$  &  $0$  &  $-1$  &  Yes     \\ 
  &&&  Unstable if $z_c>\frac{3}{3+\alpha}$ &&& &    \\ \hline
 $A_6\left(x_c,\frac{3}{3+\alpha}\right)$  &  $w=-1$ & $\left(0,-\frac{3\alpha x_c}{2(3+\alpha)}\right)$  &  {\bf Stable}  &  $1-x_c$  &  $x_c$  &  $-1$  &  Yes     \\  
\hline
$A_{7}\left(x_c,0\right)$  &  $w=0$ & $\left(0,\frac{3}{2}\right)$  &  Unstable  &  $1-x_c$  &  $x_c$  &  $\frac{1}{2}$  &  No     \\ \hline\hline
\end{tabular}%
% }
\caption{Summary of the critical points, their existence, stability and the values of the cosmological parameters at those points for the dynamical system (\ref{RG-DS-1}) with the matter creation rate $\Gamma=\Gamma_0$. 
	   }
	\label{first-table}
\end{table*}
\subsection{Model: $\Gamma=\Gamma_0$}\label{model-1}

In this section we shall discuss the effects of the constant particle creation rate $\Gamma = \Gamma_0$ considering two-fluid system. 

For the two-fluid system, considering the dimensionless variables, $x$, $z$ defined in eqn. (\ref{dimensionless-variables}), the constraint from the Friedmann equation (\ref{friedmann-1A}) with $k=0$ is, $\Omega_{\rm dm}^{\rm eff} + \Omega_f =1$ where $\Omega_f =  \kappa^2 \rho_f/(3H^2)$ and $\Omega_{\rm dm}^{\rm eff} = \kappa^2 \rho_{\rm dm}^{\rm eff}/(3H^2)$.\footnote{Note that $\Omega_{f0}$, $\Omega_{\rm dm,0}^{\rm eff}$ are the present day values of the density parameters. }  Now, using these dimensionless variables, one can write down the autonomous system as follows 
\begin{subequations} \label{DS-1}
\begin{align}
    x'=& x(1-x) \left(3w+\frac{\alpha z}{1-z}\right),   \label{DS-1-x} \\
    z'=& \frac{3}{2}z(1-z)\left[1+w(1-x)-\frac{\alpha x z}{3(1-z)} \right], \label{DS-1-z} 
\end{align}
\end{subequations}
where prime denotes the derivative with respect to $N = \ln (a)$\footnote{From now on, prime attached to any variable will refer to the derivative with respect to $N$. } and 
$\alpha=\Gamma_0/H_0$ is a parameter which takes only positive values. To address the singularity at $z=1$ of the dynamical system (\ref{DS-1}), we introduce a new time variable $\tau$, defined as $dN=(1-z)d\tau$. Consequently, the reduced autonomous system which is topologically equivalent \cite{Halder:2024aan,Halder:2024gag,Hussain:2022dhp,Agrawal:2022zno,Bahamonde:2017ize,perko2013differential,dumortier2006qualitative,meiss2007differential} to the autonomous system (\ref{DS-1}) can be read as:   
\begin{subequations} \label{RG-DS-1}
\begin{align}
    \frac{dx}{d\tau}=& x(1-x) \left(3w(1-z)+\alpha z\right),   \label{RG-DS-1-x} \\
    \frac{dz}{d\tau}=& \frac{3}{2}z(1-z)\bigg[\bigg(1+w(1-x)\bigg)(1-z)-\frac{\alpha x z}{3} \bigg], \label{RG-DS-1-z} 
\end{align}
\end{subequations}
From the above autonomous system, one can easily conclude that $x=0,~x=1,~z=0$ and $z=1$ are invariant manifolds which indicate that our physical region is positively invariant, i.e. if we take any orbit from ${\bf R}$, it never leaves the domain. Here, the decelerating parameter defined in (\ref{two-f-dec}) is simplified to $q=\frac{1}{2}\left[1+3w(1-x)-\frac{\alpha x z}{1-z}\right]$. From this relation, one can quickly derive that in order to obtain the accelerating expansion at present time, i.e. at $z = 1/2$,  
the viable orbits must satisfy the relation $(1-\alpha)x+(1+3w)(1-x)<0$.

Now, in order to understand the phase space dynamics of the autonomous system (\ref{RG-DS-1}), we find the critical points and investigate their stability conditions. The critical points and their characteristics are summarized in Table~\ref{first-table}.  In this case, we find five discrete critical points ($A_0$, $A_1$, $A_2$, $A_3$, and $A_4$) and three critical lines ($A_5$, $A_6$ and $A_7$).  Since the EoS of the second fluid, $w$, is involved and as $w$ can take any value (either non-negative or negative), therefore, in order to understand how the nature of the second fluid affects the architecture of the phase space dynamics, we have paid special attention to the cases with $w \geq 0$ indicating the normal fluid and $w < 0$ representing the hypothetical fluid. In the following we describe the qualitative features of the above dynamical system for different values of $w$.    

\begin{figure*}
    \centering
    \includegraphics[width=0.33\textwidth]{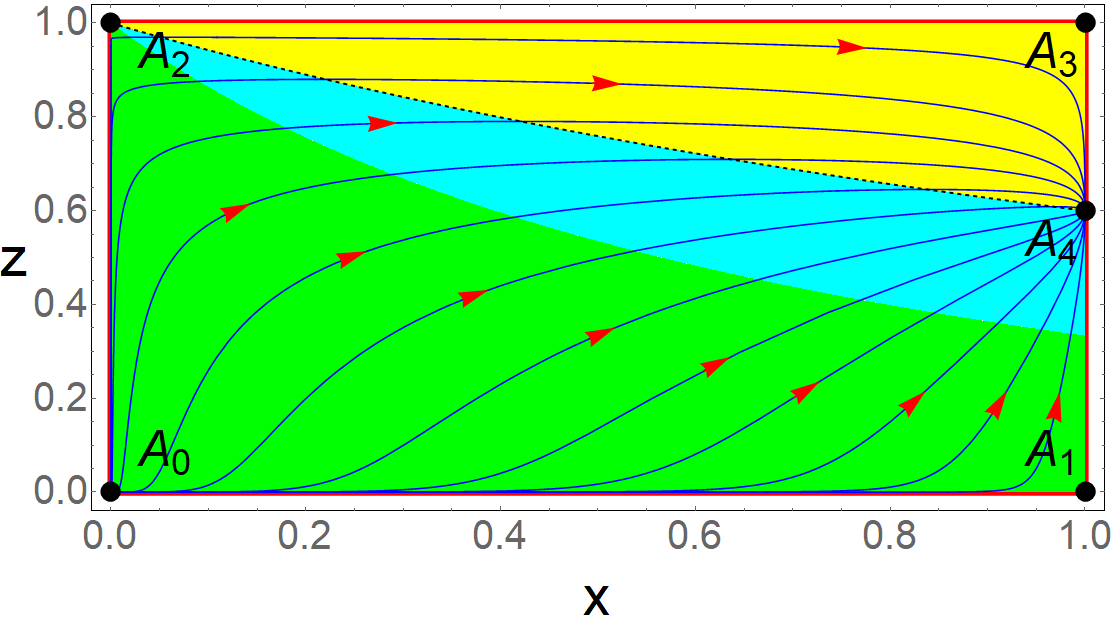}
    \includegraphics[width=0.33\textwidth]{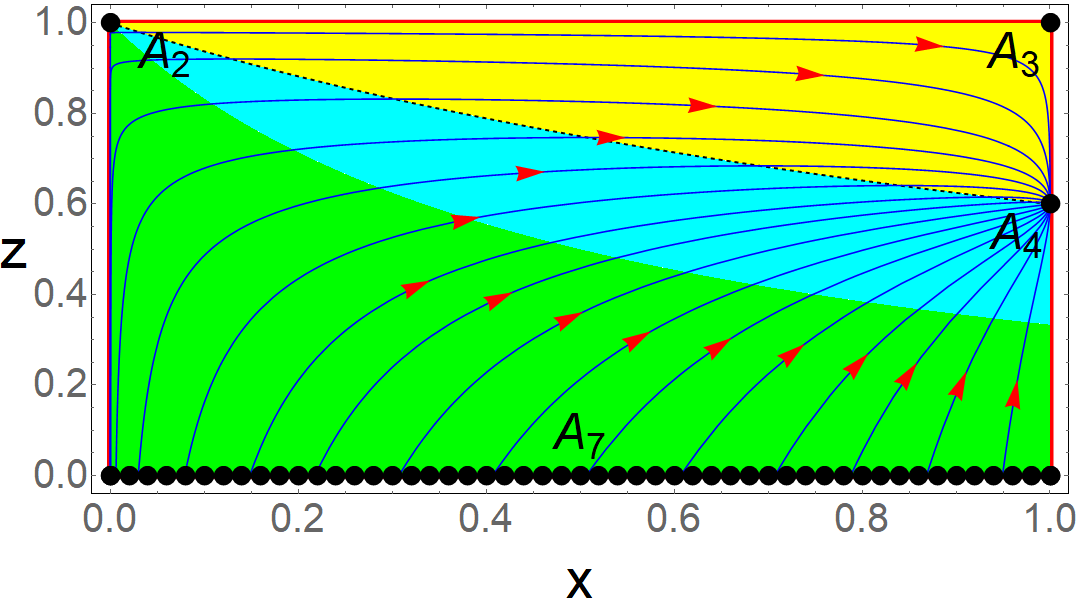}
    \includegraphics[width=0.32\textwidth]{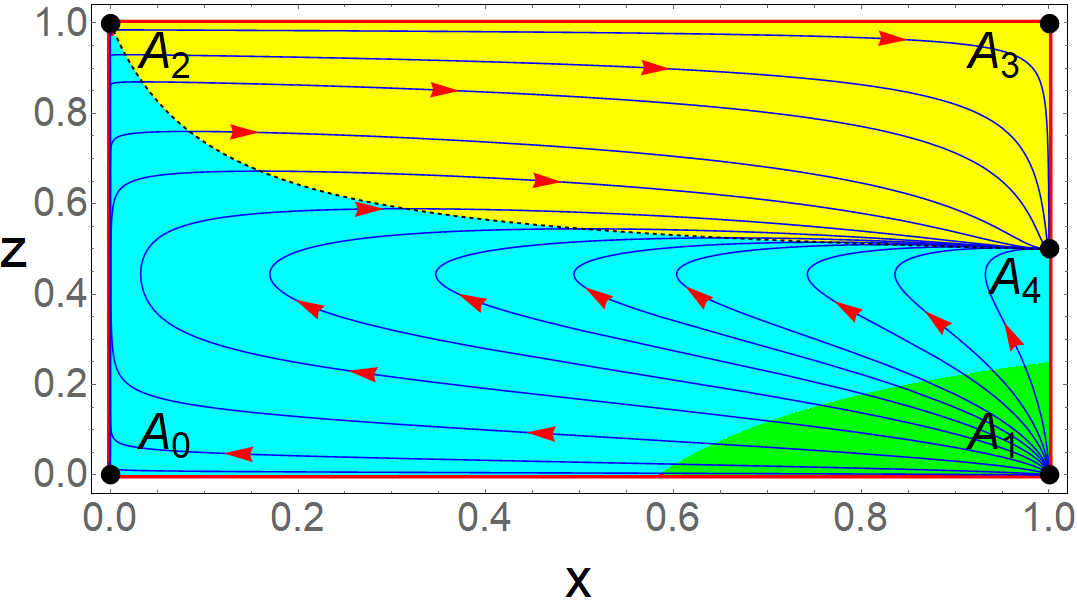}\\
    \includegraphics[width=0.33\textwidth]{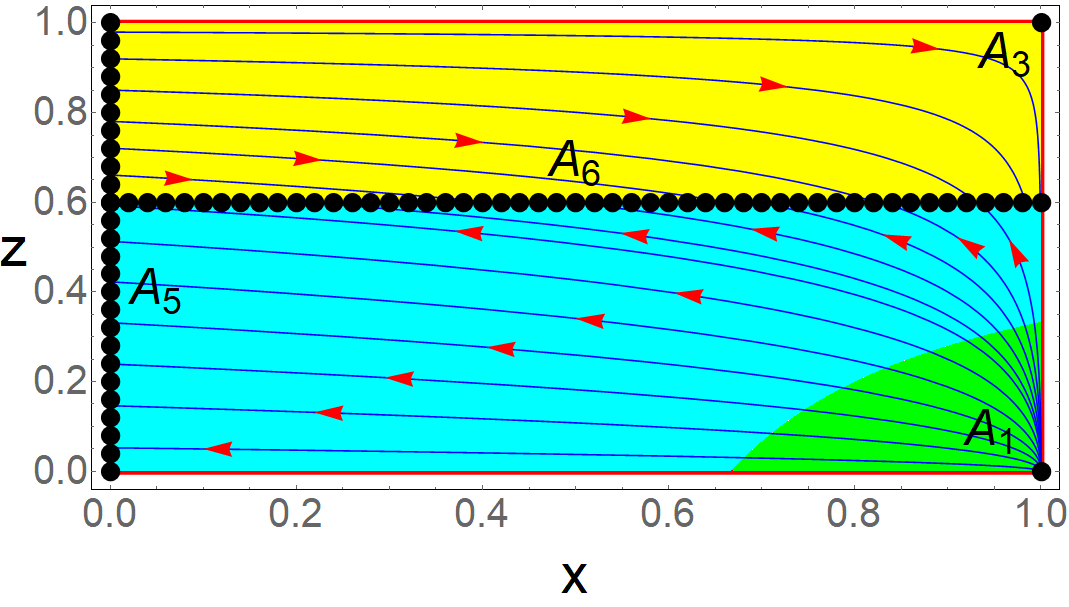} 
    \includegraphics[width=0.33\textwidth]{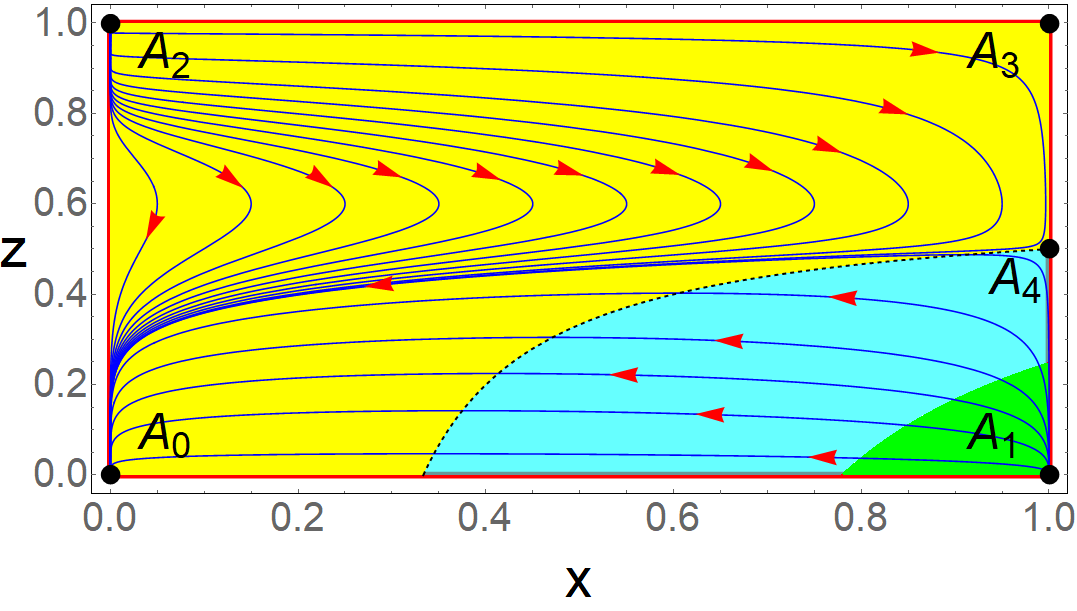}
    \caption{Description of the phase space of the matter creation scenarios in which the matter creation rate is constant, i.e. $\Gamma = \Gamma_0$. {\bf Upper Left Plot:} The phase plot of the system (\ref{RG-DS-1}) when we have assumed $w=0.1$ and $\alpha=2$. For other values of $w>0$ and $\alpha>0$, we can also obtain similar phase space structure.  {\bf Upper Middle Plot:} The phase space of the system (\ref{RG-DS-1}) when the EoS $w$ takes the value $0$. Here we use $\alpha =2$ but any positive value of $\alpha$ gives same type of phase portrait. {\bf Upper Right Plot:} The phase plot of the system ($\ref{RG-DS-1}$) considering $w=-0.8$ and $\alpha=3$. Also, we can get similar type of graphics for any positive value of $\alpha$ and negative value of $w$ in the interval $(-1,0)$. {\bf Lower Left Plot:} The phase space of the system (\ref{RG-DS-1}) when the EoS $w$ takes the value $-1$. Here we use $\alpha =3$ but any positive value of $\alpha$ gives similar type of phase portrait. {\bf Lower Right Plot:} The phase plot of the system (\ref{RG-DS-1}) when we assume $w=-1.5$ and $\alpha=3$. For other values of $w<-1$ and $\alpha>0$, we can also obtain similar graphics. In all five two-dimensional plots, green region corresponds to the decelerating phase ($q >0$), cyan region represents the accelerating phase with $-1<q<0$ and the yellow region corresponds to the super accelerating phase (i.e. $q <-1$). The black dotted curve separating the cyan and yellow regions (in the left plot, this separating curve is not visible because of the line of critical points) corresponds to $q= -1$.  }
    \label{fig1A}
\end{figure*}

\begin{figure*}
   \centering   
\includegraphics[width=0.497\textwidth]{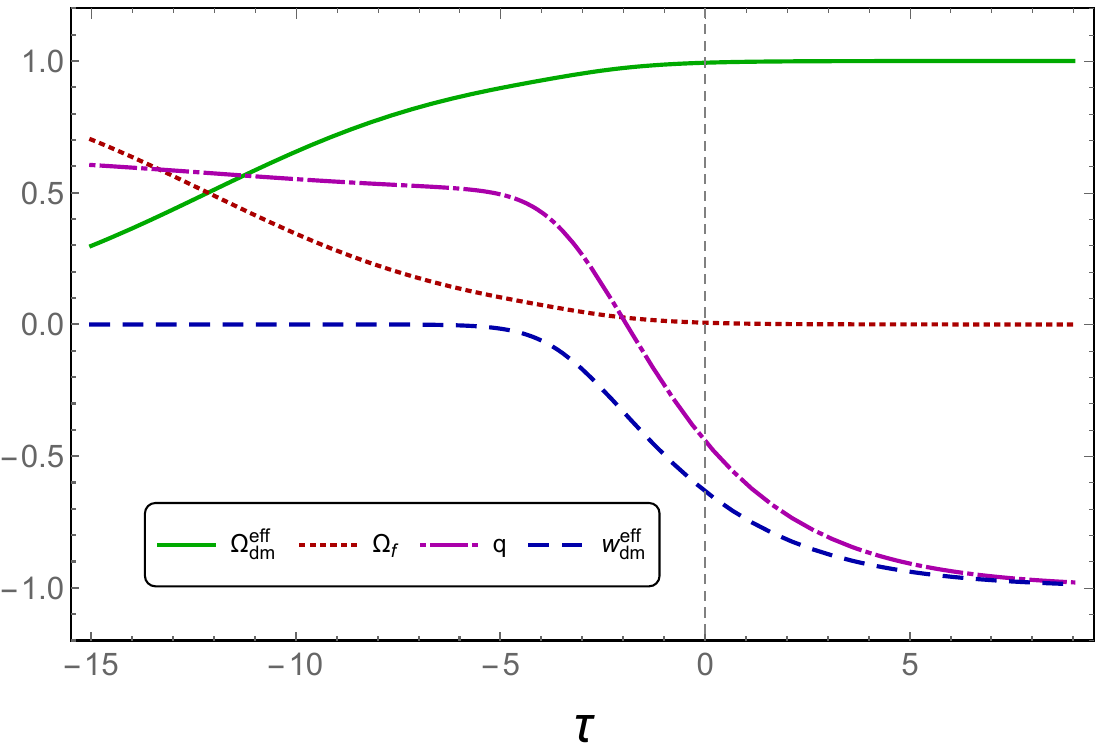} 
\includegraphics[width=0.497\textwidth]{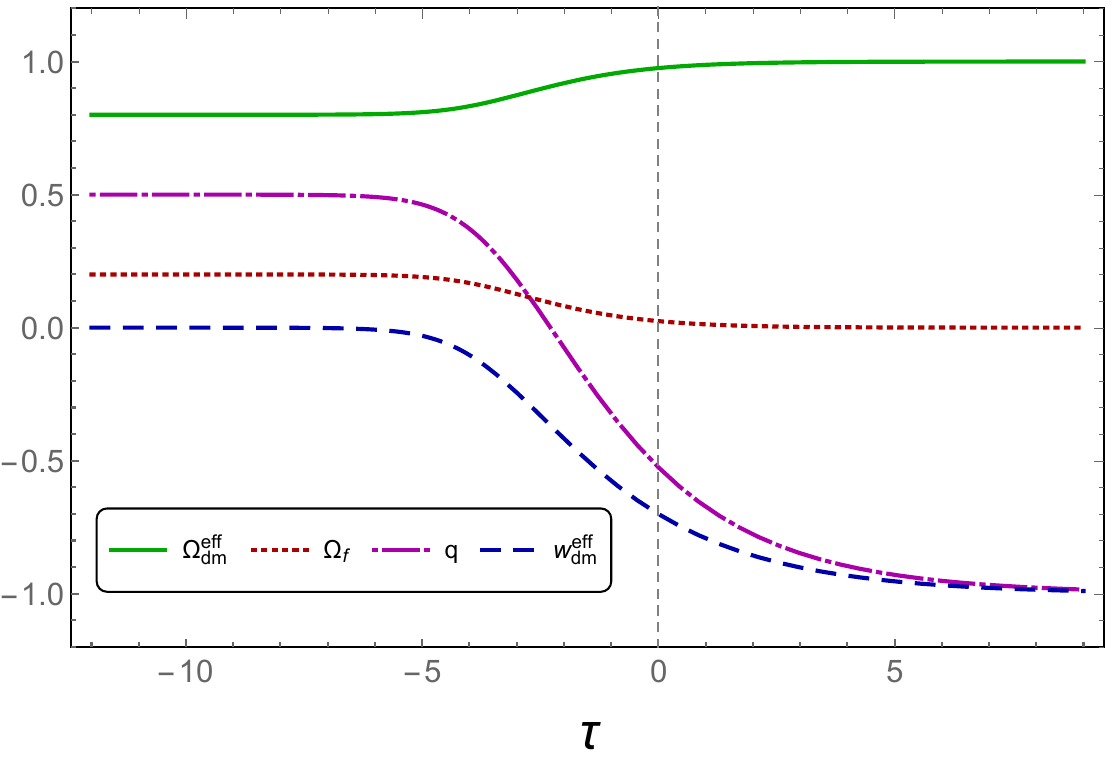}
\includegraphics[width=0.497\textwidth]{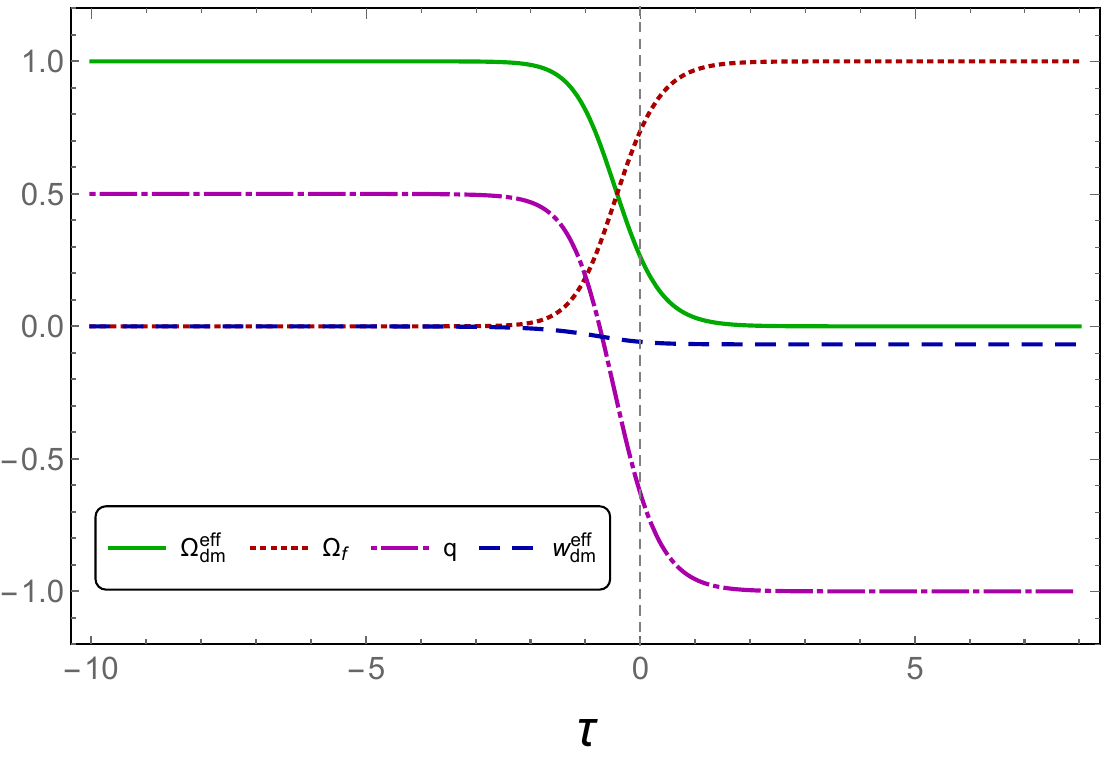} 
\includegraphics[width=0.497\textwidth]{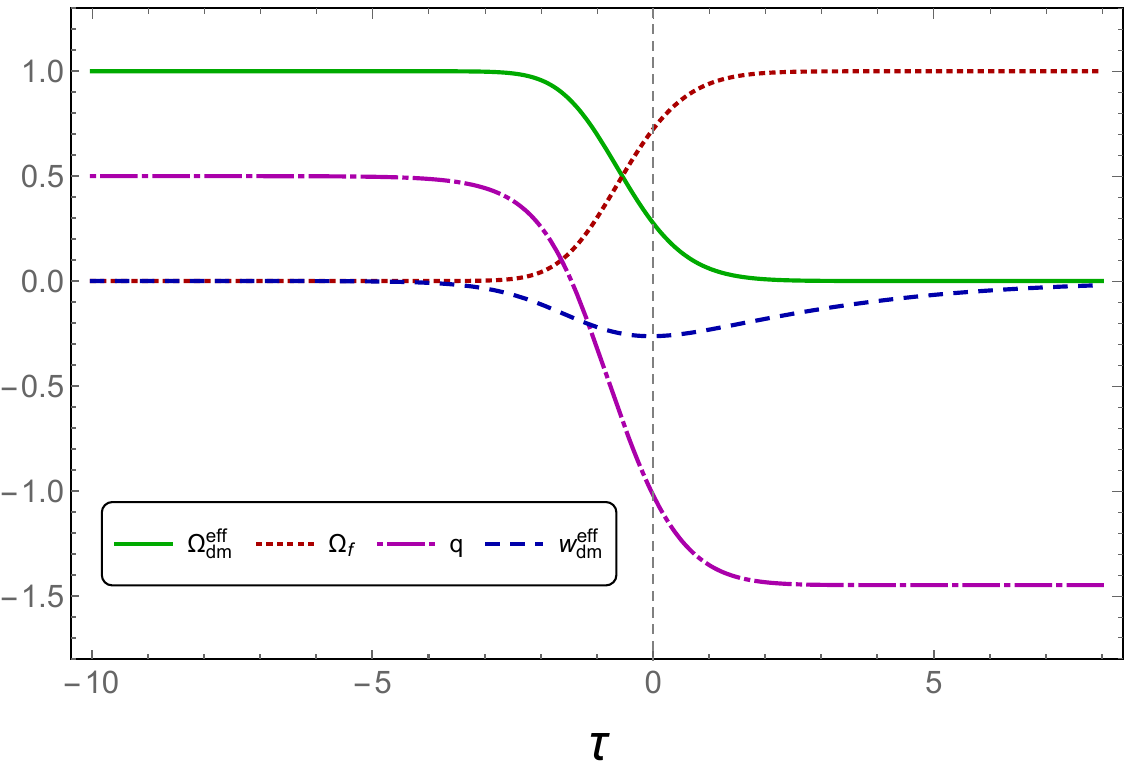}
    \caption{We provide the evolution plots for the dynamical system (\ref{RG-DS-1}), showing the behavior of the DM density parameter $\Omega_{\rm dm}^{\rm eff}$, the second fluid density parameter $\Omega_f$, the deceleration parameter $q$, and the effective equation of state of DM $w_{\rm dm}^{\rm eff}$ under the constant particle creation rate $\Gamma=\Gamma_0$. {\bf Upper Left Plot:} This figure is drawn with the model parameters $w=0.1$, $\alpha=1$, and it shows that the final fate of the universe is late time accelerated evolution in cosmological constant era dominated by DM. {\bf Upper Right Plot:} This figure is produced for the parameters values $w=0$ and $\alpha=1$. Here also, the final fate of the universe is late time accelerated evolution in cosmological constant era dominated by DM. {\bf Lower Left Plot:} We have used $w=-1$ and $\alpha=1$. This plot depicts that the ultimate fate of the universe is found to be late time accelerated evolution in cosmological constant epoch which is achieved after the occurrence of a decelerated DM dominated phase. {\bf Lower Right Plot:} We have taken $w=-1.3$ and $\alpha=1$. This plot highlights that the ultimate fate of the universe is found to be late time DE dominated accelerated evolution in phantom epoch which is achieved after the occurrence of a decelerated DM dominated phase. 
}
    \label{fig-evo-1}
\end{figure*}
\begin{figure}
    \centering
    \includegraphics[width=0.497\textwidth]{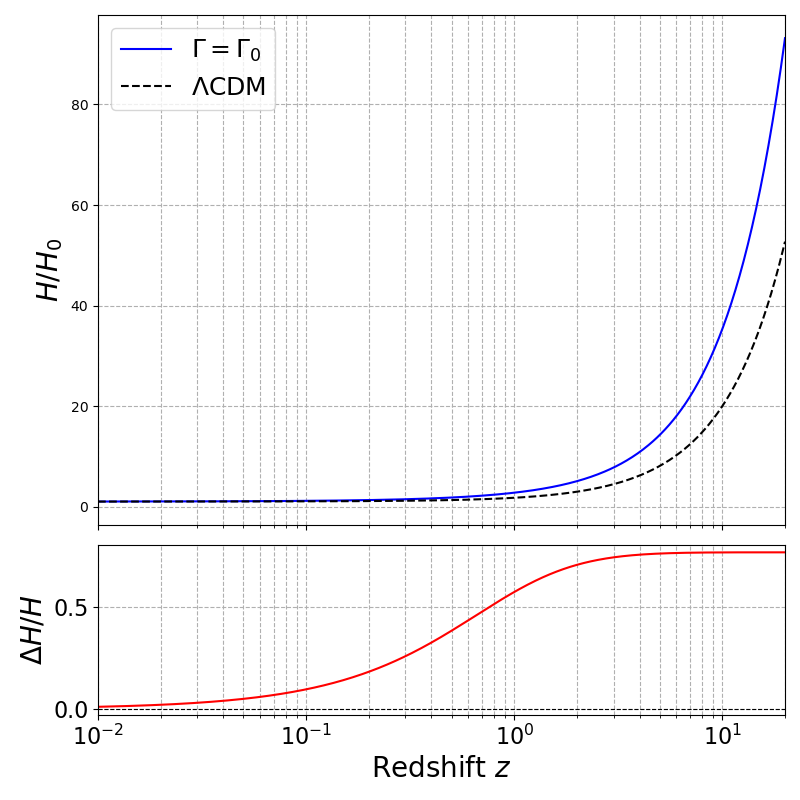}
    \caption{The redshift evolution of $H/H_0$ for the matter creation model $\Gamma= \Gamma_0$ and the $\Lambda$CDM model (upper panel) and the fractional difference $\Delta H/H = (H_{\rm Model} - H_{\Lambda {\rm CDM}})/H_{\Lambda {\rm CDM}}$ (lower panel) have been shown. For the matter creation model we have taken $w= 0$, $\alpha~(= \Gamma/H_0) = 0.1$, $\Omega_{f0} =0.04$, while for the $\Lambda$CDM model we set the matter density parameter at present $\Omega_{m0} =0.3$. }    
    \label{fig:exp-hist-M1}
\end{figure}

{\bf When $w > 0$}, we obtain five critical points, namely, $A_0$, $A_1$, $A_2$, $A_3$ and $A_4$ which always exist in our region $R$. Eigenvalues evaluated at $A_0$ are $3w$, $\frac{3}{2}(1+w)$ which are positive, hence, this indicates that $A_0$ is an unstable critical point dominated by the normal fluid  and it represents a decelerating phase of the universe. At $A_1$, the eigenvalues are $-3w$, $\frac{3}{2}$, 
that means one eigenvalue is positive while the other is negative (since $w >0$).
So, the critical point $A_1$ is saddle in nature which corresponds to the DM dominated decelerating phase. The critical point $A_2$ has one eigenvalue $0$. So, we cannot say about stability by applying the linear stability theory and that is why we have applied the center manifold theory. The center manifold is $x=0$ which coincides with the eigenvector ($z$ axis) corresponding to a zero eigenvalue. Clearly, $z'$ is positive along $z$ axis which gives $A_2$ another saddle point. As the decelerating parameter is undefined at $A_2$, hence, we are unable to predict whether this represents an accelerating phase or the reverse. 
Since, $A_3$ has two eigenvalues $-\alpha$, $\frac{\alpha}{2}$ and they are of opposite signs, we can conclude that it is a saddle point. This critical point is a DM dominated accelerated solution. The most interesting critical point is $A_4$ which has $\Omega_{dm}^{\rm eff}=1$ and $q=-1$, i.e. DM dominated accelerating solution. As both the eigenvalues are negative, the critical point $A_4$ is the only globally stable point in our domain which is shown in the upper left plot of Fig. \ref{fig1A} {where all orbits leave the decelerating phase, then enter into an accelerating phase and finally end in a DM dominated accelerating phase $(q=-1)$, as evidenced in the upper left plot in Fig. \ref{fig-evo-1}. Besides these features we observe that a set of orbits crosses the $q=-1$ curve, enter into a super accelerating phase $(q<-1)$ and finally finish with $q=-1$ phase, which indicates the  slowing down of the cosmic acceleration. Additionally, the effective EoS of DM, $w_{\rm dm}^{\rm eff}$ as displayed in the upper left graph of Fig. \ref{fig-evo-1}, becomes negative and approaches $-1$. That means, it behaves like a DE fluid.

{\bf For $w = 0$}, besides the three isolated critical points $A_2,~A_3$ and $A_4$  given in Table~\ref{first-table}, we find one critical line $A_{7}(x_c,0)$ for which the critical points $A_0$, $A_1$ lie in this critical line. In the phase space $R$, one can obtain $z'>0$ and $z'<0$ for $z<\frac{3}{3+\alpha x}$ and for $z>\frac{3}{3+\alpha x}$ respectively, and $x'$ takes always positive value (since $\alpha>0$). By the above argument, one can conclude that the critical line $A_{7}$ shows unstable nature, $A_1,~A_2$ behave like saddle points and $A_3$ is the only globally stable point. Here, $A_3,~A_4$ correspond to accelerating phase and $A_{7}$ gives decelerating phase, while decelerating parameter is undefined at $A_2$. The phase space diagram is highlighted in the upper middle plot of the Fig. \ref{fig1A} where one can notice that the universe exits from its past decelerating phase and enters into the DM dominated accelerating phase, as depicted in the upper right plot in Fig. \ref{fig-evo-1}. We also notice that $w_{\rm dm}^{\rm eff}$ becomes negative and approaches $-1$, as noticed in the earlier case with $w>0$. Along with these results, a set of orbits traverses the decelerating phase $(q>0)$, then enters into an accelerating phase with $-1<q<0$ but successively these orbits  enter into a super accelerating phase $(q<-1)$, and finally the acceleration slows down and it continues with $q=-1$.

{\bf When $-\frac{1}{3} \leq  w < 0$,} in this interval of $w$, the dynamical system (\ref{RG-DS-1}) gives five critical points, namely, $A_0,~A_1,~A_2,~A_3$ and $A_4$. Now we can conclude about the stability of the critical points via investigation of the sign of the eigenvalues and the direction of the flow on the boundary of the unit square. It is clear that $A_0,~A_2,~A_3$ are saddle by behavior and $A_1$ is unstable. Here, $A_4$ is the only globally stable point in the phase space. The critical points $A_3,~A_4$ correspond to matter dominated accelerating phases, while $A_1$ represents the matter dominated decelerating phase. Note that, $A_0,~A_2$ both are second fluid  dominated critical points but again $A_0$ shows deceleration and we can not say about the acceleration at $A_2$. The phase space trajectories are similar to the upper right plot of the Fig. \ref{fig1A}. In this case, the universe quits its past DM dominated decelerating phase and closes in accelerating DM dominated phase. Again, some special orbits tracing accelerating phase $(-1<q<0)$ and super accelerating phase $(q<-1)$, finish in the phase where $q=-1$. The slowing down of cosmic acceleration is also realized in this case. 

{\bf When $-1< w < -\frac{1}{3}$,} this case is similar to the above case where $w$ lies in the interval $\left[-\frac{1}{3},0\right)$. The only difference with the above case is in the decelerating parameter where in the above case we get $q\geq 0$ at $A_0$ but here, $q<0$ at $A_0$ which represents acceleration. The upper right plot of Fig. \ref{fig1A} exhibits the qualitative features of the critical points. 

{\bf When $w = -1$,} in this case, we have two critical lines $A_5(0,z_c)$ and $A_6\left(x_c,\frac{3}{3+\alpha}\right)$ along with the critical points $A_0,~A_1,~A_2,~A_3,~A_4$ depicted in the Table \ref{first-table}. Here, $A_0,~A_2$ belong to the critical line $A_5$ and $A_6$ contains the point $A_4$. In the phase space above the $z=3/(3+\alpha)$ line, $x'$, $z'$ are positive and negative respectively, and below the line $z=3/(3+\alpha)$, $x$ and $z$ are decreasing and increasing respectively. Again, the separatrix $z=\frac{3}{\alpha}(1-z)\sqrt{1-x}$ connecting the critical points $A_1$ and $\left(0,3/(3+\alpha)\right)$ divide the phase space below the line $z= 3/(3+\alpha)$ into two regions: the orbits below the separatrix converge to the part of the critical line $A_5$ where $ z < \frac{3}{3+\alpha}$ and the orbits above the separatrix approach the critical line $A_6$. As a result, $A_6$ behaves like an accelerating late-time attractor where DM and DE both may co-exist. So, the critical point $A_6$ has the potentiality to alleviate the coincidence problem. Again, the critical line $A_5$ where $z>\frac{3}{3+\alpha}$ are unstable in nature but the part of the critical line $A_5$ with $z<\frac{3}{3+\alpha}$ represent late time accelerating attractor dominated by the DE only. The point $A_3$ is saddle which shows matter dominated accelerated phase, while $A_1$ corresponds to matter dominated decelerating unstable point. The qualitative behavior are highlighted in the lower left plot of Fig. \ref{fig1A}. Here, we have two possibilities: one is, above the separatrix, our universe shows the transition from its past DM dominated decelerating phase to the accelerating scaling solutions and on the other hand, below the separatrix, the universe follows DM dominated decelerating phase to completely DE dominated accelerating phase. The lower left plot in Fig. \ref{fig-evo-1} highlights the dynamic behavior of the DM density parameter $\Omega_{\rm dm}^{\rm eff}$, DE density parameter $\Omega_{f}$, decelerating parameter $q$ and $w_{\rm dm}^{\rm eff}$ as functions of $e$-folding number $N=ln(a)$, reflecting key features of the model’s expansion history. One can notice that the effective EoS of DM being slightly negative during the late-evolution of the universe stays close to zero.

{\bf When $w < -1$,} we again obtain five critical points which are shown in Table~ \ref{first-table}, i.e. $A_0,~A_1,~A_2,~A_3$ and $A_4$. Since the eigenvalues are both negative at $A_0$, this is a stable point that exhibits dark-energy dominated accelerated universe. Being two eigenvalues, positive at $A_1$, this point is unstable which leads to the matter dominated decelerating phase of the universe. Clearly, the points $A_0,~A_2$ properly describe the phase of the universe. Again, as eigenvalues are opposite in sign evaluated at $A_3,~A_4$, these two critical points are saddle by nature representing acceleration with completely matter domination. Clearly, $A_2$, dark-energy dominated point is unstable because one can easily check that $x'$, $z'$ are both positive on $z=1$ and $x=0$, respectively. The lower right plot of Fig. \ref{fig1A} exhibits the phase space behavior  where a class of orbits leave the decelerating phase and end in the super accelerating phase which is expected in the phantom cosmology, consistent with the trend shown in the lower right plot in Fig. \ref{fig-evo-1}. Here we notice that $w_{\rm dm}^{\rm eff}$ has a Gaussian like feature. In particularly,  $w_{\rm dm}^{\rm eff}$ attains a minimum around the present epoch and then approaches $0$ with the evolution of the universe.

Finally, in Fig. \ref{fig:exp-hist-M1} we present the redshift evolution of the normalized Hubble parameter $H/H_0$ for the matter creation model with constant creation rate $\Gamma=\Gamma_0$, alongside the corresponding prediction from the standard $\Lambda{\rm CDM}$ cosmology (upper panel). The fractional difference, quantified as $\Delta H/H = (H_{\rm Model} - H_{\Lambda {\rm CDM}})/H_{\Lambda {\rm CDM}}$, is also displayed in the lower panel of the Fig. \ref{fig:exp-hist-M1} for clarity. We notice that in the low-redshift regime, the constant matter creation model is very close to the $\Lambda$CDM prediction, however, deviations from the $\Lambda$CDM model appear at higher redshifts, where the effect of particle creation becomes significant. This behavior highlights the potential of matter creation models to leave distinguishable imprints on the cosmic expansion history.  

\subsection{Model: $\Gamma=\Gamma_0 H$}\label{model-2}

Here, we consider the particle creation rate, $\Gamma=\Gamma_0 H$ where $\Gamma_0$ is a positive constant. In the following we analyze its effects on the universe evolution for the two-fluid system.
 
For this model, using the cosmological equations and the dimensionless variables $x$, $z$ which are mentioned in (\ref{dimensionless-variables}), we obtain the autonomous dynamical system in terms of the dimensionless variables as
\begin{subequations} \label{DS-2}
\begin{align}
    x'=& \left(3w+\Gamma_0\right)x(1-x),   \label{DS-2-x} \\
    z'=& \frac{3}{2}z(1-z)\bigg[1+w(1-x)-\frac{\Gamma_0}{3}x \bigg], \label{DS-2-z} 
\end{align}
\end{subequations}
where $\Gamma_0$ is a positive constant. The dynamical system is free from singularity. One can clearly see that the above system has four invariant manifolds $x=0,~x=1,~z=0$ and $z=1$ which make the phase space domain ${\bf R}$, a positively invariant set. In this model the decelerating parameter takes the form $q=\frac{1}{2}\left[1+3w(1-x)-\Gamma_0 x\right]$ which does not contain the $z$ variable, and for  $\left(1-\Gamma_0\right)x+(1+3w)(1-x)<0$, the accelerating expansion of the universe is realized.  In the Table \ref{second-table}, we properly present the inherent property of the dynamical system (\ref{DS-2}) (i.e. critical points and different values of the cosmological parameters at these critical points). Since the evolution of $x$ variable significantly depends on the factor $\left(3w+\Gamma_0\right)$ for which we carry out the phase space analysis in three possible ways, such as: 

{\bf When $3w+\Gamma_0<0$}, in this case, one can notice from Table~\ref{second-table} that there are four isolated critical points $B_0,~B_1,~B_2,~B_3$ and two critical lines $B_4$(if $w=-1$), $B_5$ ~(if $\Gamma_0=3$). Here, at late times, $x$ goes to zero, and then, at late times, we will have $z'=\frac{3}{2}z(1-z)(1+w)$. Thus, the following sub-cases will appear: (a) when $w<-1,$ in this case, we can see that, for $\Gamma_0\neq 3,$ the system has only four isolated critical points, namely, $B_0,~B_1,~B_2,~B_3$. At late times, $z$ goes to zero, concluding that $B_0$ is a global attractor. Again, examining the eigenvalues, one can conclude that $B_1$ is saddle for $\Gamma_0>3$ and unstable for $\Gamma_0<3$, $B_2$ is always saddle in nature, $B_3$ is unstable if $\Gamma_0>3$ and saddle by behavior if $\Gamma_0<3$. Here, $B_0,~B_2$ always lie in the accelerated region while $B_1,~B_3$ remain in the accelerated phase if $\Gamma_0>1$. So we can conclude that $B_0$ is the late time DE dominated accelerating global attractor and universe undergoes from decelerating to accelerating phase if we choose the model parameter $\Gamma_0<1$. This scenario is depicted in the lower right plot of the Fig. \ref{fig2B} and in the right plot of the Fig. \ref{fig-evo-2}. On the other hand, for $\Gamma_0=3,$ along with two isolated critical points $B_0,~B_2$, we have one critical line $B_5$, containing the critical points $B_1,~B_3$. As one eigenvalue is positive and other is zero, $B_5$ behaves like unstable point. All the critical points and line are representing accelerating solutions. Once again, $B_0$ is the only late time DE dominated global attractor which we have seen in the previous paragraph.
When the EoS of the second fluid coincides with cosmological constant $(w=-1)$, one can get two critical points and one critical line, namely, $B_1,~B_3$ and $B_4$. Now, we have $z'=\frac{3}{2}z(1-z)x\left( 1-\frac{\Gamma_0}{3}\right)$. Taking into account that $1-\frac{\Gamma_0}{3}>0$, we deduce that $z$ is an increasing function, concluding that $B_4$ behaves like stable in nature. The line $B_4$ is always accelerated while $B_1,~B_3$ are accelerated for $\Gamma_0>1$ otherwise decelerated. Thus, one may conclude that the universe leaves early time matter dominated decelerated phase for $\Gamma_0<1$ and ends with completely DE dominated accelerated phase (see lower left plot of Fig. \ref{fig2B} and middle plot of Fig. \ref{fig-evo-2}). When  $-1<w<0$, in this interval of $w$, the system (\ref{DS-2}) provides four isolated critical points $B_0,~B_1,~B_2$ and $B_3$. In this situation, at late times, $z$ goes to $1$. Thus, $B_2$ is a global attractor. Here, one can deduce that $\Gamma_0<3$ for which, the nature of eigenvalues indicate that $B_0$ is saddle point, $B_1$ is unstable in $\Gamma_0<3$, and $B_3$ is saddle in $\Gamma_0<3$. The point $B_0,~B_2$ correspond to accelerating epoch for $w<-1/3$ and $B_1,~B_3$ represents accelerating epoch in $\Gamma_0>1$. If $w$ mimics the quintessential DE EoS (i.e. $-1<w<-1/3$) with $\Gamma_0<1$, we obtain the proper evolution of the universe. 
   
{\bf When $3w+\Gamma_0=0$,} $x'=0$, meaning that the trajectories are the straight lines $x=$ constant. The equation for $z$ becomes $z'=\frac{3}{2}z(1-z)(1+w)$ and the decelerating parameter takes the form $q=(1+3w)/2$ which implies that all points in the phase space are accelerated for $w<-1/3$ and decelerated for $-1/3<w<0$. When $w<-1,$ in this phantom region of $w$, we have extracted two critical lines, namely, $B_6$, $B_7$, where DE and DM coexist. Here, $z$ is decreasing, deducing that for $w<-1$, $z$ goes to $0$. Thus $B_6$ gives stable character and $B_7$ shows unstable qualitative behavior, which are verified using the eigenvalues (see Table \ref{second-table}).  When $w=-1$, i.e. $\Gamma_0=3$, we we have $z'=0$, that is, all the points of the phase space are critical points and they are accelerated. When $-1<w<0,$ once again we have found two critical lines $\left(B_6,B_7\right)$, having both DE and DM. From the evolution of $z$, one can conclude, $z$ is increasing i.e. $z$ goes to $1$. Therefore, $B_7,~B_6$ behave like an attractor and a repeller respectively. These are also checked by eigenvalues (see Table~\ref{second-table}). The phase plot is shown in the upper right plot of Fig. \ref{fig2B}.   

{\bf When $3w+\Gamma_0>0$,} in that case, at late times, $x$ goes to $1$, and, at late times, the equation for $z$ will become: $z'=\frac{3}{2}z(1-z)\left(1-\frac{\Gamma_0}{3}\right)$. Here, it will appear the following situations, such as, $0<\Gamma_0<3:$ Here, one can verify that $w>-1$. In that case, the dynamical system (\ref{DS-2}) produces four critical points $B_0,~B_1,~B_2,~B_3$. As $z'>0$, $z$ goes to $1$, and thus, $B_3$ is a global attractor. Again, $B_0$ is unstable and $B_1,~B_2$ are saddle by behavior. If $-1/3<w$ and $\Gamma_0>1$, the evolution scenario is at early phase, orbits leaves from second fluid dominated decelerating phase, reflecting the absence of a past DM dominated era, and ends in DM dominated accelerating phase (see upper left plot of Fig. \ref{fig2B} and left plot of Fig. \ref{fig-evo-2}). For $\Gamma_0=3$, one can easily find two isolated critical points $B_0,~B_2$ and one critical line $B_5$. Now, we will have $z'=\frac{3}{2}z(1-z)(1+w)(1-x)$. Taking into account that, $w>-1$, we conclude that,  $B_5$ behaves like a global attractor. From the sign of eigenvalues, we can determine the stability of $B_0,~B_2$ ($B_0$ is unstable and $B_2$ is saddle). Therefore, the final state of the universe is DM dominated accelerating phase. For $\Gamma_0>3$,  In that case, once again, there are four critical points $B_0,~B_1,~B_2,~B_3$ in the phase space. Since, $z$ goes to $0$, and thus, $B_1$ is a global attractor. From the point of view of the nature of eigenvalues, $B_3$ is always saddle, $B_0$ is unstable where $B_2$ is saddle, and $B_2$ is unstable where $B_0$ is saddle. Here, we can observe that $B_0,~B_2$ for $w<-1/3$ and $B_1,~B_3$ always, lie in the accelerating phase.

For this model, the effective EoS of DM is given by $w_{\rm dm}^{\rm eff}=-\Gamma_0/3$, which remains a constant negative value throughout the evolution (see Fig. \ref{fig-evo-2}). For $\Gamma_0=2$, it behaves like dark energy, whereas for $\Gamma_0=0.2$ and $0.1$, it remains slightly negative but close to zero. 

Fig. \ref{fig:exp-hist-M2} shows the redshift evolution of the normalized Hubble parameter $H/H_0$ for the matter creation model $\Gamma=\Gamma_0 H$, together with the standard $\Lambda{\rm CDM}$ prediction (upper panel). The fractional deviation, defined as $\Delta H/H = (H_{\rm Model} - H_{\Lambda {\rm CDM}})/H_{\Lambda {\rm CDM}}$, is plotted in the lower panel. We find that both models exhibit nearly identical behavior at low redshifts. At higher redshifts, however, the matter creation scenario departs significantly from $\Lambda{\rm CDM}$. These deviations illustrate how matter creation models can produce observationally testable signatures in the cosmic expansion history.

%%%%%%%   Table 2  %%%%%%%%
\begin{table*}[t]
\centering
%\resizebox{0.9\textwidth}{!}{%
	\begin{tabular}{|c c c c c c c c|}\hline\hline
{\bf Critical point} & {\bf Existence} & {\bf Eigenvalue} & {\bf Stability} & $\mathbf{\Omega_f}$ & $\mathbf{\Omega_{\rm dm}^{\rm eff}}$ & $\mathbf{q}$ & {\bf Acceleration} \\ \hline
%   &&&&&     \\

$B_0(0,0)$  & Always & $\left(3w+\Gamma_0,\frac{3}{2}(1+w)\right)$  & {\bf Stable} if $3w+\Gamma_0<0,~w\leq-1$   &  $1$  &  $0$  &  $\frac{1}{2}(1+3w)$  &  $w<-\frac{1}{3}$  \\ 
 &&& and if $3w+\Gamma_0=0,~w<-1$;  &&&&  \\ 
  &&& Otherwise Unstable or Saddle &&&&  \\ \hline
  
$B_1(1,0)$  & Always & $\left(-(3w+\Gamma_0),\frac{3-\Gamma_0}{2}\right)$   &  {\bf Stable} if $3w+\Gamma_0 > 0,~\Gamma_0\geq 3$    & $0$   &  $1$  &  $\frac{1}{2}(1-\Gamma_0)$  & $\Gamma_0>1$ \\
 &&& and if $3w+\Gamma_0 = 0,~\Gamma_0>3$; &&&&  \\   &&& Otherwise Unstable or Saddle &&&&  \\  \hline

 $B_2(0,1)$  & Always & $\left(3w+\Gamma_0,-\frac{3}{2}(1+w)\right)$   &  {\bf Stable} if $3w+\Gamma_0 < 0,~w\geq -1$    & $1$   &  $0$  &   $\frac{1}{2}(1+3w)$   & $w<-\frac{1}{3}$ \\
 &&& and if $3w+\Gamma_0 = 0,~w>-1$;  &&&&  \\  
 &&& Otherwise Unstable or Saddle &&&&  \\  \hline

 $B_3(1,1)$  & Always &  $\left(-(3w+\Gamma_0),\frac{\Gamma_0-3}{2}\right)$  &  {\bf Stable} if $3w+\Gamma_0 > 0,~\Gamma_0\leq 3$   & $0$   &  $1$  &  $\frac{1}{2}(1-\Gamma_0)$  & $\Gamma_0>1$ \\
 &&& and if $3w+\Gamma_0 = 0,~\Gamma_0<3$;  &&&&  \\ 
  &&& Otherwise Unstable or Saddle &&&&  \\  \hline

 $B_4\left(0,z_c\right)$  & $w=-1$ & $\left(\Gamma_0-3,0\right)$   &  {\bf Stable} if $\Gamma_0 < 3$ and   & $1$   &  $0$  &  $-1$  & Yes \\
 &&& Unstable if $\Gamma_0 > 3$ &&&&  \\  \hline

 $B_5\left(1,z_c\right)$  & $\Gamma_0=3$ & $\left(-3(1+w),0\right)$   &  {\bf Stable} if $w > -1$ and   & $0$   &  $1$  &  $-1$  & Yes \\
 &&& Unstable if $w < -1$ &&&&  \\  \hline

 $B_6\left(x_c,0\right)$  & $3w+\Gamma_0=0$ & $\left(0,\frac{3-\Gamma_0}{2}\right)$   &  {\bf Stable} if $\Gamma_0 > 3$ and   & $1-x_c$   &  $x_c$  &  $\frac{1}{2}(1+3w)$  & $w<-\frac{1}{3}$ \\
 &&& Unstable if $\Gamma_0 < 3$ &&&&  \\  \hline

 $B_7\left(x_c,1\right)$  & $3w+\Gamma_0=0$ & $\left(0,\frac{\Gamma_0-3}{2}\right)$   &  {\bf Stable} if $\Gamma_0 < 3$ and   &  $1-x_c$  &  $x_c$  &  $\frac{1}{2}(1+3w)$  & $w<-\frac{1}{3}$ \\
 &&& Unstable if $ \Gamma_0 > 3$ &&&&  \\  \hline

 $B_8\left(x_c,z_c\right)$  & $w=-1,~\Gamma_0=3$ & $(0,0)$   &  All points in the phase   & $1-x_c$   &  $x_c$  &  $-1$  & Yes \\
 &&& space are critical points &&&&  \\  \hline
%   &&&&&  \\

\hline\hline
\end{tabular}%
% }
\caption{Summary of the critical points, their existence, stability and the values of the cosmological parameters at those points for  the dynamical system (\ref{DS-2}) with the matter creation rate $\Gamma=\Gamma_0 H$. }
	\label{second-table}
\end{table*}

\begin{figure*}
    \centering
    \includegraphics[width=0.497\textwidth]{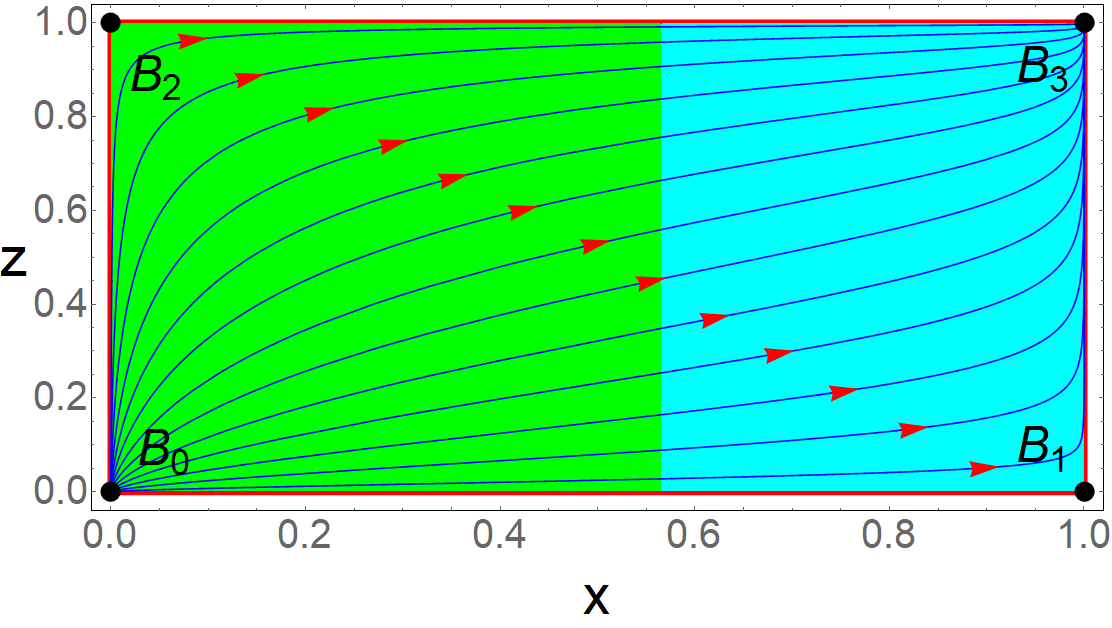}
    \includegraphics[width=0.497\textwidth]{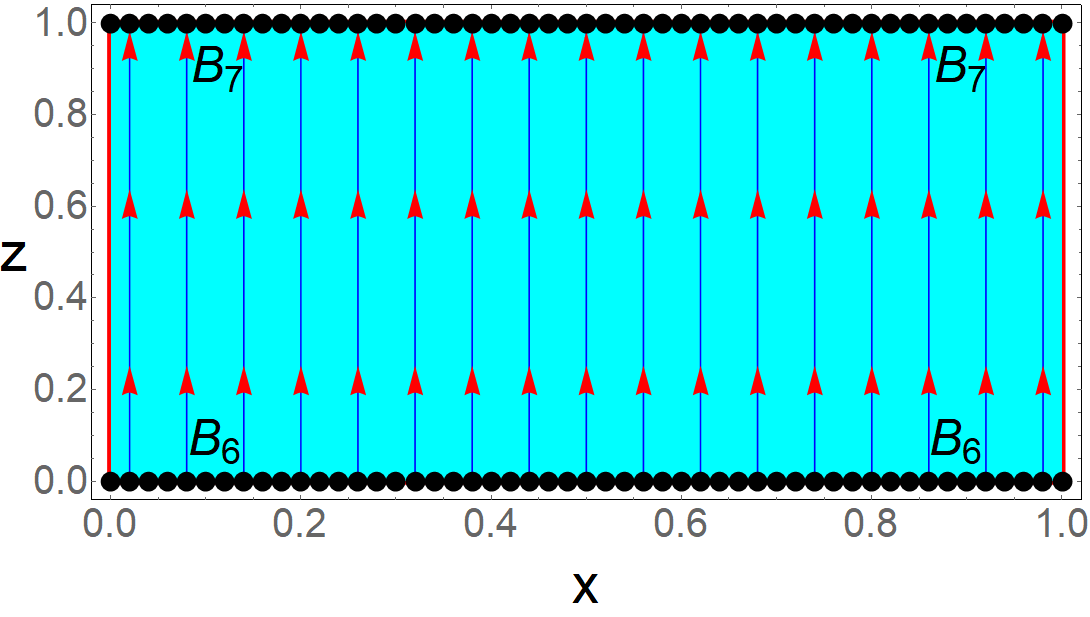} \\
    \includegraphics[width=0.497\textwidth]{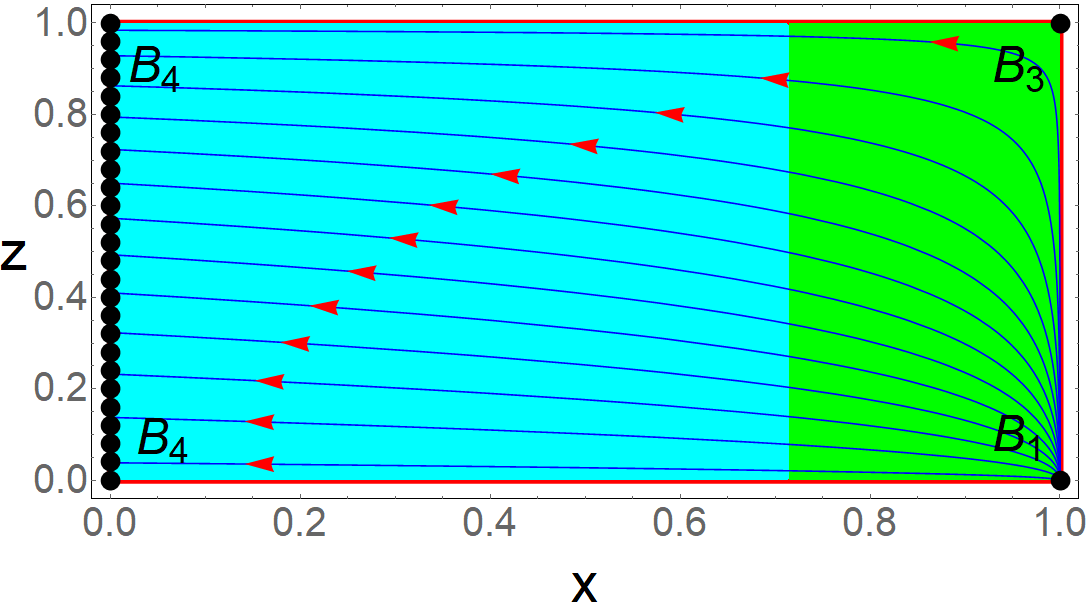}
    \includegraphics[width=0.497\textwidth]{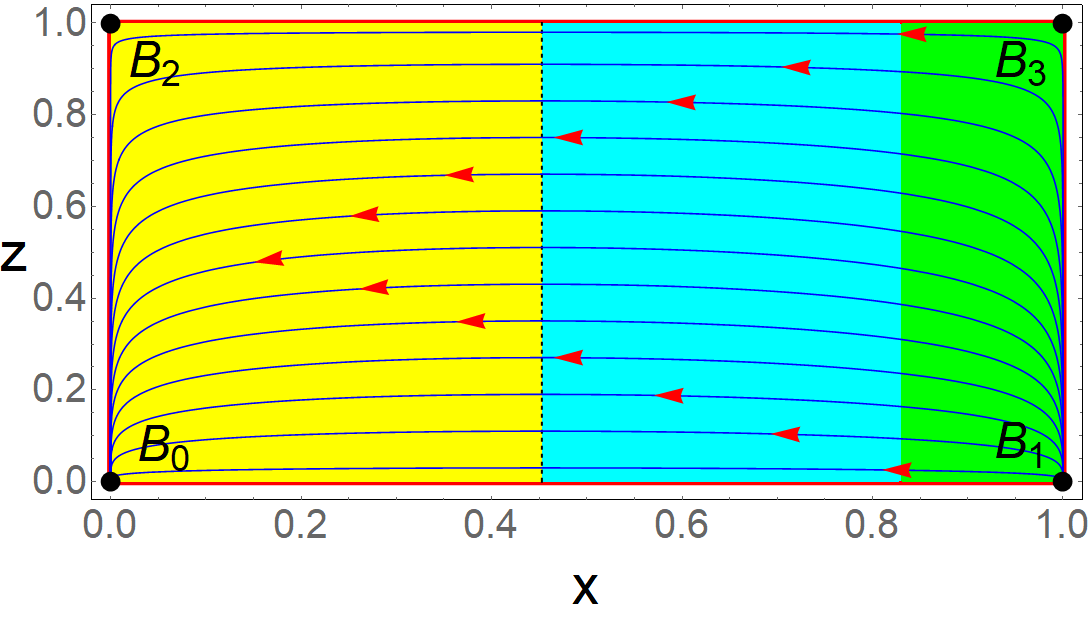}
    \caption{Description of the phase space of the two-fluid system for the matter creation rate $\Gamma = \Gamma_0 H$. {\bf Upper Left Plot:} The phase plot of the system (\ref{DS-2}) assuming $w=0.1$ and $\Gamma_0=2$. If we take any value of $w$ from $w>-1$ and $\Gamma_0$ in the region $0<\Gamma_0<3$, satisfying $3w+\Gamma_0>0$, we can get similar plot. {\bf Upper Right Plot:} The phase plot of the system (\ref{DS-2}) for $-1 <w <0$ satisfying $3w+\Gamma_0=0$. Here when we have assumed $w=-0.8$ and $\Gamma_0=2.4$, however, any value of $w$ lying in $(-1, 0)$ and any positive value of $\Gamma_0$ satisfying the above condition will give similar plot.  {\bf Lower Left Plot:} The phase plot of the system (\ref{DS-2}) when the EoS of the second fluid $w$ adopts the value $w=-1$. In this context, we choose $\Gamma_0=0.2$ but any positive value of $\Gamma_0$ satisfying $3w+\Gamma_0<0$ produces similar type of phase space diagram. {\bf Lower Right Plot:} This is the phase plot of the system (\ref{DS-2}) for $w <-1$ satisfying $3w+\Gamma_0<0$. Here we use  $w=-1.8$ and $\Gamma_0=0.1$ for drawing the graphics, however, any typical value of $w < -1$ and any positive value of $\Gamma_0$ $(\Gamma_0\neq 3)$ will give similar plot.  Here the green region corresponds to the decelerating phase ($q >0$), cyan region represents the accelerating phase with $-1<q<0$ and the yellow region corresponds to the super accelerating phase ($q <-1$). In the lower right graph, the vertical black dotted line corresponds to $q =-1$.   } 
    \label{fig2B}
\end{figure*}
\begin{figure*}
   \centering   
\includegraphics[width=0.33\textwidth]{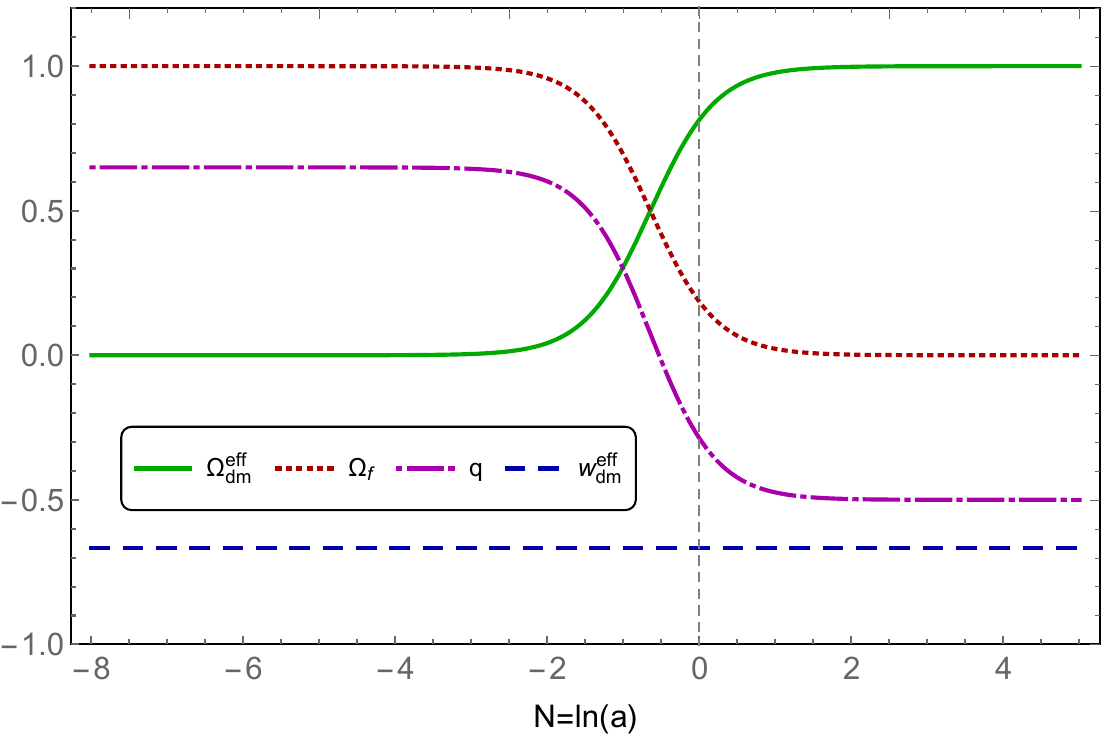} 
\includegraphics[width=0.33\textwidth]{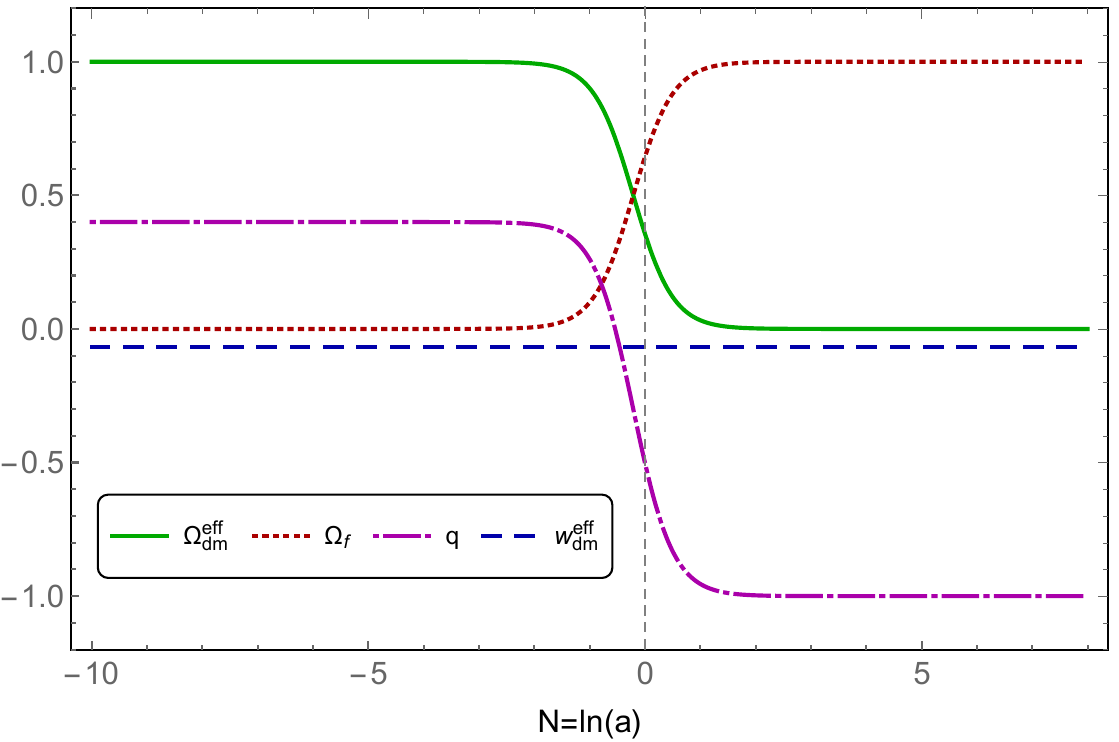}
\includegraphics[width=0.32\textwidth]{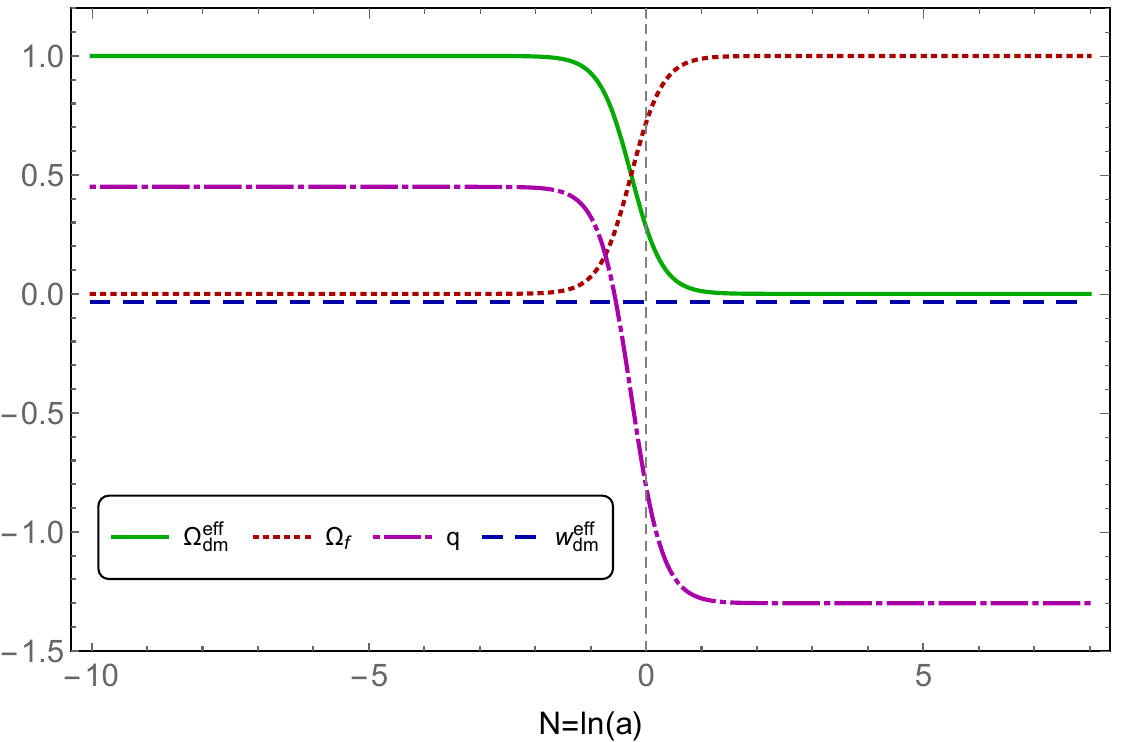} 
 \caption{We display the evolution plots of the DM density parameter $\Omega_{\rm dm}^{\rm eff}$, the second fluid density parameter $\Omega_f$, the deceleration parameter $q$, and the effective equation of state of DM $w_{\rm dm}^{\rm eff}$ corresponding to the dynamical system (\ref{DS-2}), with the matter creation rate $\Gamma=\Gamma_0 H$. {\bf Left Plot:} We have drawn this plot using the model parameters $w=0.1$ and $\Gamma_0=2$. Here, the trajectories end in DM dominated accelerated quintessence era. In this case, the past DM-dominated era can not be realized since no decelerating DM-dominated saddle point appears (see the upper left plot of Fig. \ref{fig2B}). {\bf Middle Plot:} This plot is produced adopting the values of the model parameters $w=-1$ and $\Gamma_0=0.2$ where the late time evolution of the universe is attracted by DE dominated cosmological constant era connecting through DM dominated decelerated era. {\bf Right Plot:} For this plot, we take $w=-1.2$ and $\Gamma_0=0.1$ where one can see that trajectories are attracted by DE dominated phantom era at late-times joining through a DM dominated decelerated phase.} 
    \label{fig-evo-2}
\end{figure*}
\begin{figure}
    \centering
    \includegraphics[width=0.497\textwidth]{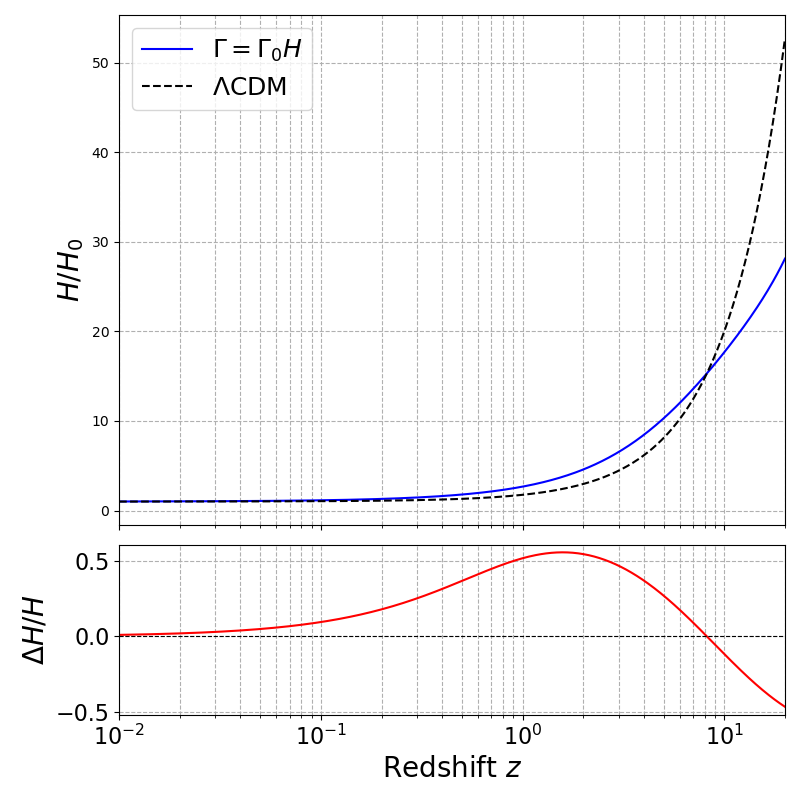}
    \caption{The redshift evolution of $H/H_0$ for the matter creation model $\Gamma= \Gamma_0 H$ and the $\Lambda$CDM model (upper panel) and the fractional difference $\Delta H/H = (H_{\rm Model} - H_{\Lambda {\rm CDM}})/H_{\Lambda {\rm CDM}}$ (lower panel) have been shown. For the matter creation model we have taken $w= 0$, $\Gamma_0 = 0.1$, $\Omega_{f0} =0.04$ while for the $\Lambda$CDM model we set the matter density parameter at present $\Omega_{m0} =0.3$.} 
    \label{fig:exp-hist-M2}
\end{figure}

\subsection{Model: $\Gamma=\Gamma_0 H^2$}\label{model-3}

In this case, we have constructed the model considering the particle creation rate $\Gamma=\Gamma_0 H^2$ where $\Gamma_0$ is a constant having dimension of the inverse of the Hubble rate. We shall also discuss how $\Gamma$ influences the universe's evolution.

In this case, the autonomous system takes the form
\begin{subequations} \label{DS-3}
\begin{align}
    x'=& x(1-x) \left(3w+\frac{\beta (1-z)}{z}\right),   \label{DS-3-x} \\
    z'=& \frac{3}{2}z(1-z)\left[1+w(1-x)-\frac{\beta x (1-z)}{3z} \right], \label{DS-3-z} 
\end{align}
\end{subequations}
where $\beta=\Gamma_0 H_0$ is a dimensionless and positive constant. To resolve the singularity at $z=0$ in the dynamical system (\ref{DS-3}), we redefine the time variable by setting $dN=zd\tau$. With this transformation, the reduced autonomous system, which remains topologically equivalent to system (\ref{DS-3}), takes the form: 
\begin{subequations} \label{RG-DS-3}
\begin{align}
    \frac{dx}{d\tau}=& x(1-x) \left(3w z+\beta (1-z)\right),   \label{RG-DS-3-x} \\
   \frac{dz}{d\tau}=& \frac{3}{2}z(1-z)\left[\left(1+w(1-x)\right)z-\frac{\beta x (1-z)}{3} \right]. \label{RG-DS-3-z} 
\end{align}
\end{subequations}
 Now, for the dynamical system (\ref{RG-DS-3}), it follows that $x=0,~x=1,~z=0,~z=1$ are invariant manifolds, which leads to the fact that the domain ${\bf R}$ is positively invariant.  Being $q=\frac{1}{2}\left[1+3w(1-x)-\frac{\beta x (1-z)}{z}\right]$ in this case, the requirement for realizing an accelerating phase at the present moment (i.e. $z= 1/2$), the physical trajectories should satisfy the relation $(1-\beta)x+(1+3w)(1-x)<0$. In the following we describe the qualitative features of the above dynamical system for different values of $w$. 

%%%%%%%   Table 3  %%%%%%%%
\begin{table*}[t]
\centering
%\resizebox{0.9\textwidth}{!}{%
	\begin{tabular}{|c c c c c c c c|}\hline\hline
{\bf Critical point} & {\bf Existence} & {\bf Eigenvalue} & {\bf Stability} & $\mathbf{\Omega_f}$ & $\mathbf{\Omega_{\rm dm}^{\rm eff}}$ & $\mathbf{q}$ & {\bf Acceleration} \\ \hline
%   &&&&&     \\

$C_{0}(0,0)$  & Always & $(\beta,0)$  & Unstable if $w > -1$;  &  1  &  0  & Undefined  & Undetermined   \\ 
 &&& Non-hyperbolic Saddle if $w<-1$  &&&&  \\  \hline
 
$C_{1}(1,0)$  & Always & $\left(-\beta,-\frac{\beta}{2}\right)$   &  {\bf Stable}   & 0   &  1  &  $-\infty$   & Yes \\  \hline
%   &&&&&  \\
$C_{2}(0,1)$  & Always  & $\left(3w,-\frac{3}{2}(1+w)\right)$   & {\bf Stable} if $-1 < w < 0$;   & 1   &  0  &  $\frac{1}{2}(1+3w)$  & $w<-\frac{1}{3}$ \\
 &&& Saddle if $w<-1$ and if $w>0$  &&&&  \\  \hline
%   &&&&&  \\
$C_{3}(1,1)$  & Always  & $\left(-3w,-\frac{3}{2}\right)$  & {\bf Stable} if $w > 0$;   &  0  & 1   &  $\frac{1}{2}$   & No \\
  &&& Saddle if $w<0$  &&&  &  \\ \hline
%   &&&&&  \\ 
$C_{4}\left(1,\frac{\beta}{\beta+3}\right)$  &  $\beta>0$ & $\left(-\frac{3\beta(1+w)}{3+\beta},\frac{3\beta}{2(3+\beta)}\right)$  &  Saddle if $w > -1$    &  0  &  1  &  $-1$  &  Yes     \\ 
  &&&  Unstable if $w < -1$ &&& &    \\ \hline
%   &&&&& \\
 $C_5\left(0,z_c\right)$  &  $w=-1$ & $\left(\beta-(\beta+3)z_c,0\right)$  &  {\bf Stable} if $z_c>\frac{\beta}{\beta+3}$  &  $1$  &  $0$  &  $-1$  &  Yes     \\ 
  &&&  Unstable if $z_c<\frac{\beta}{\beta+3}$ &&& &    \\  \hline
  
 $C_6\left(x_c,\frac{\beta}{\beta+3}\right)$  &  $w=-1$ & $\left(0,\frac{3\beta x_c}{2(3+\beta)}\right)$  &  Unstable  &  $1-x_c$  &  $x_c$  &  $-1$  &  Yes     \\ 
 \hline
 
 $C_{7}\left(x_c,1\right)$  &  $w=0$ & $\left(0,-\frac{3}{2}\right)$  &  {\bf Stable}  &  $1-x_c$  &  $x_c$  &  $\frac{1}{2}$  &  No     \\ \hline \hline
\end{tabular}%
% }
\caption{Summary of the critical points, their existence, stability and the values of the cosmological parameters at those points for the dynamical system (\ref{RG-DS-3}) with the matter creation rate $\Gamma=\Gamma_0 H^2$.
	   }
	\label{third-table}
\end{table*}

\begin{figure*}
    \centering
    \includegraphics[width=0.33\textwidth]{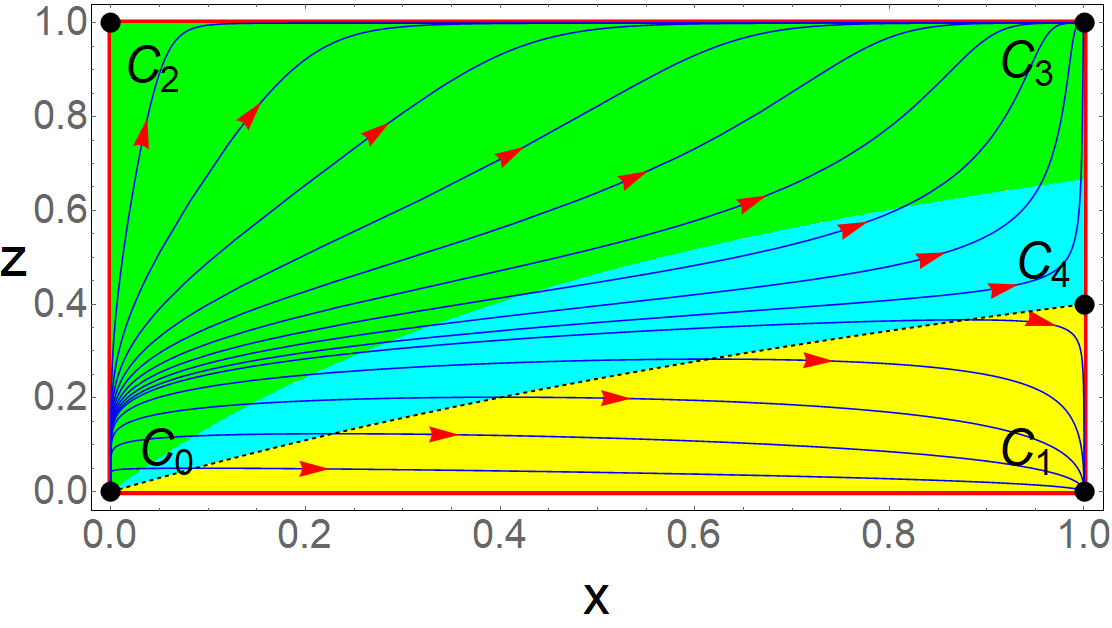}
    \includegraphics[width=0.33\textwidth]{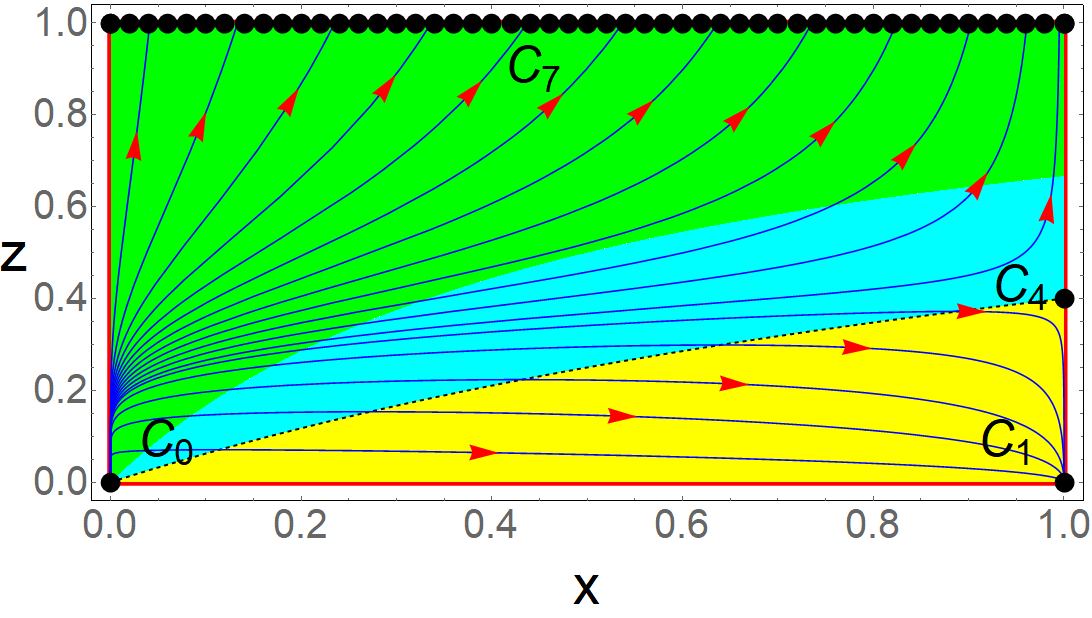}
    \includegraphics[width=0.32\textwidth]{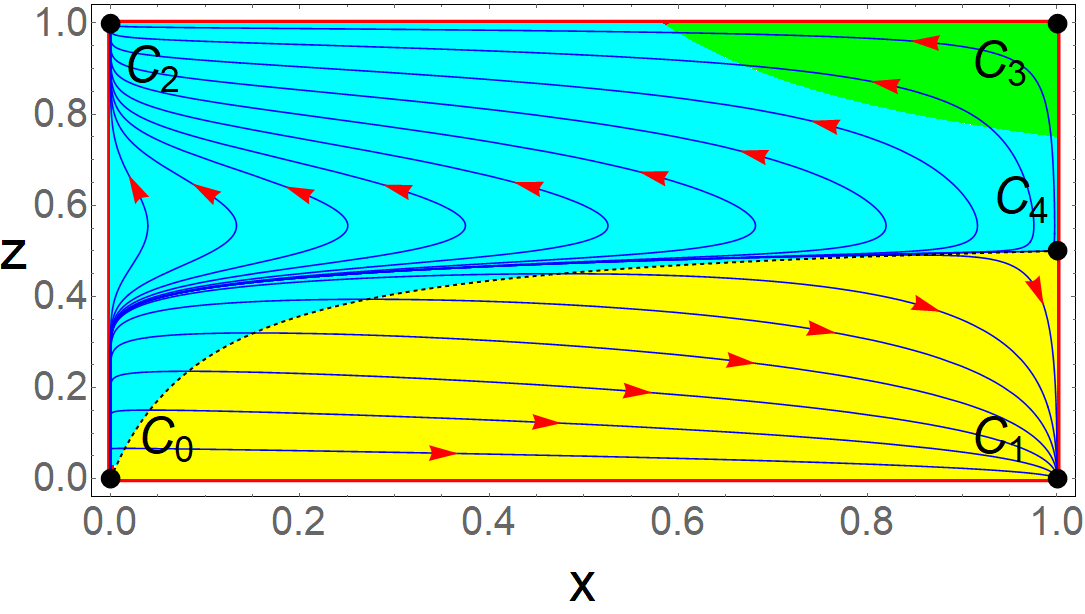}\\
    \includegraphics[width=0.33\textwidth]{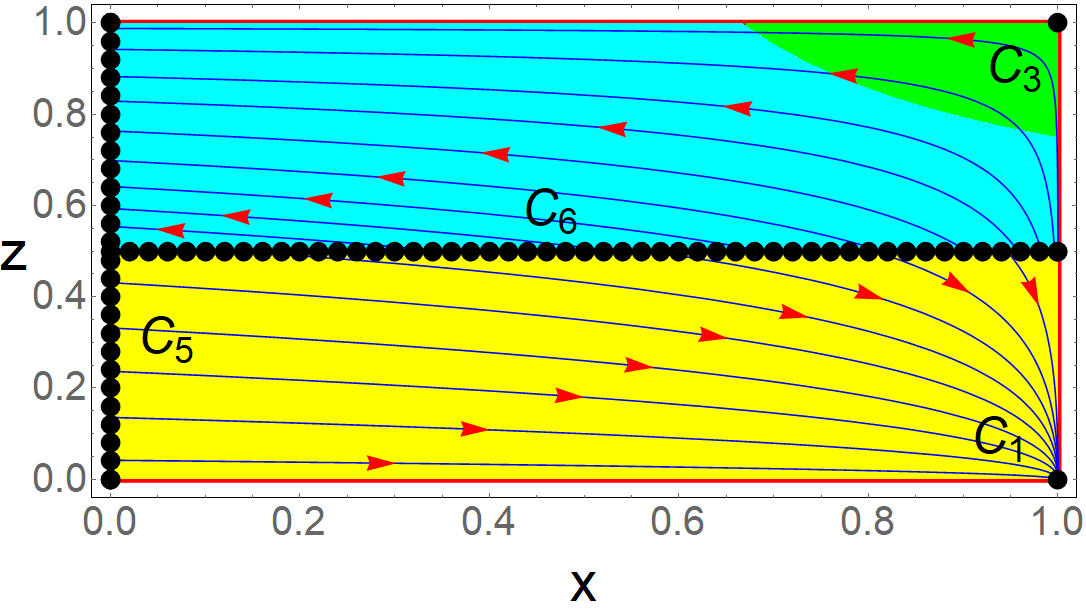}
    \includegraphics[width=0.33\textwidth]{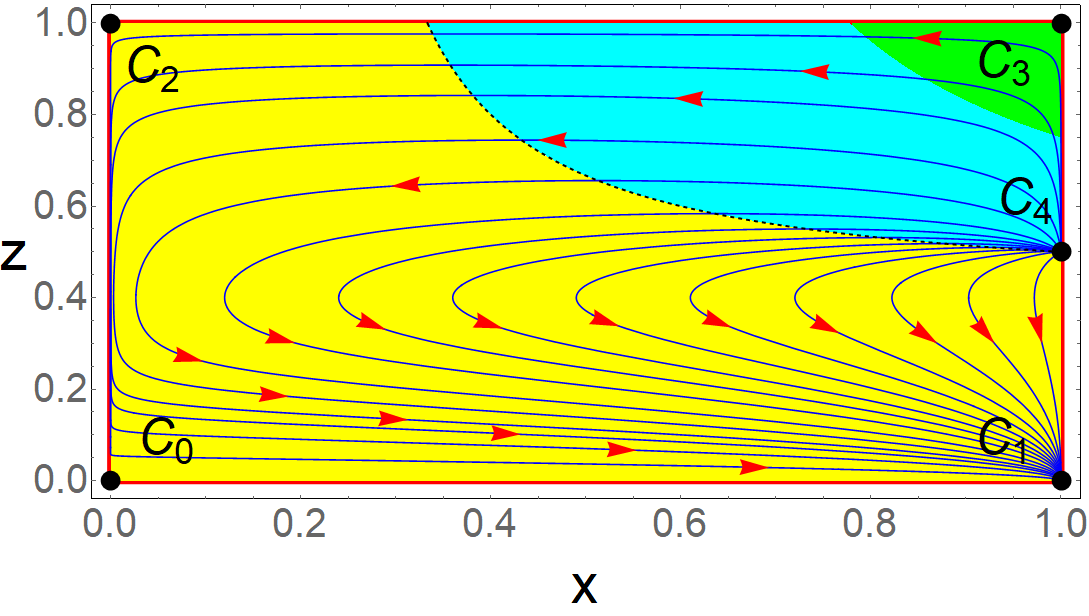}
    \caption{Description of the phase space for the matter creation rate $\Gamma = \Gamma_0 H^2$. {\bf Upper Left Plot:} The phase plot of the system (\ref{RG-DS-3}) when we have assumed $w=0.1$ and $\beta=2$. For other values of $w>0$ and $\beta>0$, we can also obtain similar phase space structure.  {\bf Upper Middle Plot:} The phase space of the system (\ref{RG-DS-3}) when the EoS $w$ takes the value $0$. Here we use $\beta =2$ but any positive value of $\beta$ gives same type of phase portrait. {\bf Upper Right Plot:} The phase plot of the system ($\ref{RG-DS-3}$) considering $w=-0.8$ and $\beta=3$. Also, we can get similar type of graphics for any positive value of $\beta$ and negative value of $w$ in the interval $(-1,0)$. {\bf Lower Left Plot:} The phase space of the system (\ref{RG-DS-3}) when the EoS $w$ takes the value $-1$. Here we use $\beta =3$ but any positive value of $\beta$ gives similar type of phase portrait. {\bf Lower Right Plot:} The phase plot of the system (\ref{RG-DS-3}) when we assume $w=-1.5$ and $\beta=3$. For other values of $w<-1$ and $\beta>0$, we can also obtain similar graphics. Here the green region corresponds to the decelerating phase ($q >0$), cyan region represents the accelerating phase with $-1<q<0$ and the yellow region corresponds to the super accelerating phase (i.e. $q <-1$). In all five graphs, the black dotted curve (in the lower left graph this curve is not visible because of the presence of the critical line) separating the cyan and yellow regions corresponds to $q =-1$.    }
    \label{fig3A}
\end{figure*}

\begin{figure*}
   \centering   
\includegraphics[width=0.497\textwidth]{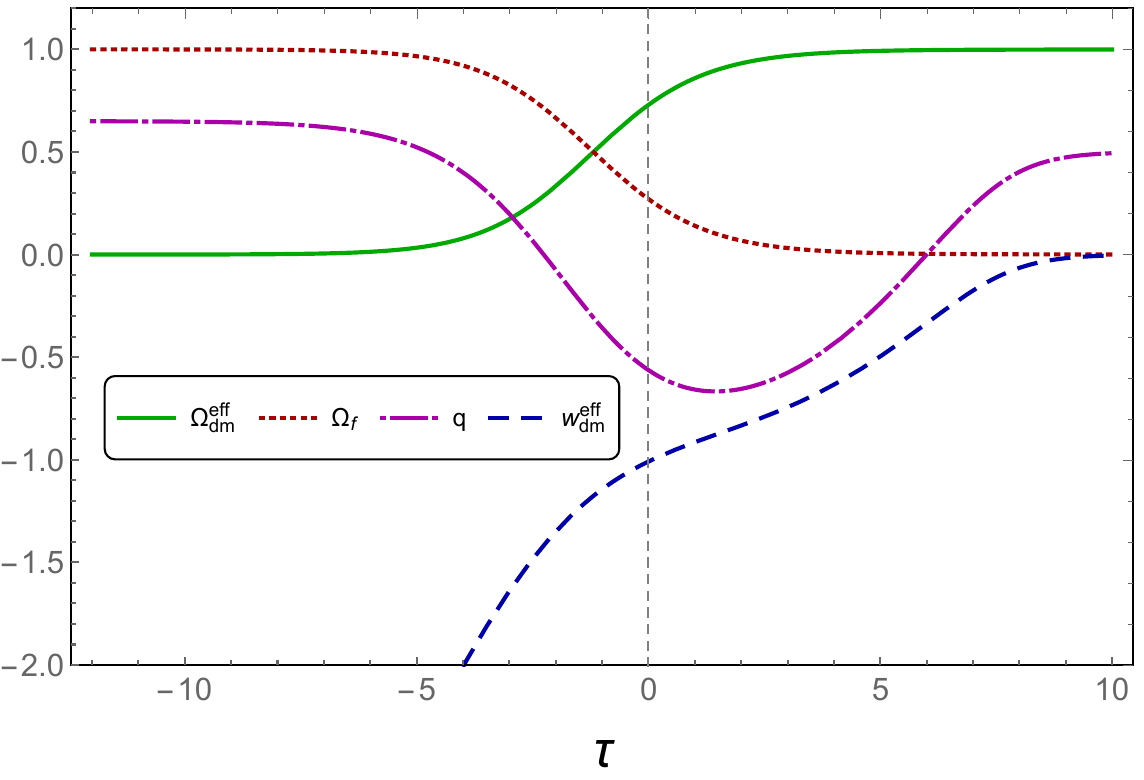} 
\includegraphics[width=0.497\textwidth]{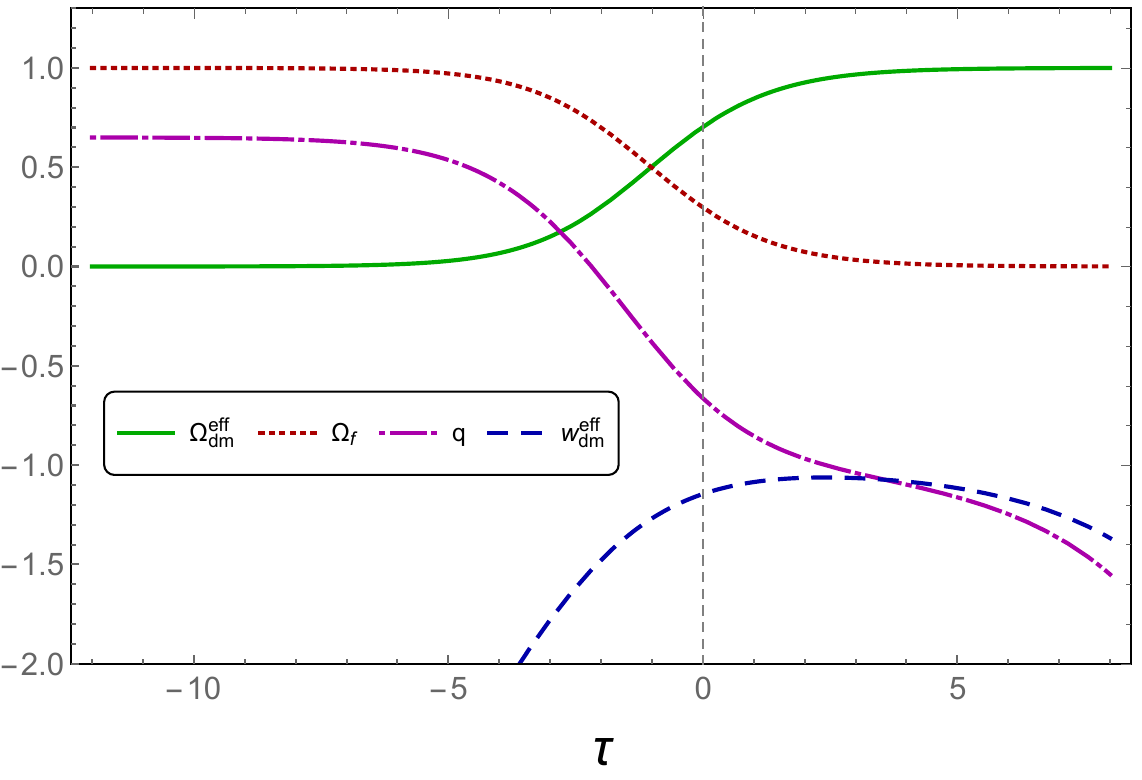}\\
\includegraphics[width=0.497\textwidth]{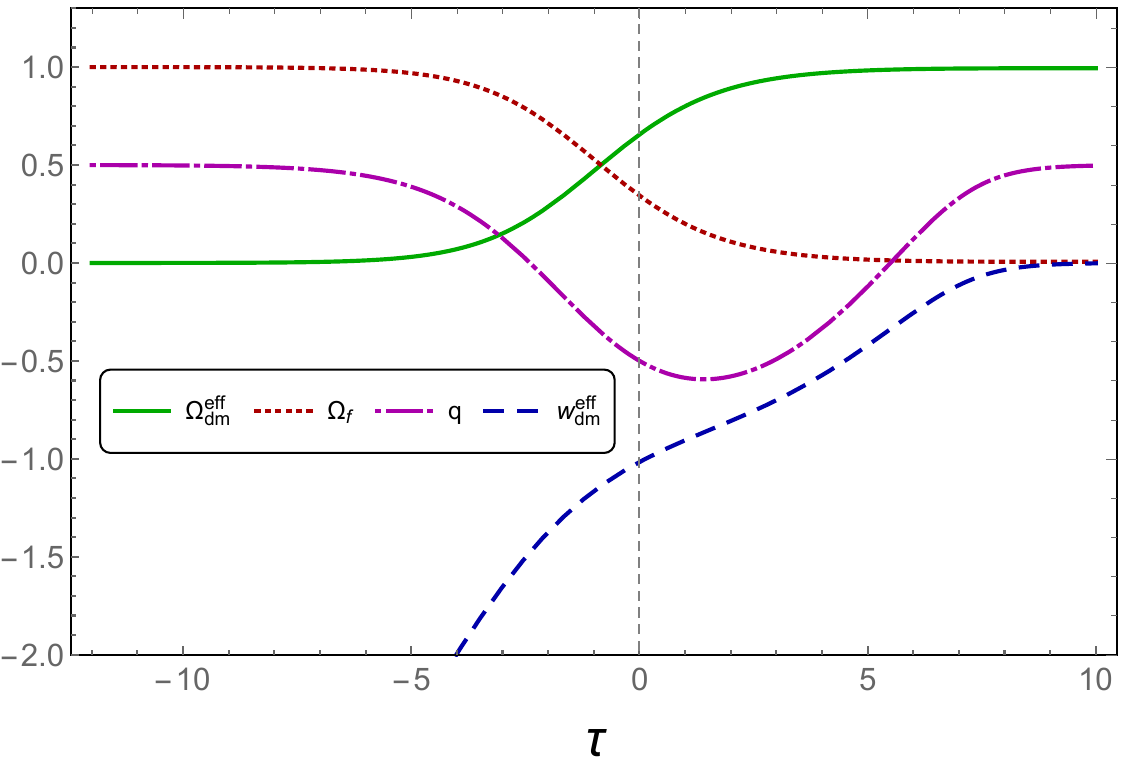} 
\includegraphics[width=0.497\textwidth]{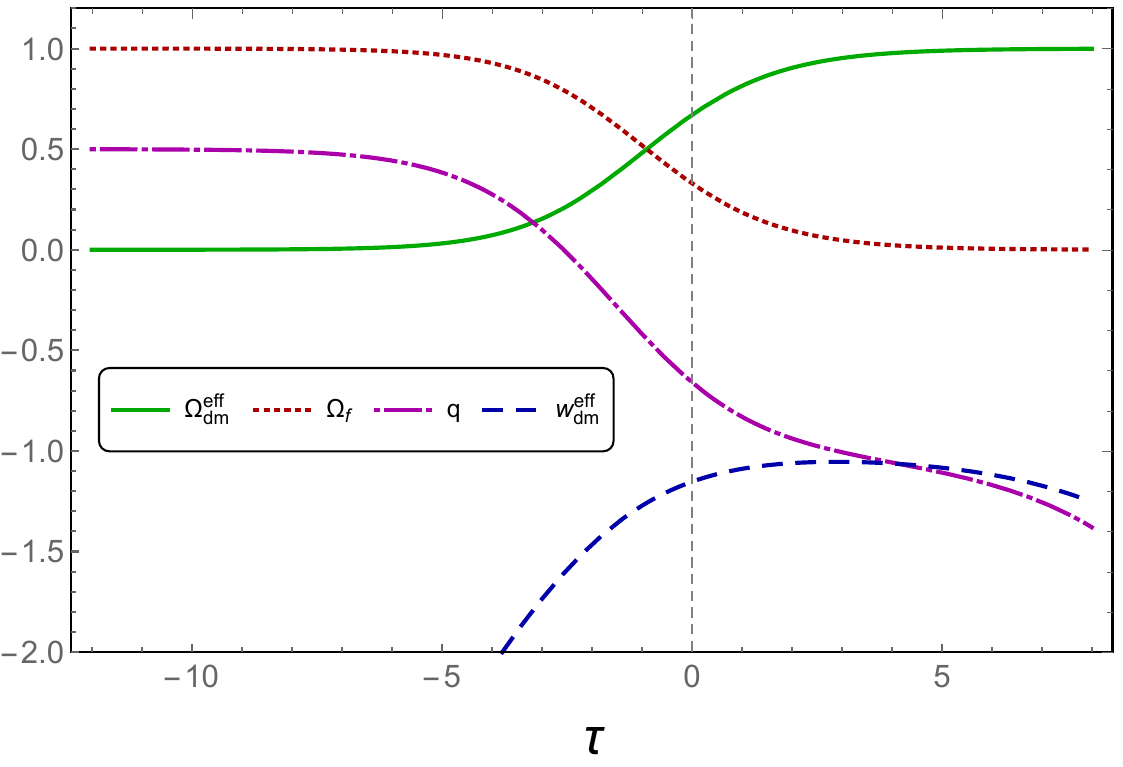}
    \caption{We have presented the evolution of key cosmological parameters namely, the DM density parameter $\Omega_{\rm dm}^{\rm eff}$, the second fluid density parameter $\Omega_f$, the decelerating parameter $q$, and the effective equation of state of DM $w_{\rm dm}^{\rm eff}$ for the dynamical system (\ref{RG-DS-3}) with the matter creation rate $\Gamma=\Gamma_0 H^2$. {\bf Upper Left Plot:} This plot is appeared for the model parameter $w=0.1$ and $\beta=1$. Numerical simulation shows that the final fate of the universe is attracted by DM dominated decelerated phase. {\bf Upper Right Plot:} This plot is also appeared for $w=0.1$, $\beta=1$, but the trajectories finish in a DM dominated phantom phase. {\bf Lower Left Plot:} For this plot, we have taken the model parameters values $w=0$ and $\beta=1$. Numerical simulation depicts that the ultimate fate of the universe is attracted in a decelerated phase. {\bf Lower Right Plot:} For this plot, we have also taken the model parameters values $w=0$, $\beta=1$, but the trajectories finish in a DM dominated phantom phase. In these scenarios, no decelerating DM-dominated saddle point is found (see upper left and middle plots of Fig. \ref{fig3A}), and hence the past DM-dominated phase can not be realized. Again, each orbit originates from the critical point $C_0$, where $w_{\rm dm}^{\rm eff}$ diverges to $-\infty$. Consequently, $w_{\rm dm}^{\rm eff}$ diverges during the early phase.
}
    \label{fig-evo-3}
\end{figure*}
\begin{figure}
    \centering
    \includegraphics[width=0.497\textwidth]{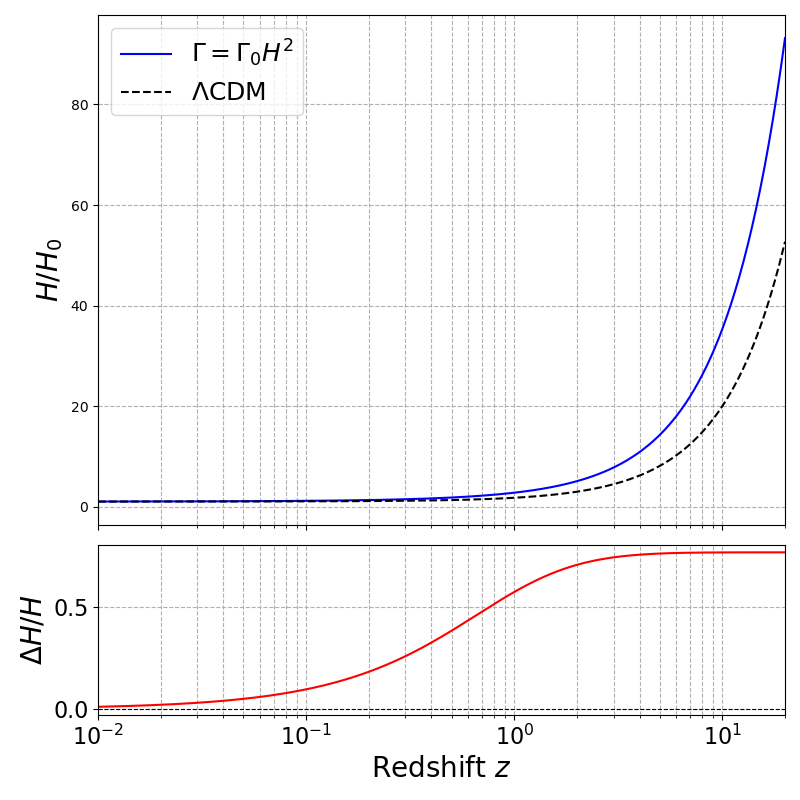}
    \caption{The redshift evolution of $H/H_0$ for the matter creation model $\Gamma= \Gamma_0 H^2$ and the $\Lambda$CDM model (upper panel) and the fractional difference $\Delta H/H = (H_{\rm Model} - H_{\Lambda {\rm CDM}})/H_{\Lambda {\rm CDM}}$ (lower panel) have been shown. For the matter creation model we have taken $w= 0$, $\beta~(= \Gamma_0H_0) = 0.1$,  $\Omega_{f0} =0.04$ while for the $\Lambda$CDM model we set the matter density parameter at present $\Omega_{m0} = 0.3$. }
    \label{fig:exp-hist-M3}
\end{figure}
    
{\bf When $w > 0$}, one can get five isolated critical points, namely, $C_0,~C_1,~C_2,~C_3,~C_4$ which are highlighted in the Table~\ref{third-table}. Now investigation of eigenvalues assure that $C_1,~C_3$ are stable points and $C_2,~C_4$ are saddle points. Again, eigenvectors corresponding to the eigenvalues $\beta,~0$ at the point $C_0$ lie on $x$ and $z$ axis respectively. Since, $x$ is increasing on $z$ axis and $z$ is increasing on $x$ axis, $C_0$ behaves like an unstable point. Here, $C_1,~C_4$ belong to the accelerating phase and $C_2,~C_3$ belong to the decelerating phase, while decelerating parameter is undefined at $C_0$. Now, the condition to be a viable trajectories are  the ones  that at $z=1/2$ the deceleration parameter should be negative. This implies 
$x\left(1- {\beta}    \right)+(1+3w)(1-x)<0
\Longrightarrow x>\frac{1+3w}{\beta+3w}
.$ The upper left plot of the Fig. \ref{fig3A} depicts the complete evolution of the phase space orbits. Here, note that one can trace a set of orbits for which the universe leaves the past decelerating phase, enters in the present (or at finite time) accelerating epoch and ends again in a DM dominated decelerating phase or a DM dominated super accelerating phase, which we have demonstrated in the upper left and right plots in Fig. \ref{fig-evo-3}. As seen in this plot, there is no past DM dominated epoch, and the evolution of $w_{\rm dm}^{\rm eff}$ which diverges in the early phase.  

{\bf When $w=0$}, we obtain three isolated critical points $C_0,~C_1,~C_4$ which are mentioned earlier and one critical line $C_{7}(x_c,1)$. Again, one can obtain $z'>0$, $z'<0$ and $x'>0$ according to $z>\frac{\beta x}{3+\beta x}$, $z<\frac{\beta x}{3+\beta x}$ and on $z=0$ line, respectively if we choose $\beta$ as positive parameter. Therefore, the above arguments indicate that $C_{7}$ gives stable like nature, $C_0$ is unstable, $C_1$ is stable and $C_4$ is saddle point. Note that $C_1,~C_4$ are accelerated points, but $C_{7}$ represents decelerating phase. The upper middle plot of the Fig. \ref{fig3A} highlights the qualitative behavior of this model. It is also noted that the universe leaves its past decelerating phase and then it enters into the current (or at finite time) accelerating phase and finally it again enters into a decelerating phase admitting scaling solutions or a super accelerating phase dominated by DM. This scenario is clearly noticed in the lower left and right plots in the Fig. \ref{fig-evo-3}. It is evident from this plot that no past DM dominated era occurs, and $w_{\rm dm}^{\rm eff}$ becomes divergent in the early phase.   
   
{\bf When $-\frac{1}{3} \leq  w < 0$},  five isolated critical points ($C_0,~C_1,~C_2,~C_3,~C_4$) always belong to our domain $R$. From the sign of the eigenvalues, one can conclude that $C_1,~C_2$ act like stable points and $C_3,~C_4$ are saddle by nature. As before with similar reason, $C_0$ shows unstable character. Depending on the decelerating parameter, $q$, the critical points $C_1,~C_4$ lie in the accelerating phase and $C_2,~C_3$ also lie in the decelerating phase, while we can say nothing about the acceleration of the point $C_0$. The phase plot is similar to the upper right plot of the Fig. \ref{fig3A}. Thus, the final state of the universe is second fluid dominated decelerating phase or DM dominated super accelerating epoch.

{\bf When $-1<w < -\frac{1}{3}$}, the qualitative behavior of the critical points given in the Table \ref{third-table} are exactly similar to the above scenario where $w$ lies in $-\frac{1}{3} \leq w <0$, which are exhibited in the upper right plot of the Fig. \ref{fig3A}. Note that, here the point $C_2$ represents acceleration. Here, the fate of the universe is completely DE dominated accelerating epoch  or DM dominated super accelerating phase.

{\bf When $w=-1$}, the dynamical system (\ref{DS-3}) gives two isolated critical points $C_1$, $C_3$ which are described in Table \ref{third-table}, and two extra critical lines, namely, $C_5(0,z_c)$ and $C_6\left(x_c,\frac{\beta}{\beta+3}\right)$. There are no restriction on existence of the critical points $C_1,~C_3$ and the critical line $C_5$ but the critical line $C_6$ belongs to our domain $R$ only when $\beta$ is taken to be positive. If we look into the evolution of $z$, one can easily find that $z'$ is positive and negative when we choose $z > \frac{\beta}{\beta+3}$ and $z<\frac{\beta}{\beta+3}$, respectively. Again, $x$ is increasing below the line $z=\frac{\beta}{\beta+3}$ and it is decreasing above the line $z=\frac{\beta}{\beta+3}$. As a result, $C_1$ is stable, $C_3$ is always saddle, $C_6$ gives unstable behavior, $C_5$ is stable for $z_c>\frac{\beta}{\beta+3}$ otherwise this line is unstable. Clearly, the critical point $C_1$ and the critical line $C_5,~C_6$ lead to the accelerated solution, while $C_3$ gives decelerated solution. The phase space structure are clearly visible in the lower left plot of Fig. \ref{fig3A}. Therefore, $C_5$ with $z_c>\frac{\beta}{\beta+3}$ represents completely DE dominated accelerating solution.  

{\bf When $w < -1$},  the critical points for the dynamical system (\ref{RG-DS-3}) are $C_0,~C_1,~C_2,~C_3,~C_4$ which all are isolated. Inspection of the eigenvalues and the direction of the flow on the $x,~z$ axes, we can easily say that $C_0,~C_2,~C_3$ are saddle by behavior, $C_4$ is unstable and $C_1$ is the only globally stable point. Here, $C_1,~C_3,~C_4$ are matter dominated solutions and $C_0,~C_2$ correspond to completely DE dominated solution. Again, $C_1,~C_2,~C_4$ are accelerated critical points and $C_3$ gives deceleration. The lower right plot of Fig. \ref{fig3A} shows the correct phase portrait. Although, we assume that the second fluid is phantom-like, however, we do not notice any DE dominated accelerating late time stable point, rather we find DM dominated super accelerating $(q<-1)$ late time stable point.

In Fig. \ref{fig:exp-hist-M3}, the normalized Hubble parameter $H/H_0$ is plotted for the matter creation model ($\Gamma=\Gamma_0H^{2}$) alongside $\Lambda{\rm CDM}$ (upper panel). The fractional deviation $\Delta H/H$ is shown below. We observe that the two models agree at low redshift, but at higher-redshift, deviations reveal the influence of particle creation.

\subsection{Model: $\Gamma=\Gamma_0 H^{-1}$}\label{model-4}
In this series of matter creation models, one of the interesting matter creation rate is $\Gamma=\Gamma_0 H^{-1}$, where $\Gamma_0$ is constant having dimension equal to the dimension of the square of the Hubble rate. In the following we describe the influence of this model on the two-fluid systems. 

For this matter creation model, the two dimensional autonomous system becomes  
\begin{subequations} \label{DS-4}
\begin{align}
    x'=& x(1-x) \left(3w+\frac{\gamma z^2}{(1-z)^2}\right),   \label{DS-4-x} \\
    z'=& \frac{3}{2}z(1-z)\left[1+w(1-x)-\frac{\gamma x z^2}{3(1-z)^2} \right], \label{DS-4-z} 
\end{align}
\end{subequations}
where $\gamma=\Gamma_0/H_0^2$ is  a dimensionless parameter and it is positive. In order to remove the singularity at $z=1$ of the above dynamical system, we introduce a new time variable $\tau$, defined through $dN=(1-z)^2d\tau$. The resulting autonomous system is topologically equivalent to (\ref{DS-4}) and is given by 
\begin{subequations} \label{RG-DS-4}
\begin{align}
    \frac{dx}{d\tau}=& x(1-x) \left(3w (1-z)^2+\gamma z^2\right),   \label{RG-DS-4-x} \\
    \frac{dz}{d\tau}=& \frac{3}{2}z(1-z)\left[\left(1+w(1-x)\right)(1-z)^2-\frac{\gamma x z^2}{3} \right]. \label{RG-DS-4-z} 
\end{align}
\end{subequations}
 If we look into the system (\ref{RG-DS-4}), we find that $x=0,~x=1,~z=0,~z=1$ are invariant manifolds. So, the domain ${\bf R}$ is positively invariant. For this model we obtain the decelerating parameter as $q=\frac{1}{2}\left[1+3w(1-x)-\frac{\gamma x z^2}{(1-z)^2}\right]$. Thus, at present time (i.e. $z=1/2$) to obtain an accelerating phase of the universe the viable orbits should maintain the relation $(1-\gamma)x+(1+3w)(1-x)<0$. The critical points and different cosmological parameter at these points are depicted in the Table~\ref{fourth-table}. Now we shall analyze the qualitative features of the system (\ref{RG-DS-4}) for different values of $w$.  
%%%%%%%   Table 4  %%%%%%%%
\begin{table*}[t]
\centering
%\resizebox{0.9\textwidth}{!}{%
	\begin{tabular}{|c c c c c c c c|}\hline\hline
{\bf Critical point} & {\bf Existence} & {\bf Eigenvalue} & {\bf Stability} & $\mathbf{\Omega_f}$ & $\mathbf{\Omega_{\rm dm}^{\rm eff}}$ & $\mathbf{q}$ & {\bf Acceleration} \\ \hline
%   &&&&&     \\

$D_{0}(0,0)$  & Always & $\left(3w,\frac{3}{2}(1+w)\right)$  & {\bf Stable} if $w < -1$;  &  1  &  0  &  $\frac{1}{2}(1+3w)$  &  $w<-\frac{1}{3}$  \\ 
 &&& Saddle if $-1< w < 0$;  &&&&  \\
  &&& Unstable if $w > 0$  &&&&  \\ \hline
 
$D_{1}(1,0)$  & Always & $\left(-3w,\frac{3}{2}\right)$   &  Saddle if $w > 0$;   & 0   &  1  &  $\frac{1}{2}$  & No \\
 &&& Unstable if $w < 0$ &&&&  \\  \hline
%   &&&&&  \\
$D_{2}(0,1)$  & Always  & $(\gamma,0)$   & Non-hyperbolic Saddle if $w > -1$;   & 1   &  0  & Undefined   &  Undetermined \\
 &&& Unstable if $w<-1$ &&&&  \\     \hline
%   &&&&&  \\
$D_{3}(1,1)$  & Always  & $\left(-\gamma,\frac{\gamma}{2}\right)$  & Always Saddle  &  0  & 1   &  $-\infty$  &  Yes \\  \hline
%   &&&&&  \\ 
$D_{4}\left(1,\frac{1}{1+\sqrt{\frac{\gamma}{3}}}\right)$  &  $\gamma>0$ & $\left(-\frac{\gamma(1+w)}{\left(1+\sqrt{\frac{\gamma}{3}}\right)^2},-\frac{\gamma}{\left(1+\sqrt{\frac{\gamma}{3}}\right)^2}\right)$  &  {\bf Stable} if $w > -1$    &  0  &  1  &  $-1$  &  Yes     \\ 
  &&&  Saddle if $w<-1$ &&& &    \\ \hline
%   &&&&& \\
  $D_5\left(0,z_c\right)$  &  $w=-1$ & $\left(\gamma z_c^2-3(1-z_c)^2,0\right)$  &  {\bf Stable} if $z_c<\frac{1}{1+\sqrt{\frac{\gamma}{3}}}$     &  $1$  &  $0$  &  $-1$  &  Yes     \\ 
  &&&  and Unstable if $z_c>\frac{1}{1+\sqrt{\frac{\gamma}{3}}}$ &&& &    \\ \hline
  
 $D_6\left(x_c,\frac{1}{1+\sqrt{\frac{\gamma}{3}}}\right)$  &  $w=-1$ & $\left(0,-\frac{\gamma x_c}{\left(1+\sqrt{\frac{\gamma}{3}}\right)^2}\right)$  &  {\bf Stable}     &  $1-x_c$  &  $x_c$  &  $-1$  &  Yes     \\   \hline
 
 $D_{7}\left(x_c,0\right)$  &  $w=0$ & $\left(0,\frac{3}{2}\right)$  &  Unstable     &  $1-x_c$  &  $x_c$  &  $\frac{1}{2}$  &  No     \\ \hline \hline
\end{tabular}%
% }
\caption{Summary of the critical points, their existence, stability and the values of the cosmological parameters at those points for the dynamical system (\ref{RG-DS-4}) with the matter creation rate $\Gamma=\Gamma_0 H^{-1}$.
	   }
	\label{fourth-table}
\end{table*}
\begin{figure*}
    \centering
    \includegraphics[width=0.33\textwidth]{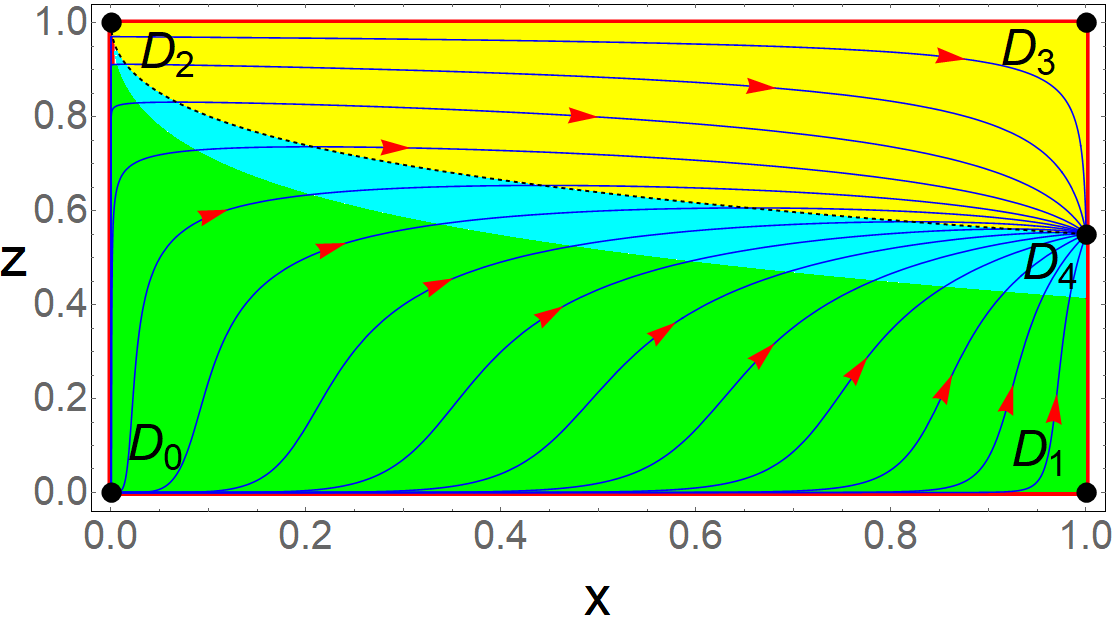}
    \includegraphics[width=0.33\textwidth]{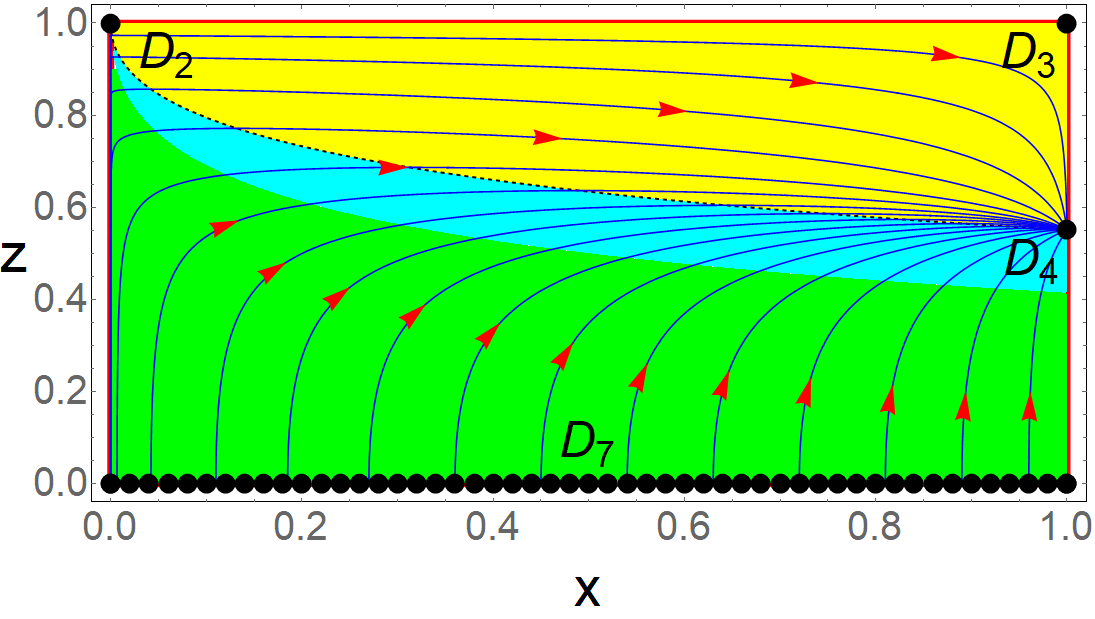}
    \includegraphics[width=0.32\textwidth]{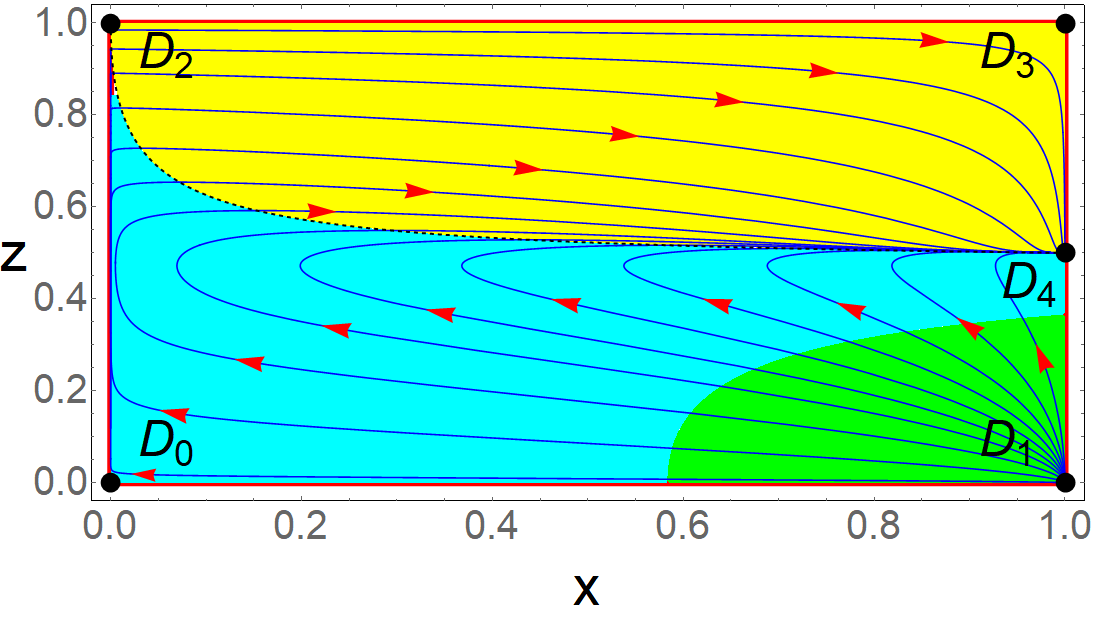}\\
    \includegraphics[width=0.33\textwidth]{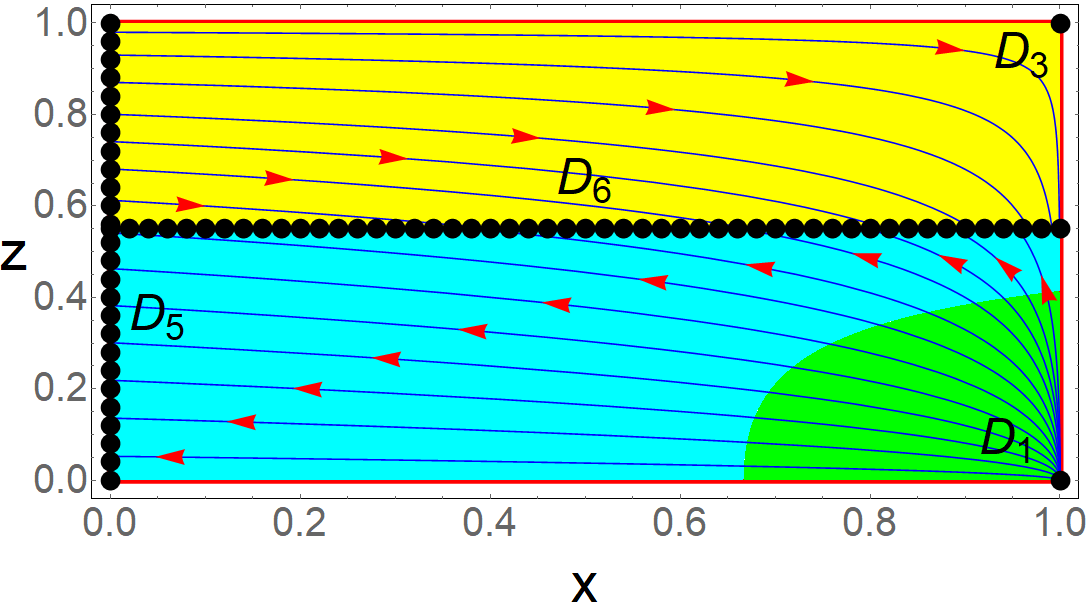}
    \includegraphics[width=0.33\textwidth]{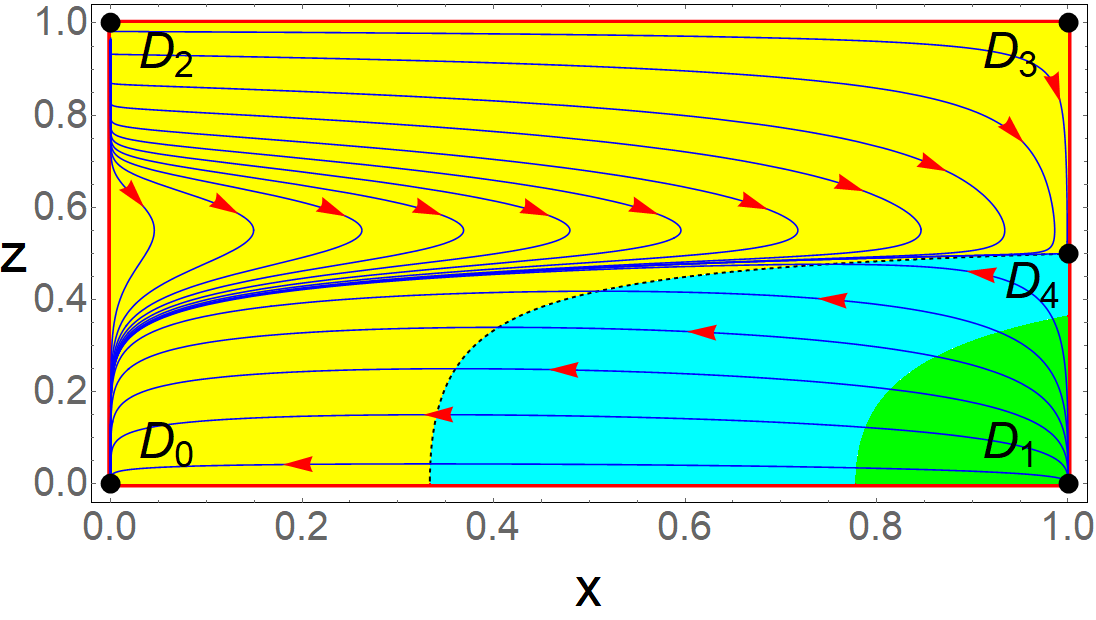}
    \caption{Description of the phase space controlled by the matter creation rate $\Gamma = \Gamma_0 H^{-1} $. {\bf Upper Left Plot:} The phase plot of the system (\ref{RG-DS-4}) when we have assumed $w=0.1$ and $\gamma=2$. For other values of $w>0$ and $\gamma>0$, we can also obtain similar phase space structure.  {\bf Upper Middle Plot:} The phase space of the system (\ref{RG-DS-4}) when the EoS $w$ takes the value $0$. Here we use $\gamma =2$ but any positive value of $\gamma$ gives same type of phase portrait. {\bf Upper Right Plot:} The phase plot of the system ($\ref{RG-DS-4}$) considering $w=-0.8$ and $\gamma=3$. Also, we can get similar type of graphics for any positive value of $\gamma$ and negative value of $w$ in the interval $(-1,0)$. {\bf Lower Left Plot:} The phase space of the system (\ref{RG-DS-4}) when the EoS $w$ takes the value $-1$. Here we use $\gamma =2$ but any positive value of $\gamma$ gives similar type of phase portrait. {\bf Lower Right Plot:} The phase plot of the system (\ref{RG-DS-4}) when we assume $w=-1.5$ and $\gamma=3$. For other values of $w<-1$ and $\gamma>0$, we can also obtain similar graphics. Here the green region corresponds to the decelerating phase ($q >0$), the cyan region represents the accelerating phase with $-1<q<0$ and the yellow region corresponds to the super accelerating phase (i.e. $q <-1$). The black dotted curve (in the lower left graph this curve is not visible because of the presence of the critical line) separating the cyan and yellow regions corresponds to $q =-1$.  }
    \label{fig4A}
\end{figure*}

\begin{figure*}
   \centering   
\includegraphics[width=0.497\textwidth]{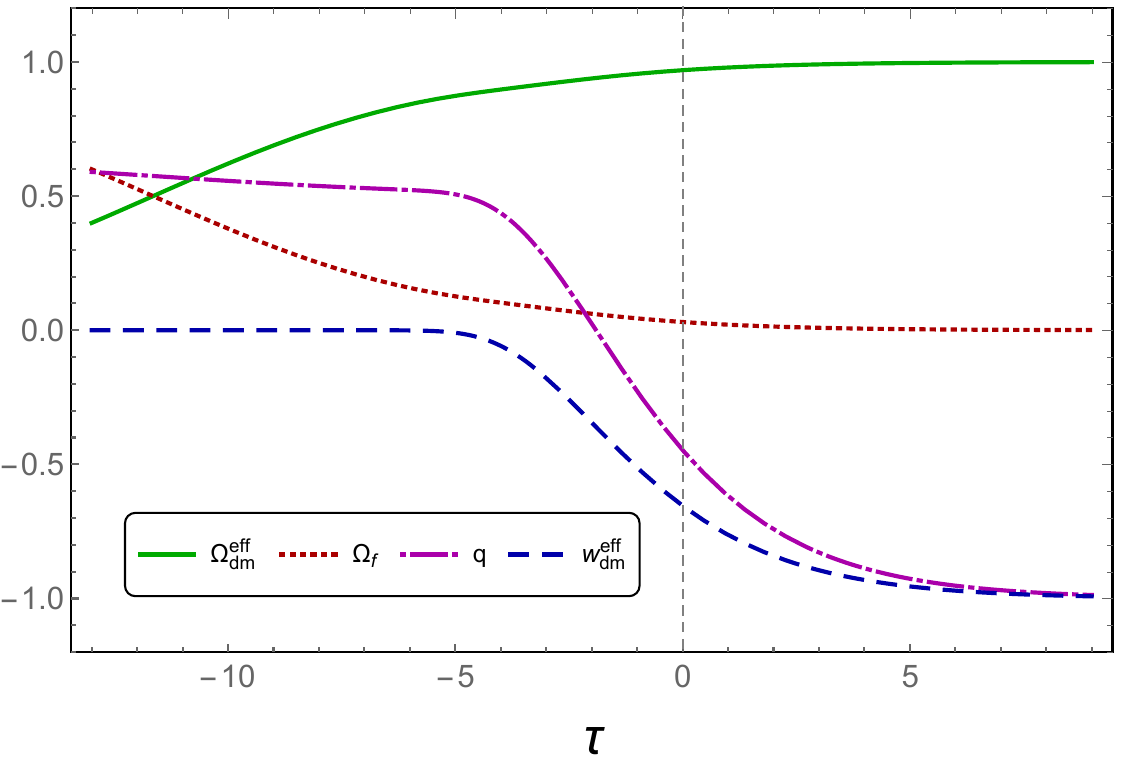} 
\includegraphics[width=0.497\textwidth]{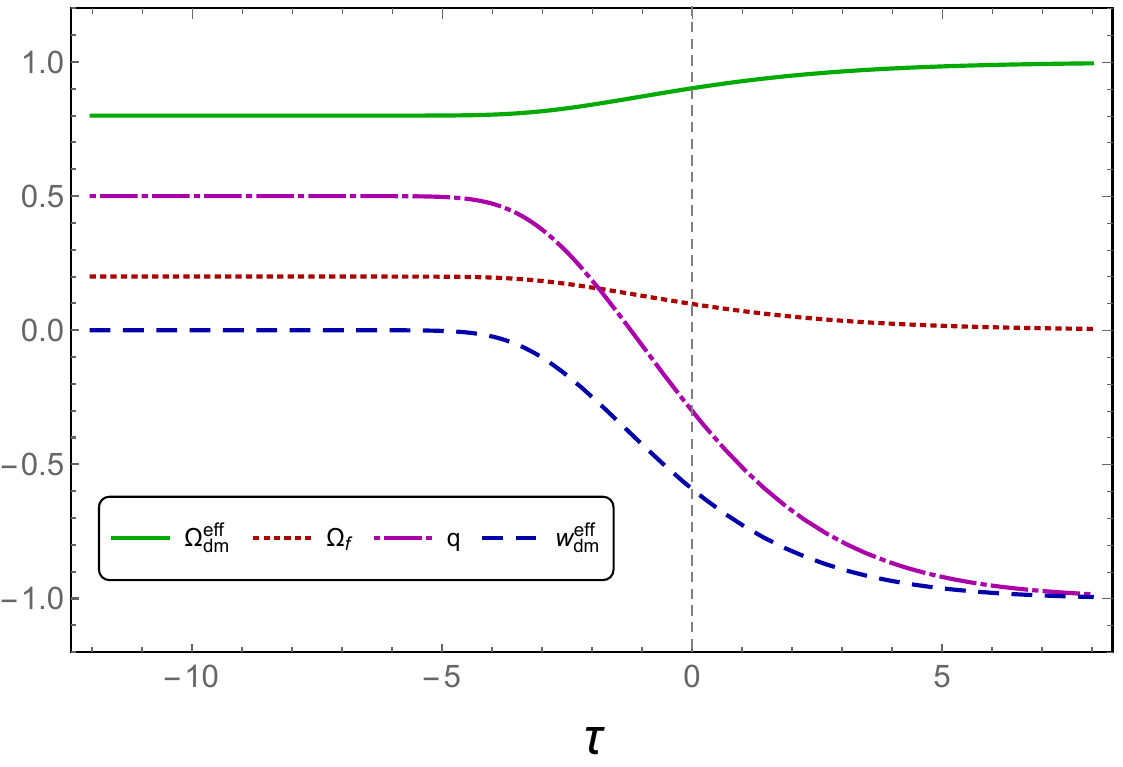}\\
\includegraphics[width=0.497\textwidth]{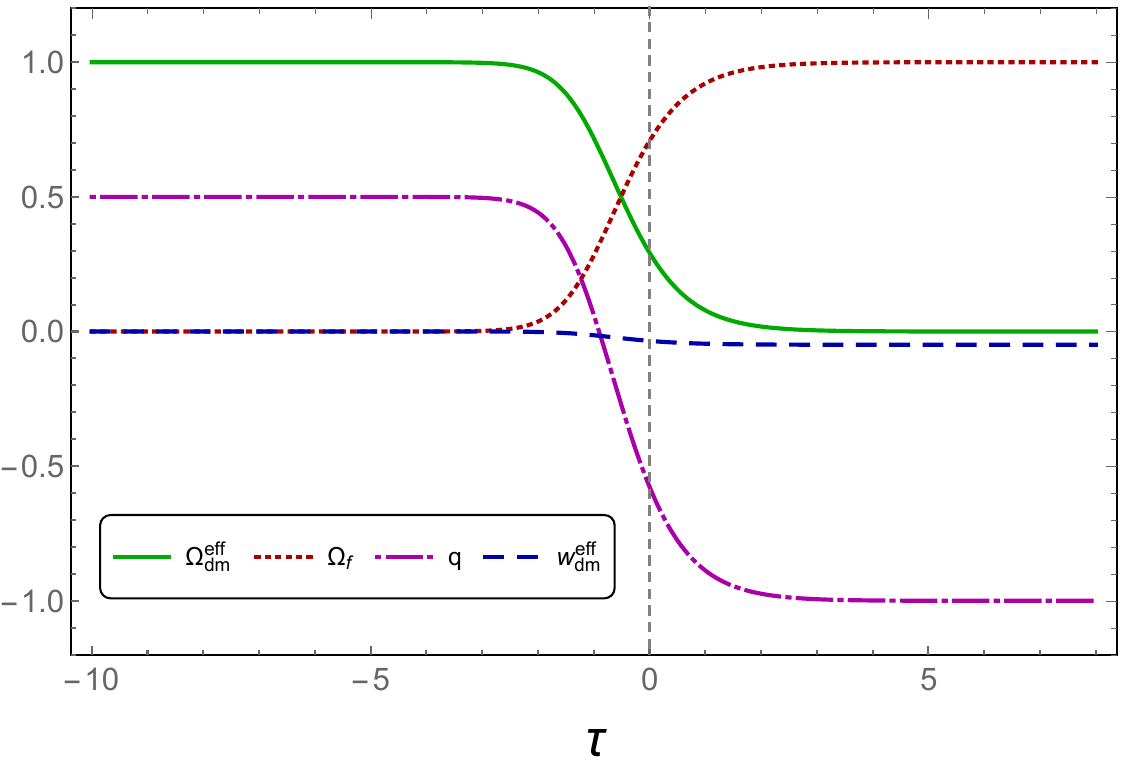} 
\includegraphics[width=0.497\textwidth]{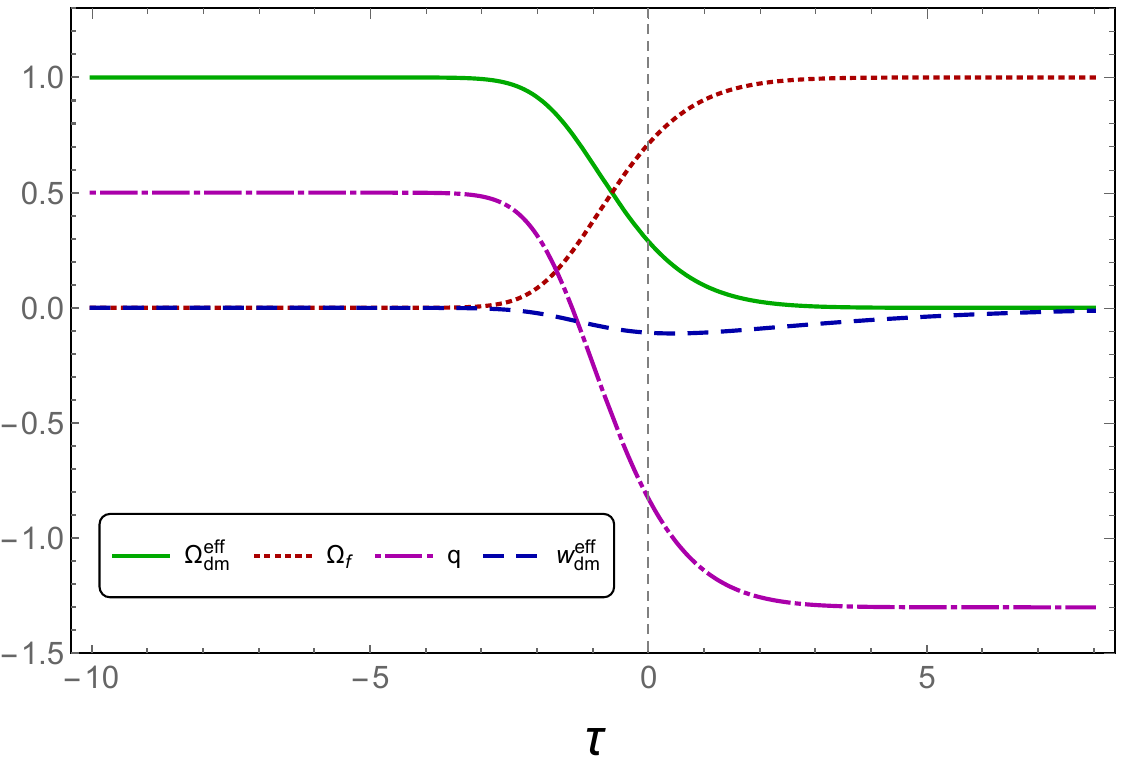}
    \caption{We display the evolution plots of the DM density parameter $\Omega_{\rm dm}^{\rm eff}$, the second fluid density parameter $\Omega_f$, the deceleration parameter $q$, and the effective equation of state of DM $w_{\rm dm}^{\rm eff}$ corresponding to the dynamical system (\ref{RG-DS-4}), with the matter creation rate $\Gamma=\Gamma_0 H^{-1}$. {\bf Upper Left Plot:} This figure is drawn with the model parameters $w=0.1$, $\gamma=1$, and it shows that the final fate of the universe is late time accelerated evolution in cosmological constant era dominated by DM.  {\bf Upper Right Plot:} This figure is produced for the parameters values $w=0$ and $\gamma=1$. Here also, the final fate of the universe is late time accelerated evolution in cosmological constant era dominated by DM. {\bf Lower Left Plot:} We have used $w=-1$ and $\gamma=1$. This plot depicts that the ultimate fate of the universe is found to be late time accelerated evolution in cosmological constant epoch which is achieved after the occurrence of a decelerated DM dominated phase. {\bf Lower Right Plot:} We have taken $w=-1.2$ and $\gamma=1$. This plot highlights that the ultimate fate of the universe is found to be late time DE dominated accelerated evolution in phantom epoch which is achieved after the occurrence of a decelerated DM dominated phase. 
}
    \label{fig-evo-4}
\end{figure*}
\begin{figure}
    \centering
    \includegraphics[width=0.497\textwidth]{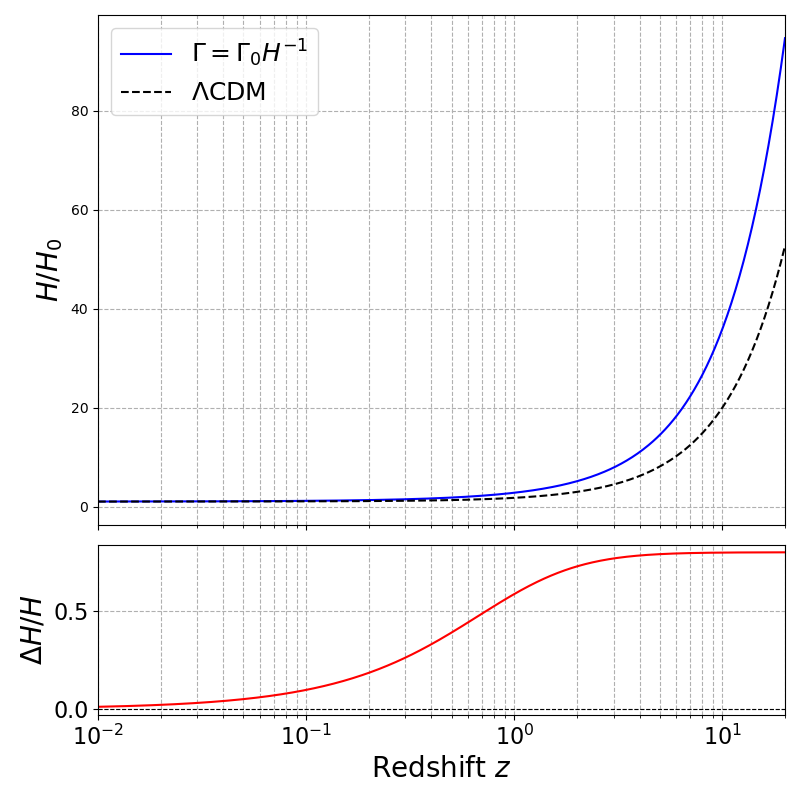}
    \caption{The redshift evolution of $H/H_0$ for the matter creation model $\Gamma= \Gamma_0H^{-1}$ and the $\Lambda$CDM model (upper panel) and the fractional difference $\Delta H/H = (H_{\rm Model} - H_{\Lambda {\rm CDM}})/H_{\Lambda {\rm CDM}}$ (lower panel) have been shown. For the matter creation model we have taken $w= 0$, $\gamma = 0.1$,  $\Omega_{f0} =0.04$ while for the $\Lambda$CDM model we set the matter density parameter at present $\Omega_{m0} = 0.3$.} 
    \label{fig:exp-hist-M4}
\end{figure}
   
{\bf When $w > 0$}, we obtain five isolated critical points, namely, $D_0,~D_1,~D_2,~D_3$ and $D_4$ in which $D_0,~D_2$ are dominated by the second fluid only ($\Omega_f =1$) and $D_1,~D_3,~D_4$ are DM dominated points ($\Omega_{\rm dm} = 1$). Inspecting the nature of eigenvalues from Table~\ref{fourth-table}, according to the linear stability analysis, we find  that $D_0$ is unstable, $D_1,~D_3$ both are saddle by nature and $D_4$ is the only globally stable point. The critical point $D_2$ at which decelerating parameter is undefined, behaves like saddle point because along $z$ axis, $z$ is increasing and along $z=1$ line $x$ is increasing. Here, the points $D_0,~D_1$ lie in the decelerating phase and $D_3,D_4$ also belong to the accelerated phase of the universe. The phase space trajectories are highlighted in the upper left plot of the Fig. \ref{fig4A},  indicating a transition of our universe from the second fluid dominated decelerating phase to a DM dominated accelerating phase where the slowing down of the cosmic acceleration is observed similar to what we have already seen for the two-fluid system of the Model $\Gamma = \Gamma_0$ with $w >0$, see section~\ref{model-1}. This is further supported by the upper left plot of Fig. \ref{fig-evo-4}, which illustrates the evolution of $\Omega_{\rm dm}^{\rm eff}$, $\Omega_f$, $q$ and $w_{\rm dm}^{\rm eff}$.

{\bf When $w = 0$}, in this case, we obtain total three isolated critical points $D_2,~D_3,~D_4$ and one critical line $D_{7}(x_c,0)$. In our phase space, $x'$ is always positive and, $z$ is increasing for $z<\frac{1}{1+\sqrt{\frac{\gamma}{3}}\sqrt{x}}$ and it is decreasing for $z>\frac{1}{1+\sqrt{\frac{\gamma}{3}}\sqrt{x}}$. Therefore, $D_{7}$ where DM and DE coexist, is always unstable and $D_2,~D_3$ give saddle type behavior. Since in our domain $D_4$ is the only stable point, we claim that it is a globally stable point. The accelerated phase includes $D_3,~D_4$ and on the other side, the decelerated phase comprises the line $D_{7}$. The phase space orbits are depicted in the upper middle plot of the Fig. \ref{fig4A},  showing an alteration of the universe from the decelerating scaling solutions (DM and second fluid both exist in the picture) to DM dominated accelerating era where the slowing down of the cosmic acceleration is noted $-$ the same feature has been observed in the model $\Gamma = \Gamma_0$  with $w=0$, see section~\ref{model-1}. As supporting evidence, the upper right plot in Fig. \ref{fig-evo-4} shows the evolution of the density parameters $\Omega_{\rm dm}^{\rm eff}$, $\Omega_f$, $q$ and $w_{\rm dm}^{\rm eff}$.  

{\bf When $-\frac{1}{3} \leq  w < 0$},  once again we can get five isolated critical points, namely, $D_0,~D_1,~D_2,~D_3$ and $D_4$ by solving the autonomous system (\ref{RG-DS-4}). Looking into the eigenvalues from the Table~\ref{third-table} and using the direction of the flow on the boundary of phase space, one can easily conclude about the stability of the critical points. Therefore, $D_0,~D_2$ and $D_3$ give saddle type character, $D_1$ shows unstable nature and as before $D_4$ corresponds to matter dominated late time globally stable point. Here also, $D_0$, $D_1$ lie in the decelerating phase and $D_3$ belongs to the accelerating phase. The evolution of the solution curves are clearly observed in the upper right plot of the Fig. \ref{fig4A}, highlighting the alteration from DM dominated decelerating epoch to DM dominated accelerating epoch with its slowing down nature, which one can meanwhile find in the Model \ref{model-1} with the case $-1/3<w<0$. The case with {\bf $-1< w < -\frac{1}{3}$} properly replicates the above case where the range of $w$ is $-\frac{1}{3}\leq w<0$. The only difference lies on the decelerating parameter at the point $D_0$ which shows acceleration. Thus, the phase plot is the upper right plot of Fig. \ref{fig4A}.

{\bf For $ w=-1 $}, we have a set of critical points which contains two isolated critical points $D_1,~D_3$ and two critical lines $D_5(0,z_c)$, $D_6\left(x_c,\frac{1}{1+\sqrt{\frac{\gamma}{3}}}\right)$. In the phase space above the line $z=\frac{1}{1+\sqrt{\frac{\gamma}{3}}}$, $x$ and $z$ show increasing and decreasing character, respectively, and below this line, $x$ is decreasing function while $z$ is increasing function. The equation of the separatrix joining the points $D_1$ and $\left(0,\frac{1}{\sqrt{\frac{\gamma}{3}}}\right)$ is $z=\sqrt{\frac{3}{\gamma}}(1-z)\sqrt{1-x}$ which separates the phase space below the line $z=\frac{1}{1+\sqrt{\frac{\gamma}{3}}}$ into two parts; one is below the separatrix and other one is the above the separatrix. In the first part, any trajectory approaches to the line $D_5$ and in the second part, trajectories converge to the line $D_6$. Therefore, $D_6$ where DM and DE coexist (except the end points), gives stable like nature. Here, $D_1$ and $D_6$ with $z>\frac{1}{1+\sqrt{\frac{\gamma}{3}}}$ show unstable nature and $D_3$ is a saddle point. The accelerating phase includes $D_3$, $D_5$, $D_6$ and $D_1$ corresponds to deceleration. The part of the line $D_5$ below $z=\frac{1}{1+\sqrt{\frac{\gamma}{3}}}$ corresponds to late time completely DE dominated stable points. On the other hand, the line $D_6$ has the ability to solve the coincidence problem. The phase space structure is properly presented in the lower left plot of Fig. \ref{fig4A} which depicts two types of transitions: the first one is DM dominated decelerating phase to accelerating scaling solutions (DM and DE both exist) and the second one is DM dominated decelerating phase to completely DE dominated accelerating phase, that is similar to the case $w=-1$ of Model \ref{model-1}. The lower left plot in Fig. \ref{fig-evo-4} supports this scenario by displaying the evolution of $\Omega_{\rm dm}^{\rm eff}$, $\Omega_f$, $q$ and $w_{\rm dm}^{\rm eff}$.       

{\bf When $w < -1$}, we again find five critical points, namely, $D_0,~D_1,~D_2,~D_3,~D_4$ which lie on the boundary of the phase space region $R$. From the sign of the eigenvalues and the direction of the vector field on the boundary of the phase space, we can lead to the conclusion that $D_1,~D_2$ are unstable critical points, $D_3,~D_4$ give saddle type nature and $D_0$ is the only global stable critical point. Here, $D_0,~D_3,~D_4$ correspond to accelerating points and $D_1$ belong to deceleration phase. As a result $D_1$ is the past matter dominated unstable critical point and $D_0$ is the late time completely DE dominated late time stable critical point. The phase space orbits are exhibited in the lower right plot of Fig. \ref{fig4A}, providing the transition from the DM dominated accelerating phase to the DE dominated super accelerating phase analogous to the case $w<-1$ of the Model \ref{model-1}. As shown in the lower right plot of Fig. \ref{fig-evo-4}, the evolution of $\Omega_{\rm dm}^{\rm eff}$, $\Omega_f$,  $q$, and $w_{\rm dm}^{\rm eff}$ reinforces this behavior.

Finally, in Fig. \ref{fig:exp-hist-M4}, we illustrate the evolution of $H/H_0$ for the matter creation scenario with $\Gamma=\Gamma_0 H^{-1}$ compared to $\Lambda{\rm CDM}$. The fractional difference $\Delta H/H$ is shown in the lower panel. %Using $w=0$, $\gamma=0.1$, and $\Omega_{\rm dm,0}=0.3$, 
We find that the models match at low redshift, while at higher-redshift, differences highlight the role of particle creation.

\subsection{Model: $\Gamma=\Gamma_0 H^{-2}$}\label{model-5}

In this class of matter creation models, a notable case is $\Gamma=\Gamma_0 H^{-2}$, where $\Gamma_0$ is a constant with the same dimension as the cube of the Hubble parameter. 

The autonomous system in this case becomes,     
\begin{subequations} \label{DS-5}
\begin{align}
    x'=& x(1-x) \left(3w+\frac{\mu z^3}{(1-z)^3}\right),   \label{DS-5-x} \\
    z'=& \frac{3}{2}z(1-z)\left[1+w(1-x)-\frac{\mu x z^3}{3(1-z)^3} \right], \label{DS-5-z} 
\end{align} 
\end{subequations}
where $\mu=\frac{\Gamma_0}{H_0^3}$ is a dimensionless parameter that adopts a positive value. To eliminate the singular behavior of the system (\ref{DS-5}) at $z=1$, we perform a time reparametrization via $dN=(1-z)^3d\tau$. This yields a reduced autonomous system, topologically equivalent to the original one, expressed as:  
\begin{subequations} \label{RG-DS-5}
\begin{align}
    \frac{dx}{d\tau}=& x(1-x) \left(3w (1-z)^3+\mu z^3\right),   \label{RG-DS-5-x} \\
    \frac{dz}{d\tau}=& \frac{3}{2}z(1-z)\left[\left(1+w(1-x)\right)(1-z)^3-\frac{\mu x z^3}{3} \right]. \label{RG-DS-5-z} 
\end{align}
\end{subequations}
 Here, one can easily conclude that ${\bf R}$ is a positively invariant set in the dynamical system (\ref{RG-DS-5}) because the boundaries of ${\bf R}$ are invariant manifolds of (\ref{RG-DS-5}). Here,  $q=\frac{1}{2}\left[1+3w(1-x)-\frac{\mu x z^3}{(1-z)^3}\right]$. Therefore, at present time (i.e. $z=1/2$) for the accelerating expansion of the universe, the physical trajectories must follow the condition $(1-\mu)x+(1+3w)(1-x)<0$. The cosmological features that are obtained from the above system are briefly given in the Table \ref{fifth-table}. Now, we split the equation of the state parameter into various parts and analyze the system (\ref{RG-DS-5}) in each part of $w$, which have been described below: 
%%%%%%%   Table 5  %%%%%%%%
\begin{table*}[t]
\centering
%\resizebox{0.9\textwidth}{!}{%
	\begin{tabular}{|c c c c c c c c|}\hline\hline
{\bf Critical point} & {\bf Existence} & {\bf Eigenvalue} & {\bf Stability} & $\mathbf{\Omega_f}$ & $\mathbf{\Omega_{\rm dm}^{\rm eff}}$ & $\mathbf{q}$ & {\bf Acceleration} \\ \hline
%   &&&&&     \\

$E_{0}(0,0)$  & Always & $\left(3w,\frac{3}{2}(1+w)\right)$  & {\bf Stable} if $ w < -1 $;  &  1  &  0  &  $\frac{1}{2}(1+3w)$  &  $w<-\frac{1}{3}$  \\ 
 &&& Saddle if $-1<w<0$;  &&&&  \\
  &&& Unstable if $w > 0$  &&&&  \\ \hline
 
$E_{1}(1,0)$  & Always & $\left(-3w,\frac{3}{2}\right)$   &  Saddle if $w>0$;   & 0   &  1  &  $\frac{1}{2}$  & No \\
 &&& Unstable if $ w < 0 $ &&&&  \\  \hline
%   &&&&&  \\
$E_{2}(0,1)$  & Always  & $(\mu,0)$   & Non-hyperbolic Saddle if $ w > -1 $;   & 1   &  0  & Undefined   &  Undetermined \\
 &&& Unstable if $w<-1$ &&&&  \\    \hline
%   &&&&&  \\
$E_{3}(1,1)$  & Always  & $\left(-\mu,\frac{\mu}{2}\right)$  & Always Saddle  &  0  & 1   &  $-\infty$  &  Yes \\ \hline
%   &&&&&  \\ 
$E_{4}\left(1,\frac{1}{1+\sqrt[3]{\frac{\mu}{3}}}\right)$  &  $\mu>0$ & $\left(-\frac{\mu(1+w)}{\left(1+\sqrt[3]{\frac{\mu}{3}}\right)^3},-\frac{3\mu}{2\left(1+\sqrt[3]{\frac{\mu}{3}}\right)^3}\right)$  &  {\bf Stable} if $ w > -1 $    &  0  &  1  &  $-1$  &  Yes     \\ 
  &&&  Saddle if $ w < -1 $ &&& &    \\ \hline
%   &&&&& \\
 $E_5\left(0,z_c\right)$  &  $w=-1$ & $\left(\mu z_c^3-3\left(1-z_c\right)^3,0\right)$  & {\bf Stable} if $z_c<\frac{1}{1+\sqrt[3]{\frac{\mu}{3}}}$ and    &  $1$  &  $0$  &  $-1$  &  Yes     \\ 
  &&&  Unstable if $z_c>\frac{1}{1+\sqrt[3]{\frac{\mu}{3}}}$ &&& &    \\ \hline 

 $E_6\left(x_c,\frac{1}{1+\sqrt[3]{\frac{\mu}{3}}}\right)$  &  $w=-1$ & $\left(0,-\frac{3\mu x_c}{2\left(1+\sqrt[3]{\frac{\mu}{3}}\right)^3}\right)$  & {\bf Stable}  &  $1-x_c$  &  $x_c$  &  $-1$  &  Yes     \\  \hline
 
 $E_{7}\left(x_c,0\right)$  &  $w=0$ & $\left(0,\frac{3}{2}\right)$  &  Unstable    &  $1-x_c$  &  $x_c$  &  $\frac{1}{2}$  &  No     \\ \hline \hline
\end{tabular}%
% }
\caption{Summary of the critical points, their existence, stability and the values of the cosmological parameters at those points for the dynamical system (\ref{RG-DS-5}) with the matter creation rate $\Gamma=\Gamma_0 H^{-2}$. }
	\label{fifth-table}
\end{table*}
\begin{figure*}
    \centering
    \includegraphics[width=0.33\textwidth]{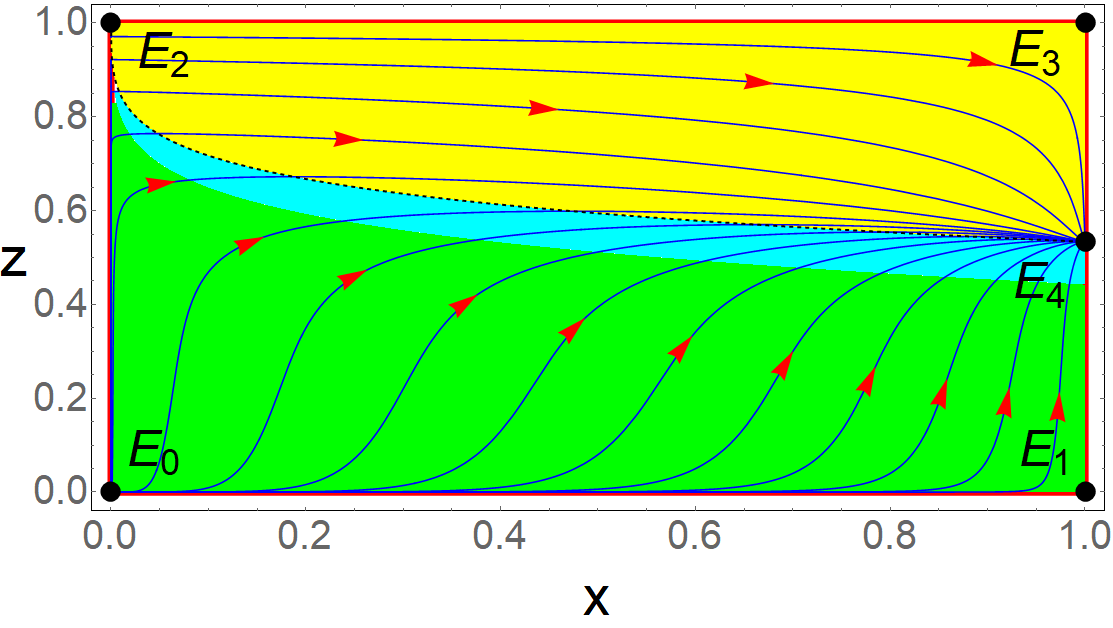}
    \includegraphics[width=0.33\textwidth]{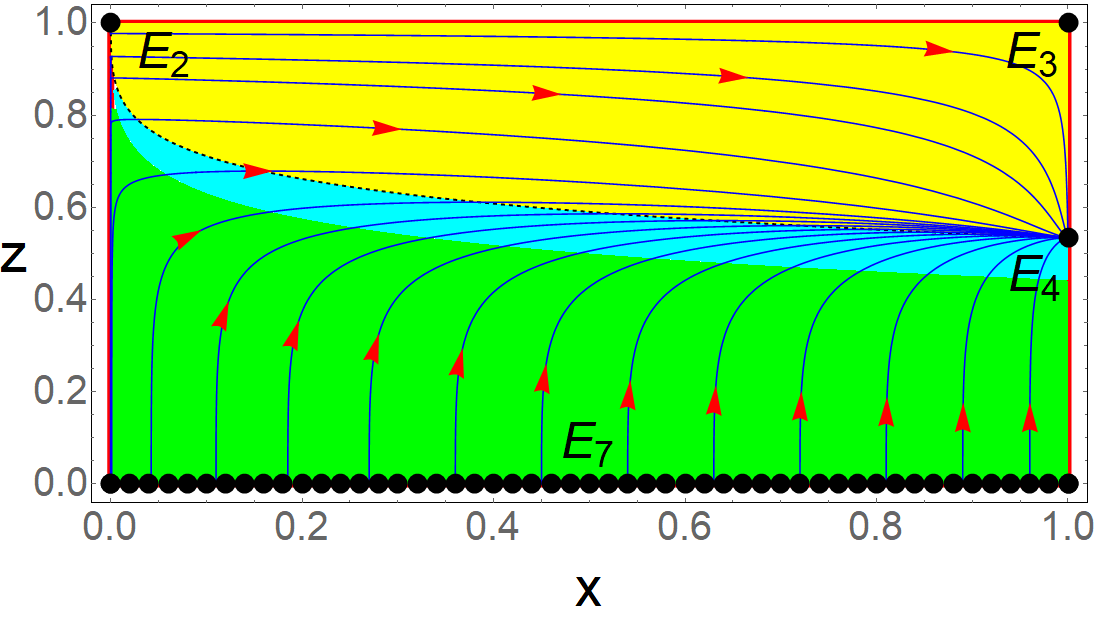}
    \includegraphics[width=0.32\textwidth]{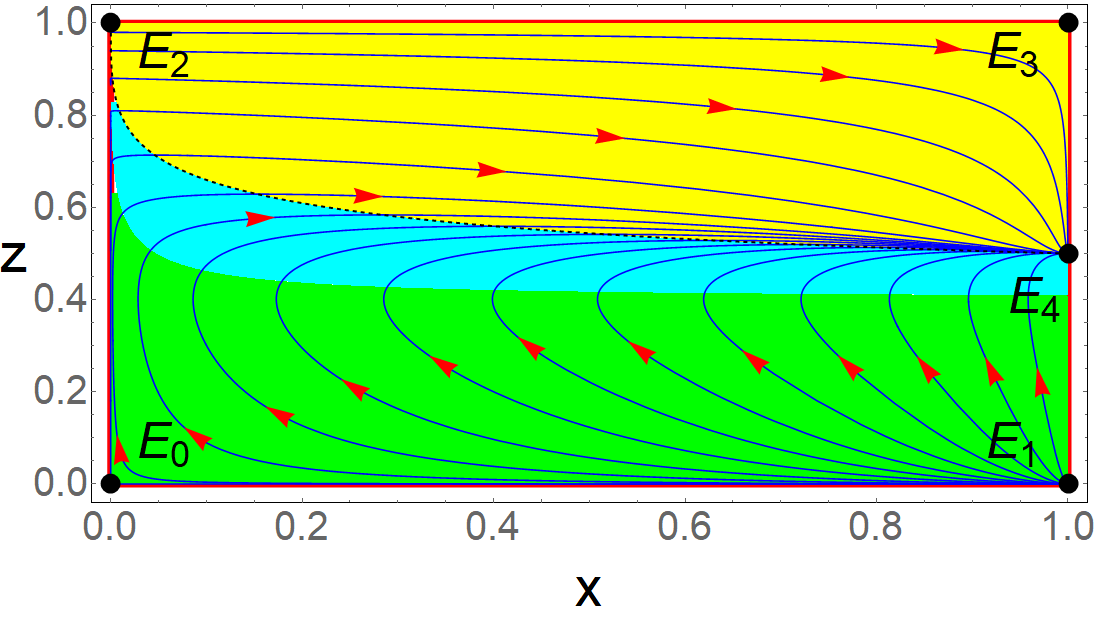}\\
     \includegraphics[width=0.33\textwidth]{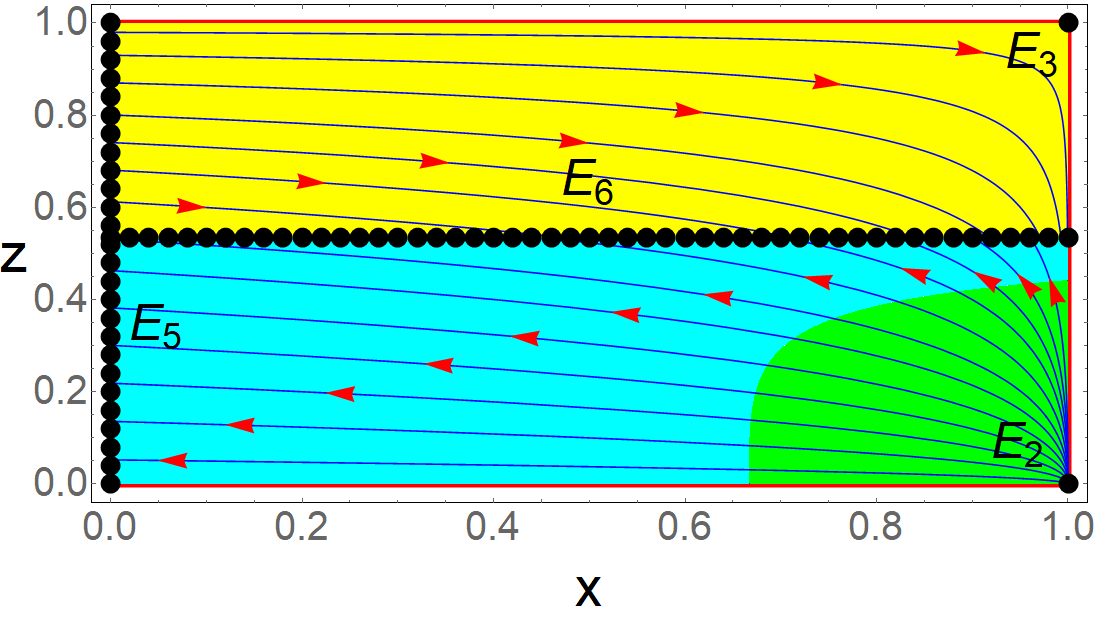}
    \includegraphics[width=0.33\textwidth]{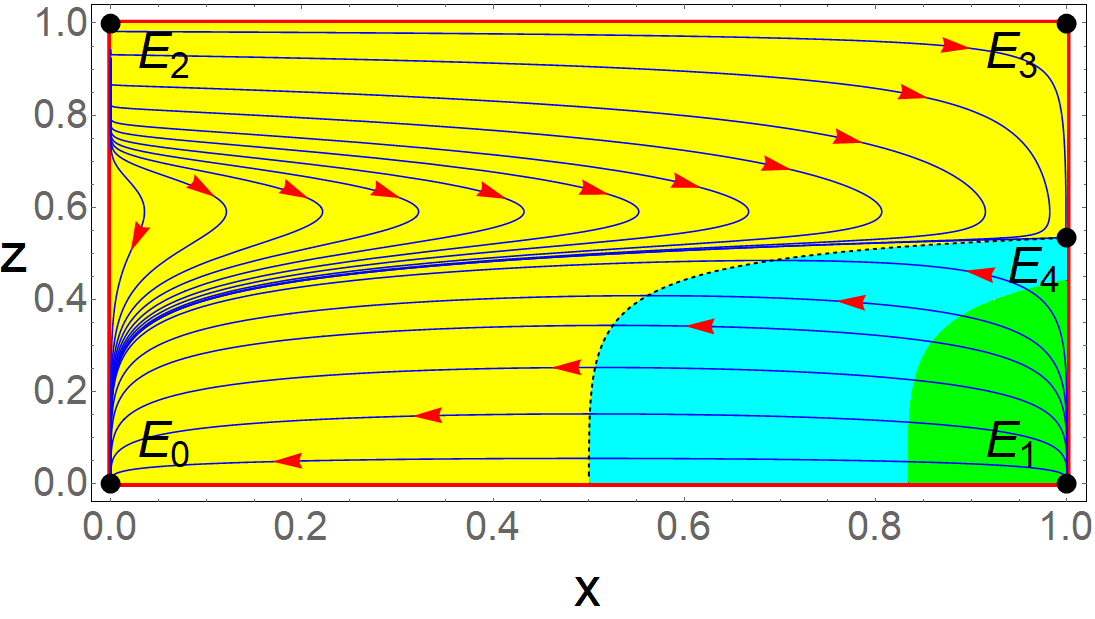}
    \caption{Description of the phase space structure for the matter creation rate $\Gamma = \Gamma_0 H^{-2}$. {\bf Upper Left Plot:} The phase plot of the system (\ref{RG-DS-5}) when we have assumed $w=0.1$ and $\mu=2$. For other values of $w>0$ and $\mu>0$, we can also obtain similar phase space structure. {\bf Upper Middle Plot:} The phase space of the system (\ref{RG-DS-5}) when the EoS $w$ takes the value $0$. Here we use $\mu =2$ but any positive value of $\mu$ gives same type of phase portrait. {\bf Upper Right Plot:} The phase plot of the system ($\ref{RG-DS-5}$) considering $w=-0.3$ and $\mu=3$. Also, we can get similar type of graphics for any positive value of $\mu$ and negative value of $w$ in the interval $(-1,0)$. {\bf Lower Left Plot:} The phase space of the system (\ref{RG-DS-5}) when the EoS $w$ takes the value $-1$. Here we use $\mu =2$ but any positive value of $\mu$ gives similar type of phase portrait. {\bf Lower Right Plot:} The phase plot of the system (\ref{RG-DS-5}) when we assume $w=-2$ and $\mu=2$. For other values of $w<-1$ and $\mu>0$, we can also obtain similar graphics. Here the green region corresponds to the decelerating phase ($q >0$), the cyan region represents the accelerating phase with $-1<q<0$ and the yellow region corresponds to the super accelerating phase (i.e. $q <-1$). The black dotted curve separating the cyan and yellow regions (in the lower left graph this curve is not clear because of the critical line) corresponds to $q =-1$.    }
    \label{fig5A}
\end{figure*}

\begin{figure*}
   \centering   
\includegraphics[width=0.497\textwidth]{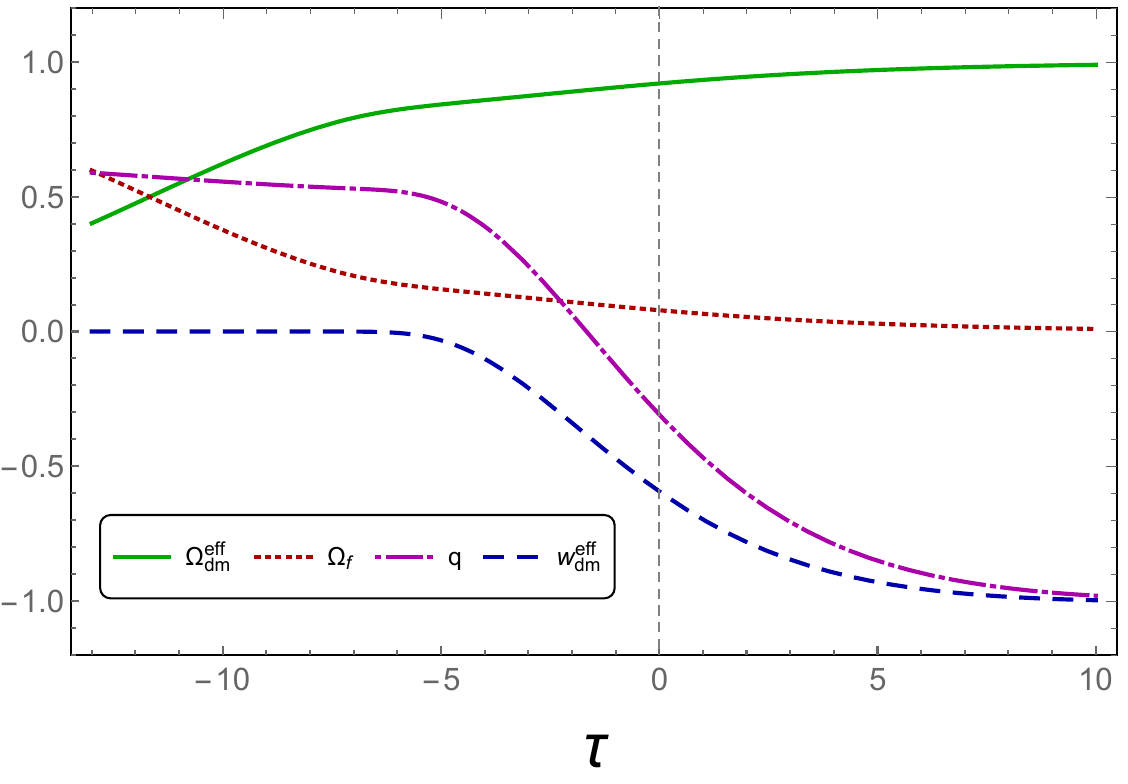} 
\includegraphics[width=0.497\textwidth]{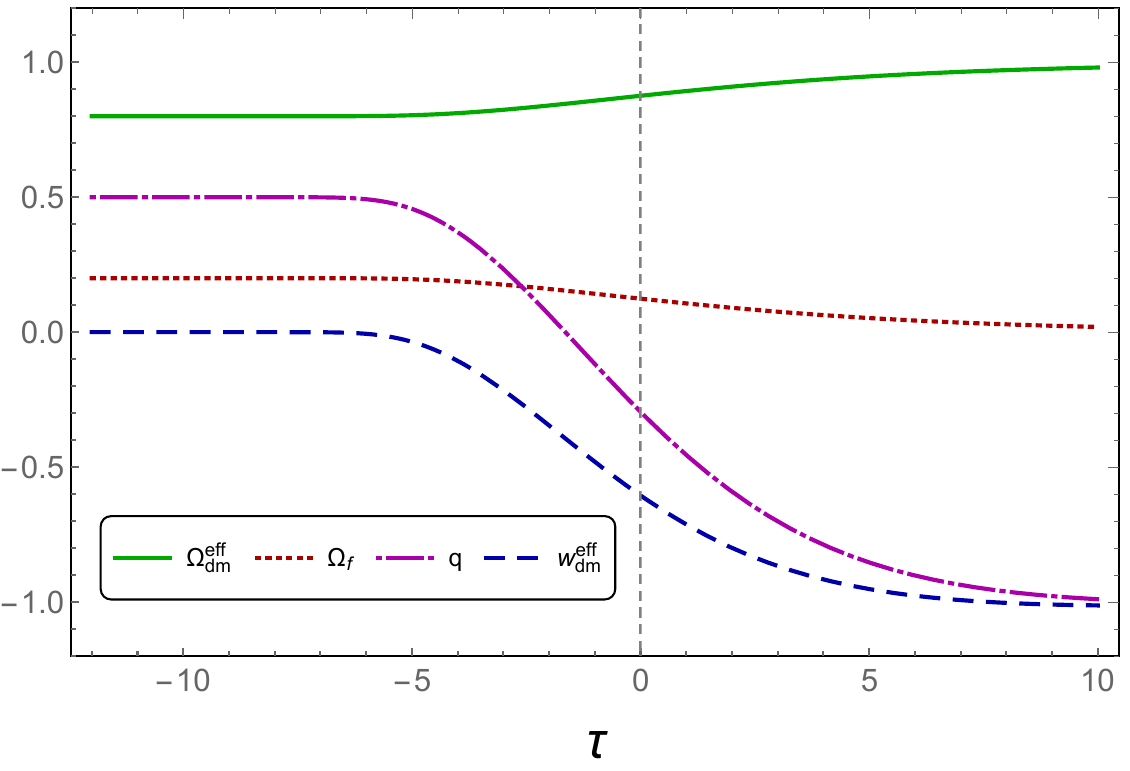}\\
\includegraphics[width=0.497\textwidth]{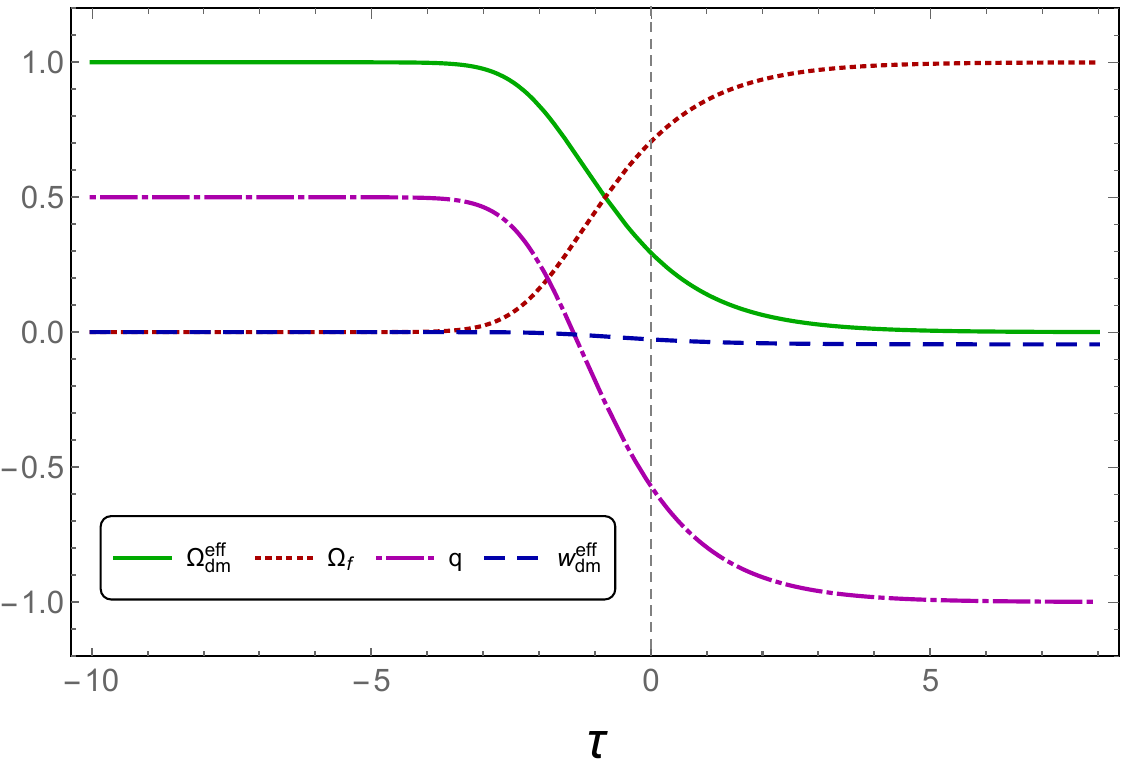} 
\includegraphics[width=0.497\textwidth]{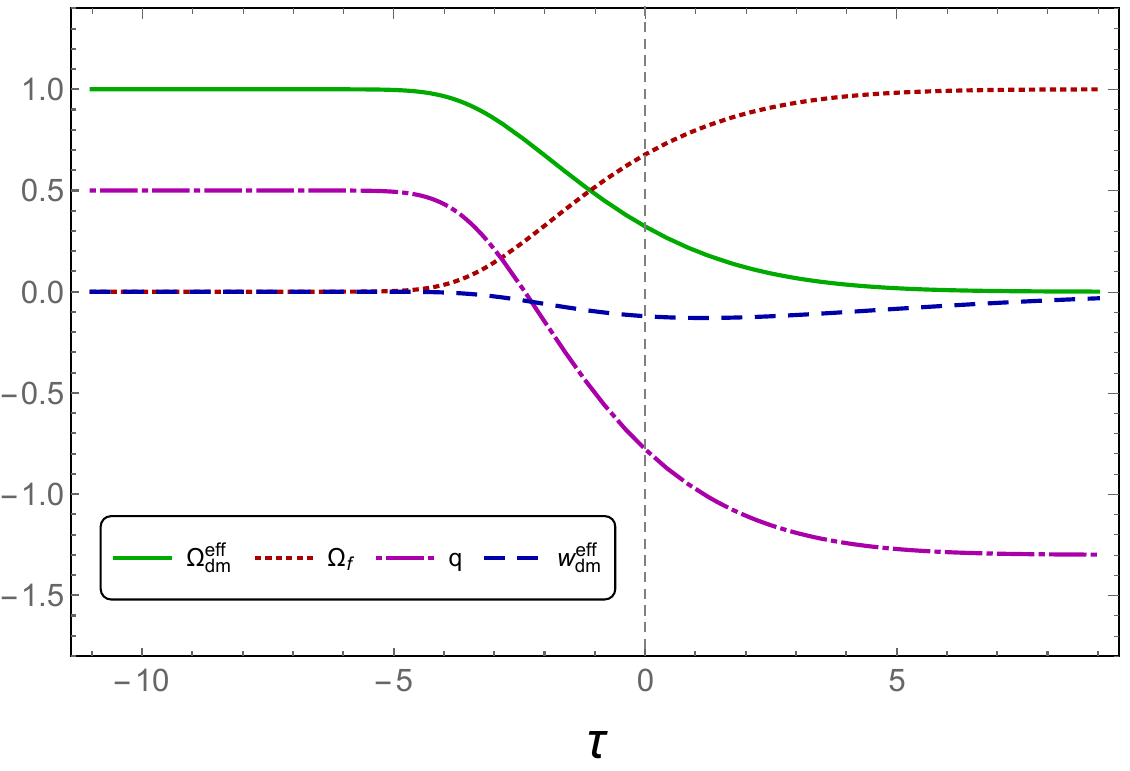}
    \caption{We have presented the evolution of key cosmological parameters namely, the DM density parameter $\Omega_{\rm dm}^{\rm eff}$, the second fluid density parameter $\Omega_f$, the decelerating parameter $q$, and the effective equation of state of DM $w_{\rm dm}^{\rm eff}$ for the dynamical system (\ref{RG-DS-5}) with the matter creation rate $\Gamma=\Gamma_0 H^{-2}$. {\bf Upper Left Plot:} This figure is drawn with the model parameters $w=0.1$, $\mu=1$, and it shows that the final fate of the universe is late time accelerated evolution in cosmological constant era dominated by DM.  {\bf Upper Right Plot:} This figure is produced for the parameters values $w=0$ and $\mu=1$. Here also, the final fate of the universe is late time accelerated evolution in cosmological constant era dominated by DM. {\bf Lower Left Plot:} We have used $w=-1$ and $\mu=1$. This plot depicts that the ultimate fate of the universe is found to be late time accelerated evolution in cosmological constant epoch which is achieved after the occurrence of a decelerated DM dominated phase. {\bf Lower Right Plot:} We have taken $w=-1.2$ and $\mu=1$. This plot highlights that the ultimate fate of the universe is found to be late time DE dominated accelerated evolution in phantom epoch which is achieved after the occurrence of a decelerated DM dominated phase. 
}
    \label{fig-evo-5}
\end{figure*}
\begin{figure}
    \centering
    \includegraphics[width=0.497\textwidth]{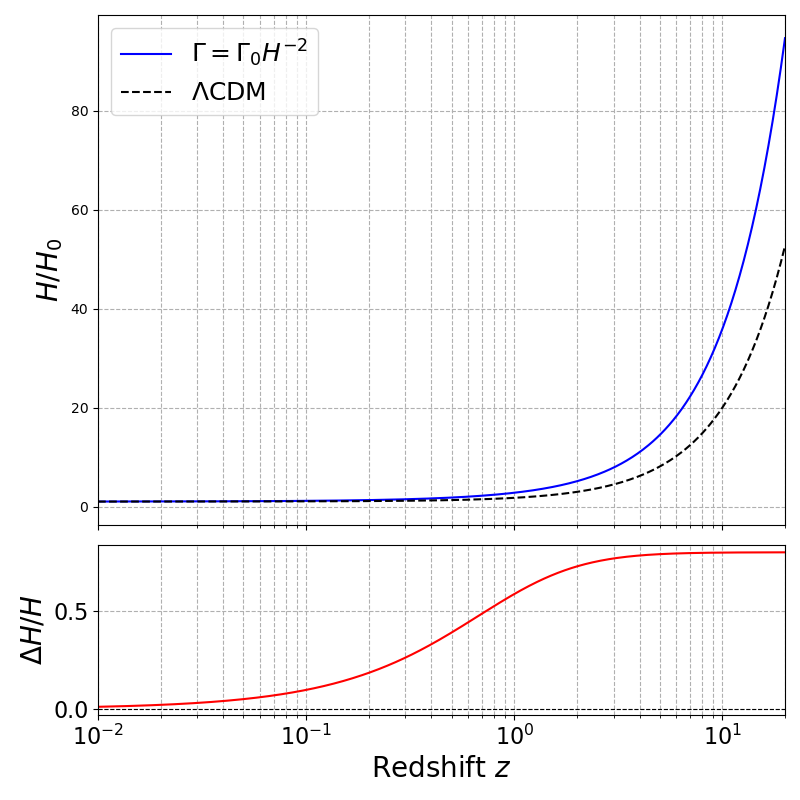}
    \caption{The redshift evolution of $H/H_0$ for the matter creation model $\Gamma= \Gamma_0H^{-2}$ and the $\Lambda$CDM model (upper panel) and the fractional difference $\Delta H/H = (H_{\rm Model} - H_{\Lambda {\rm CDM}})/H_{\Lambda {\rm CDM}}$ (lower panel) have been shown. For the matter creation model we have taken $w= 0$, $\mu = 0.1$, $\Omega_{f0} =0.04$ while for the $\Lambda$CDM model we set the matter density parameter at present $\Omega_{m0} = 0.3$. }
    \label{fig:exp-hist-M5}
\end{figure}
   
{\bf When $w > 0$}, the dynamical system (\ref{RG-DS-5}) allows five critical points $E_0,~E_1,~E_2,~E_3$ and $E_4$. Here, the accelerating phase includes the points $E_3,~E_4$ and the decelerating phase contains the critical points $E_0,~E_1$. Again, $E_1,~E_3,~E_4$ correspond to matter dominated solutions, while $E_0,~E_2$ represent completely second fluid dominated points. One can easily conclude about stability from the direction of the vector field on the boundary of the phase space and signs of the eigenvalues. Therefore, $E_0$ is unstable, $E_4$ is the only globally stable point and $E_1,~E_2,~E_3$ are saddle by nature. Note that, $E_4$ is the DM dominated late time accelerating globally stable point. The upper left plot of Fig. \ref{fig5A} gives clear picture of the qualitative behavior of critical points. The inherent features of this plot are similar to the case $w>0$ of the Models \ref{model-1} and \ref{model-4}. As further evidence, the upper left plot of Fig. \ref{fig-evo-5} displays the evolution of the DM and second fluid density parameters $\left(\Omega_{\rm dm}^{\rm eff},\Omega_f\right)$ along with the deceleration parameter $q$ and the effective EoS of DM, $w_{\rm dm}^{\rm eff}$. The evolution of these parameters is almost similar to what we have observed earlier in the models with $\Gamma = \Gamma_0$, $\Gamma = \Gamma_0 H^{-1}$.

{\bf When $w = 0$},  we obtain three isolated critical points $E_2(0,1),~E_3(1,1)$ and $E_{4}\left(1,\frac{1}{1+\sqrt[3]{\frac{\mu}{3}}}\right)$ along with one critical lines, namely, $E_{7}(x_c,0)$. Now $x$ is always increasing and, $z$ is increasing and decreasing depending on $ z $ variable chosen from $\left(0,\frac{1}{1+\sqrt[3]{\frac{\mu x}{3}}}\right)$ and $\left(\frac{1}{1+\sqrt[3]{\frac{\mu x}{3}}},1 \right)$ respectively.  Therefore $E_2,~E_3$ are saddle, $E_{7}$ is unstable and $E_4$ is a globally stable critical point, which are obtained in the upper middle plot of Fig. \ref{fig5A}. Once again $E_4$ corresponds to DM dominated late time accelerated globally stable point. In this case, the cosmological characteristics are completely identical with the case $w=0$ of the Models \ref{model-1} and \ref{model-4}. Supporting this scenario, Fig. \ref{fig-evo-5}(upper right plot) illustrates the evolution of $\Omega_{\rm dm}^{\rm eff}$, $\Omega_f$, decelerating parameter $q$ and $w_{\rm dm}^{\rm eff}$. We also noticed that the evolution of these parameters is almost similar to what we have observed earlier in the models with $\Gamma = \Gamma_0$, $\Gamma = \Gamma_0 H^{-1}$.

{\bf When $ -\frac{1}{3} \leq  w < 0$}, similar to the case where $w>0$, in this part of $w$, we acquire five isolated critical points which are depicted in the Table \ref{fifth-table}. From this table and the flow on the unit square, we can say that $E_1$ is unstable i.e. the DM dominated point $E_1$ leaves its decelerating phase, $E_0,~E_2,~E_3$ are saddle by behavior and once again, $E_4$ is the DM dominated accelerating globally stable point. Therefore, the universe ends with DM dominated accelerating phase. The orbits in the phase space are highlighted in the upper right plot of Fig. \ref{fig5A}, providing the results which replicate the case $-1/3<w<0$ for the Models \ref{model-1} and \ref{model-4}.

{\bf When $ -1 < w < -\frac{1}{3}$}, this part of $w$ is analogous to the above case where $w$ lies in the interval $\left[-\frac{1}{3},0\right)$, except decelerating parameter at the critical point $E_0$ which belongs to the accelerating region in the phase space. Therefore, the phase plot is similar to the upper right plot of Fig. \ref{fig5A}.

{\bf When $ w = -1 $},  we obtain two isolated critical points $E_1,~E_3$ which are mentioned in Table \ref{fifth-table} and two critical lines, namely, $E_5(0,z_c),~E_6\left(x_c,\frac{1}{1+\sqrt[3]{\frac{\mu}{3}}}\right)$. It is clear that $x'$ is positive and negative respectively if we choose $z$ from the interval $\left(\frac{1}{1+\sqrt[3]{\frac{\mu}{3}}},1\right]$ and the interval $\left[0,\frac{1}{1+\sqrt[3]{\frac{\mu}{3}}}\right)$. Again $z$ is increasing and decreasing below and above the line $z=\frac{1}{1+\sqrt[3]{\frac{\mu}{3}}}$ respectively. In the phase space below the line $z=\frac{1}{1+\sqrt[3]{\frac{\mu}{3}}}$, the equation of the separatrix is $z=\sqrt[3]{\frac{3}{\mu}}(1-z)\sqrt{1-x}$ which connects the critical points $E_1$ and $\left(0,\frac{1}{1+\sqrt[3]{\frac{\mu}{3}}}\right)$. As a result, $E_1$ is unstable, $E_3$ is saddle, $E_6$ is stable and, $E_5$ is stable when $z_c<\frac{1}{1+\sqrt[3]{\frac{\mu}{3}}}$ and unstable when $z_c > \frac{1}{1+\sqrt[3]{\frac{\mu}{3}}}$. The lower left plot of Fig.~\ref{fig5A} depicts qualitative features of the critical points. Here, $E_5$, lying in the accelerating phase, corresponds to completely DE domination and $E_6$, belonging also to the accelerating phase, represents the points where DM, DE coexist except the end points. Clearly, $E_6$ can solve cosmic coincidence problem. The key findings of this case are equivalent to the Models \ref{model-1} and \ref{model-4} with $w=-1$. Fig. \ref{fig-evo-5} (lower left plot) offers further support, capturing the evolving dynamics of the DM and DE density parameters, along with the deceleration parameter $q$ and $w_{\rm dm}^{\rm eff}$.  

{\bf When $ w < -1 $}, one can get two unstable critical points $\left(E_1,E_2\right)$, two saddle type critical points $\left(E_3,E_4\right)$ and $E_0$ which is the only globally stable critical point. From Table \ref{fifth-table}, it is clear that, $E_1$ is the past decelerating DM dominated solution and point $E_0$ where the universe ends in the accelerating completely DE dominated phase. Also note that, $E_3,~E_4$ lie in the accelerating phase and on the other hand, at $E_2$, acceleration can not be determined. The graphics in the lower right plot of Fig.~\ref{fig5A} properly represents the stability character of the critical points, leading to the results which are very same as the case $w<-1$ for the Models \ref{model-1} and \ref{model-4}. Additional insight is provided by the lower left plot of Fig. \ref{fig-evo-5}, where the evolution of $\Omega_{\rm dm}^{\rm eff}$, $\Omega_f$, $q$ and $w_{\rm dm}^{\rm eff}$
is clearly exhibited, further substantiating the model’s predictions.

 Finally, in Fig. \ref{fig:exp-hist-M5} we compare the normalized Hubble parameter $H/H_0$ for the matter creation rate ($\Gamma=\Gamma H^{-2}$) with the standard $\Lambda{\rm CDM}$ model (upper panel). The lower panel shows $\Delta H/H$. The evolution of the matter creation model follows $\Lambda{\rm CDM}$ at low redshift but differs at higher redshifts due to particle creation effects.

\subsection{Model: $\Gamma=2\left(1-\frac{1}{\xi}\right)\frac{\dot{H}}{H}$}\label{model-6}
In this section, we consider an inhomogeneous matter creation rate given by $\Gamma=2\left(1-\frac{1}{\xi}\right)\frac{\dot{H}}{H}$, where $\xi$ is a non-zero constant. We then analyze how this specific form of $\Gamma$ influences the evolution of the Universe.

The autonomous dynamical system, formulated using the dimensionless variables $x$ and $z$ introduced in eqn. (\ref{dimensionless-variables}), can be read as
\begin{subequations} \label{DS-6-new}
\begin{align}
    x'=& \frac{3x(1-x)(1+w-\xi)}{\xi(1-x)+x},   \label{DS-6-new-x} \\
    z'=& \frac{3\xi z(1-z)(1+w(1-x))}{2(\xi(1-x)+x)}. \label{DS-6-new-z} 
\end{align} 
\end{subequations}
The autonomous system described above may exhibit singular behavior depending on the value of the parameter $\xi$. For positive values of $\xi$, the system remains free of singularities. However, when $\xi$ is negative, singularities arise along the line $x=\frac{\xi}{\xi-1}$. For this matter creation rate, the decelerating parameter is given by $q=\frac{\xi+2(\xi-1)x+3w(1-x)}{2(\xi-(\xi-1)x)}$. We now execute a phase space stability analysis by considering two distinct regimes of the model parameter $\xi$: the positive $\xi$ region and the negative $\xi$ region. The critical points or lines, along with the corresponding values of key cosmological parameters, are summarized in Table. \ref{six-table}.  
%%%%%%%   Table 6  %%%%%%%%
\begin{table*}[t]    
\centering
\resizebox{1.0\textwidth}{!}{%
\begin{tabular}{|c|c c c c c c c c|}
 \hline
  \bf{ Parameter ($\mathbf{\xi}$)} & \bf{Critical point} & \bf{Existence} & \bf{Eigenvalues} & \bf{Stability} & $\bf{\Omega_{\rm dm}^{\rm eff}}$ & $\bf{\Omega_f}$  & $\mathbf{q}$ & \bf{Acceleration}    \\
   \hline\hline
    &$N_1\left(0,0\right)$ & Always & $\left(\frac{3(1+w-\xi)}{\xi},\frac{3(1+w)}{2}\right)$  & {\bf Stable} if $\xi>1+w$, $w<-1$;  & $0$ & $1$ & $\frac{1}{2}(1+3w)$ &  $w<-\frac{1}{3}$ \\
    &  &  &  & Unstable if $\xi<1+w$; &  &  &  &  \\
    &  &  &  & Saddle if $\xi>1+w$, $w>-1$ &  &  &  &  \\\cline{2-9}
     
  & $N_2\left(1,0\right)$ & Always  & $\left(-3(1+w-\xi),\frac{3\xi}{2}\right)$   & Saddle if $\xi<1+w$;  & $1$ & $0$  & $\frac{3\xi}{2}-1$ & $\xi<\frac{2}{3}$ \\
  &  &  &  & Unstable if $\xi>1+w$ &  &  &  &  \\\cline{2-9}
     
   & $ N_3\left(0,1\right)$ & Always & $\left(\frac{3(1+w-\xi)}{\xi},-\frac{3(1+w)}{2}\right)$  & {\bf Stable} if $\xi>1+w$, $w>-1$;  & $0$ & $1$ & $\frac{1}{2}(1+3w)$ &  $w<-\frac{1}{3}$  \\
   &  &  &  & Otherwise Saddle &  &  &  &  \\\cline{2-9}
  
 $\xi>0$& $N_4\left(1,1\right)$ & Always & $\left(-3(1+w-\xi),-\frac{3\xi}{2}\right)$  & {\bf Stable} if $\xi<1+w$; & $1$ & $0$ & $\frac{3\xi}{2}-1$ & $\xi<\frac{2}{3}$\\
 &  &  &  & Saddle if $\xi>1+w$ &  &  &  &  \\\cline{2-9} 
 
 & $N_5\left(0,z_c\right)$ & $w=-1$ & $(-3,0)$  & {\bf Stable} & $0$ & $1$ & $-1$ & Yes \\\cline{2-9}
 & $N_6\left(x_c,0\right)$ & $\xi=1+w$ & $\left(\frac{3(1+w)}{2},0\right)$  & Unstable & $x_c$ & $1-x_c$ & $\frac{1}{2}(1+3w)$ & $w<-\frac{1}{3}$ \\\cline{2-9}
  & $N_7\left(x_c,1\right)$ & $\xi=1+w$ & $\left(-\frac{3(1+w)}{2},0\right)$  & {\bf Stable} & $x_c$ & $1-x_c$ & $\frac{1}{2}(1+3w)$ & $w<-\frac{1}{3}$ \\
\hline 
  &$P_1\left(0,0\right)$ & $0\leq x<\frac{\xi}{\xi-1}$ & $\left(\frac{3(1+w-\xi)}{\xi-1},\frac{3(1+w)\xi}{2(\xi-1)}\right)$  & {\bf Stable} if $\xi<1+w$, $w<-1$; & $0$ & $1$ & $\frac{1}{2}(1+3w)$ &  $w<-\frac{1}{3}$ \\
  &  &  &  & Otherwise Saddle &  &  &  &  \\\cline{2-9}
     
  & $P_2\left(1,0\right)$ & $\frac{\xi}{\xi-1}<x\leq 1$  & $\left(\frac{3(1+w-\xi)}{\xi-1},\frac{3\xi}{2(1-\xi)}\right)$   & {\bf Stable} if $\xi<1+w$;  & $1$ & $0$  & $\frac{3\xi}{2}-1$ & $\xi<\frac{2}{3}$ \\
  &  &  &  & Saddle if $\xi>1+w$ &  &  &  &  \\\cline{2-9}
     
   & $ P_3\left(0,1\right)$ & $0\leq x<\frac{\xi}{\xi-1}$ & $\left(\frac{3(1+w-\xi)}{\xi-1},-\frac{3(1+w)\xi}{2(\xi-1)}\right)$  & {\bf Stable} if $\xi<1+w$, $w>-1$; & $0$ & $1$ & $\frac{1}{2}(1+3w)$ &  $w<-\frac{1}{3}$  \\
   &  &  &  & Unstable if $\xi>1+w$, $w<-1$; &  &  &  &  \\
   &  &  &  & Otherwise Saddle &  &  &  &  \\\cline{2-9}
  
 $\xi<0$& $P_4\left(1,1\right)$ & $\frac{\xi}{\xi-1}<x\leq 1$ & $\left(\frac{3(1+w-\xi)}{\xi-1},\frac{3\xi}{2(\xi-1)}\right)$  & Saddle if $\xi<1+w$; & $1$ & $0$ & $\frac{3\xi}{2}-1$ & $\xi<\frac{2}{3}$\\
 &  &  &  & Unstable if $\xi>1+w$ &  &  &  &  \\\cline{2-9}   
 & $P_5\left(0,z_c\right)$ & $0\leq x\leq\frac{\xi}{\xi-1}$, $w=-1$ & $\left(\frac{3\xi}{1-\xi},0\right)$  & {\bf Stable} & $0$ & $1$ & $-1$ & Yes \\\cline{2-9}
 & $P_6\left(x_c,0\right)$ & $\xi=1+w$, $x_c\neq\frac{\xi}{\xi-1}$ & $(\phi_\mp,0)$  & {\bf Stable} & $x_c$ & $1-x_c$ & $\frac{1}{2}(1+3w)$ & $w<-\frac{1}{3}$ \\\cline{2-9}
  & $P_7\left(x_c,1\right)$ & $\xi=1+w$, $x_c\neq\frac{\xi}{\xi-1}$ & $(\psi_\pm,0)$  & Unstable & $x_c$ & $1-x_c$ & $\frac{1}{2}(1+3w)$ & $w<-\frac{1}{3}$ \\
 \hline\hline 
\end{tabular}}
\caption{Summary of the critical points, their existence, stability and the values of the cosmological parameters at those points for the dynamical systems (\ref{DS-6-new}), (\ref{RG1-DS-6-new}) and (\ref{RG2-DS-6-new}) with the matter creation rate $\Gamma=2\left(1-\frac{1}{\xi}\right)\frac{\dot{H}}{H}$. Here, the eigenvalues $\phi_\mp$, $\psi_\pm$ are given by $\phi_\mp=\mp\frac{3(1+w)\left(1+w\left(1-x_c\right)\right)}{2w}$ and $\psi_\pm=\pm\frac{3(1+w)\left(1+w\left(1-x_c\right)\right)}{2w}$. Specifically, $\phi_-$ and $\psi_+$ are relevant in the region $\frac{\xi}{\xi-1}<x\leq 1$, while $\phi_+$ and $\psi_-$ correspond to $0\leq x<\frac{\xi}{\xi-1}$. }
\label{six-table}
\end{table*}
\begin{figure*}
    \centering
    \includegraphics[width=0.33\textwidth]{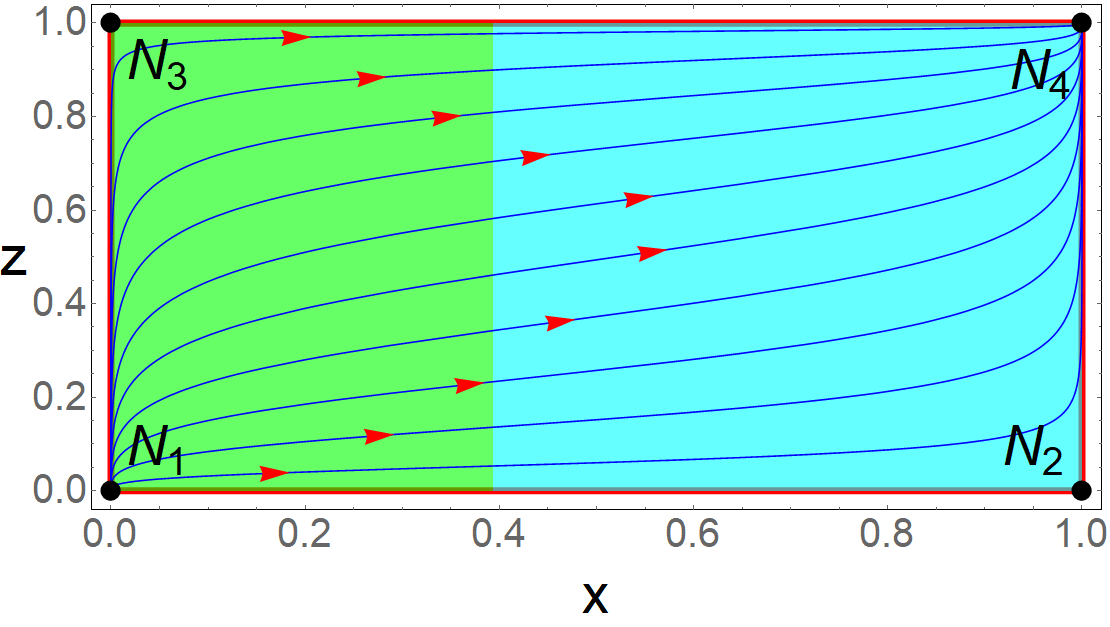}
    \includegraphics[width=0.33\textwidth]{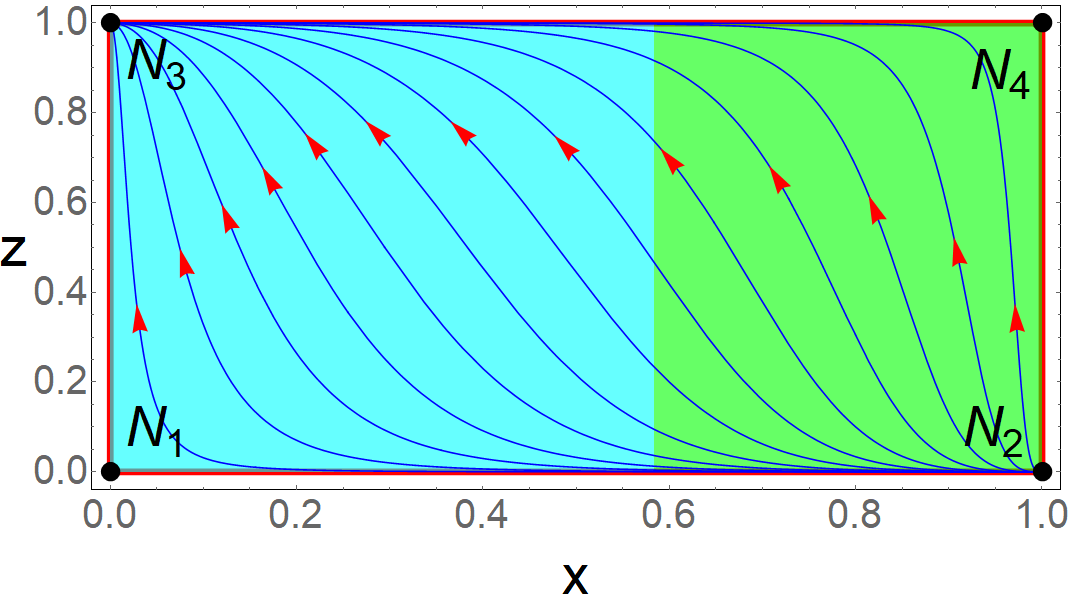}
    \includegraphics[width=0.32\textwidth]{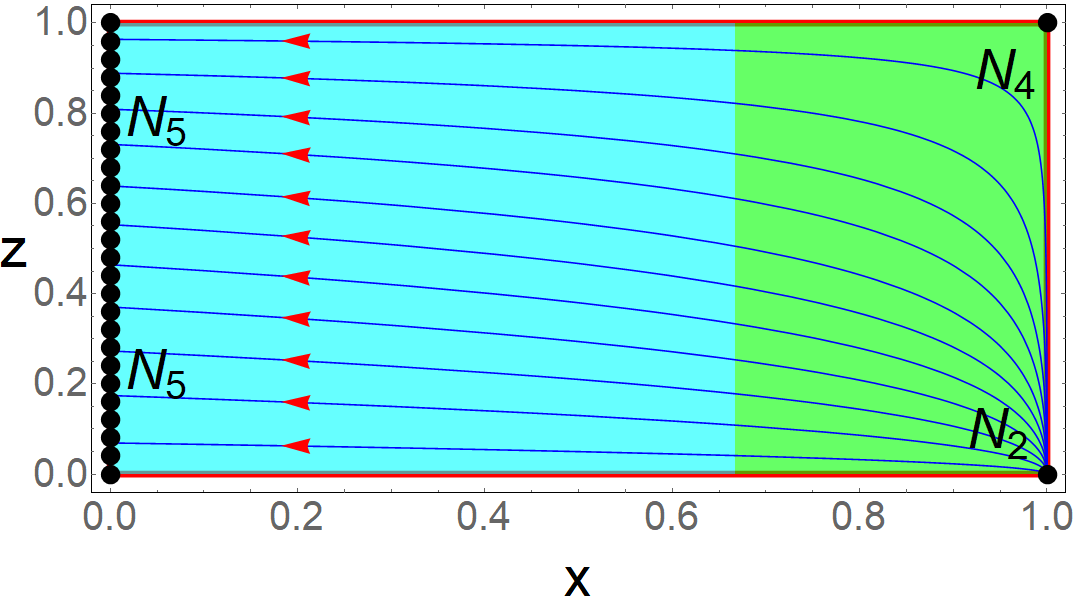}\\
    \includegraphics[width=0.33\textwidth]{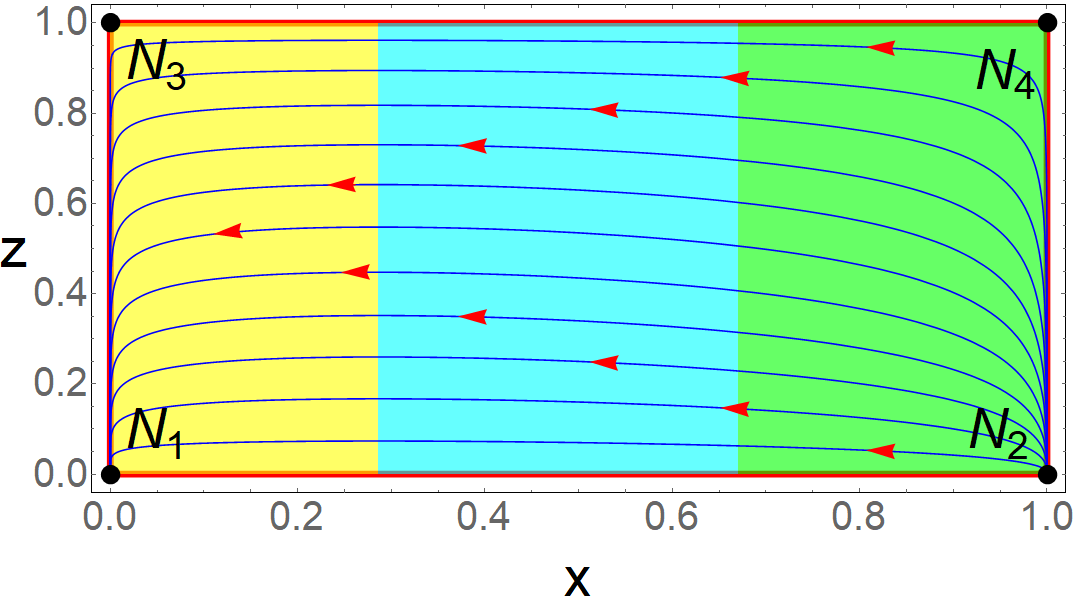}
    \includegraphics[width=0.33\textwidth]{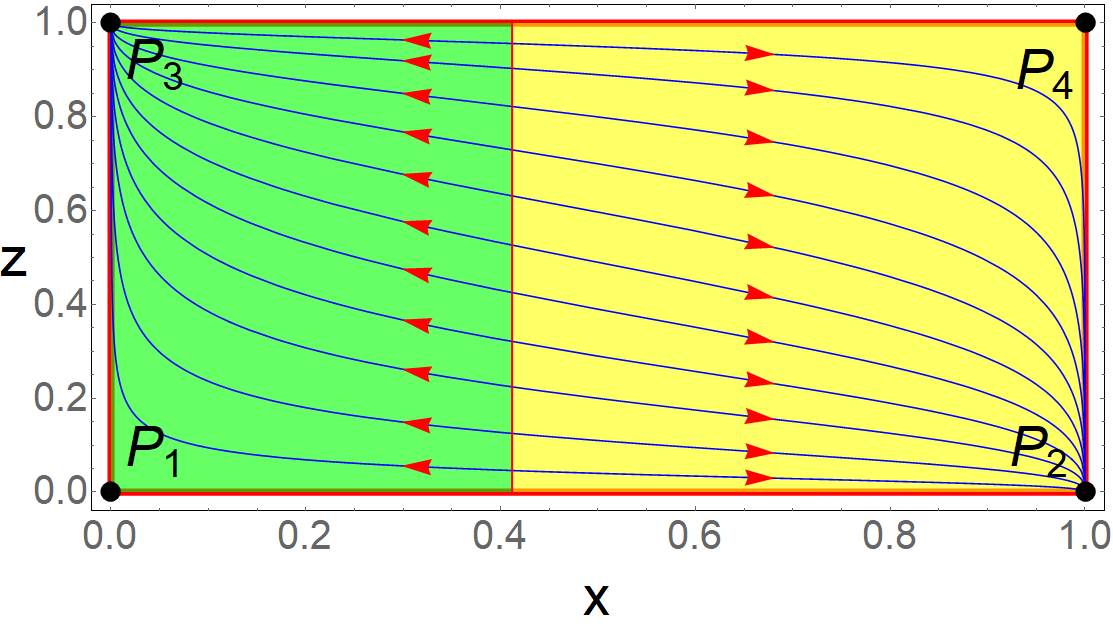}
    \includegraphics[width=0.32\textwidth]{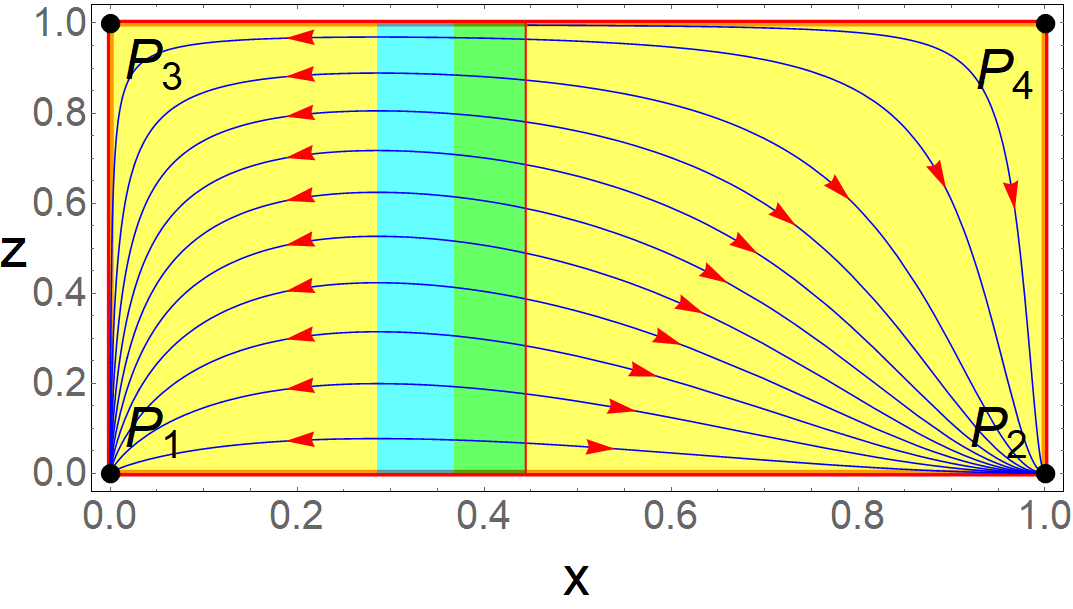}
    \caption{Description of the phase space controlled by the matter creation rate $\Gamma=2\left(1-\frac{1}{\xi}\right)\frac{\dot{H}}{H}$. {\bf Upper Left Plot:} The phase plot of the system (\ref{DS-6-new}) when we have assumed $w=0.1$ and $\xi=0.4$. {\bf Upper Middle Plot:} The phase plot of the system ($\ref{DS-6-new}$) considering $w=-0.4$ and $\xi=0.7$. {\bf Upper Right Plot:} The phase space of the system (\ref{DS-6-new}) when the EoS $w$ takes the value $-1$. Here we use $\beta =1$ but any positive value of $\beta$ gives same type of phase portrait. {\bf Lower Left Plot:} The phase plot of the system (\ref{DS-6-new}) when we assume $w=-1.4$ and $\xi=1.4$. {\bf Lower Middle Plot:} The phase plot of the systems (\ref{RG1-DS-6-new}) and (\ref{RG2-DS-6-new}) when we assume $w=0.1$ and $\xi=-0.7$. {\bf Lower Right Plot:} The phase portrait of the dynamical systems (\ref{RG1-DS-6-new}) and (\ref{RG2-DS-6-new}) when we take $w=-1.4$ and $\xi=-0.8$. Here the green region corresponds to the decelerating phase ($q >0$), the cyan region represents the accelerating phase with $-1<q<0$ and the yellow region corresponds to the super accelerating phase (i.e. $q <-1$). The red line corresponding to the singular line $x=\frac{\xi}{\xi-1}$ of the system (\ref{DS-6-new}) separates the domain ${\bf R}=[0,1]^2$ into two parts.}
    \label{fig6A}
\end{figure*}
\begin{figure*}
   \centering   
\includegraphics[width=0.33\textwidth]{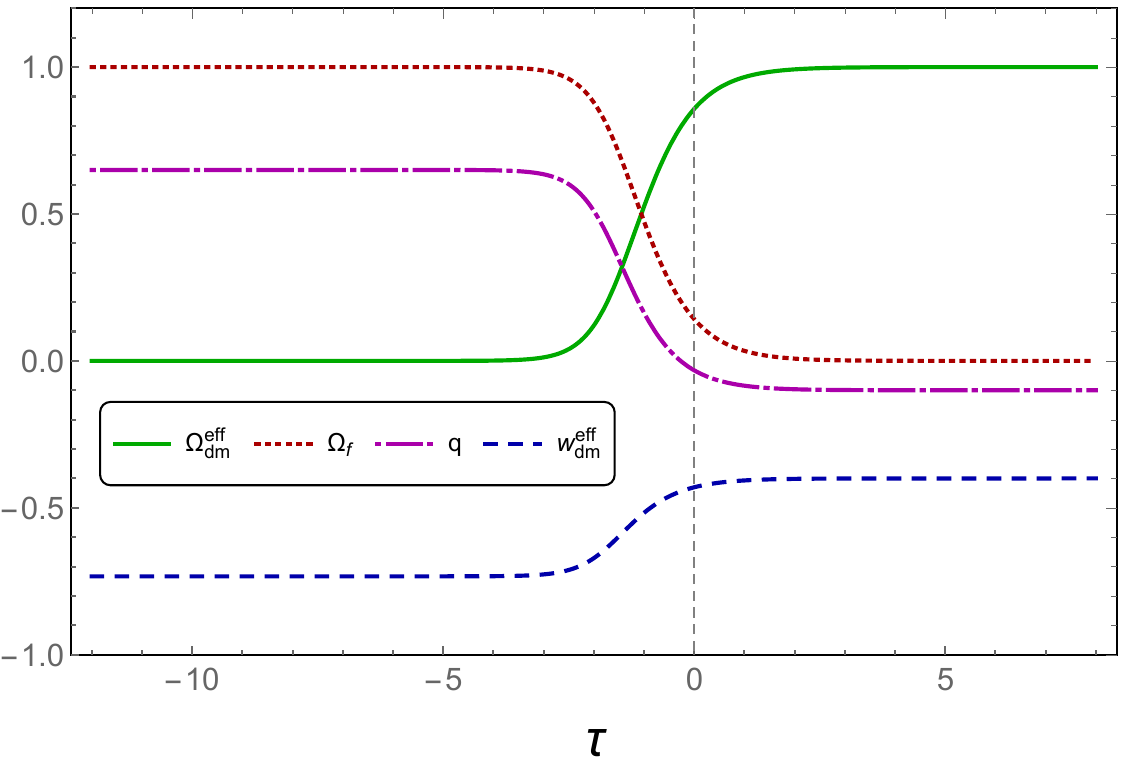} 
\includegraphics[width=0.33\textwidth]{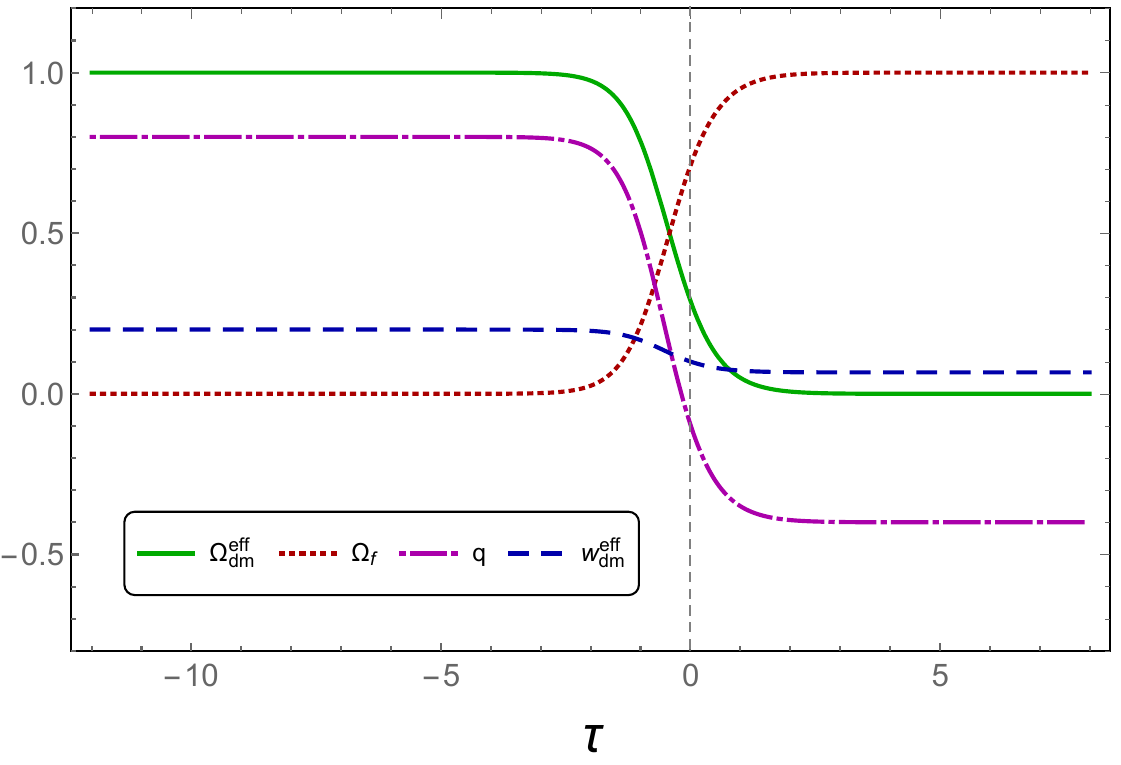}
\includegraphics[width=0.32\textwidth]{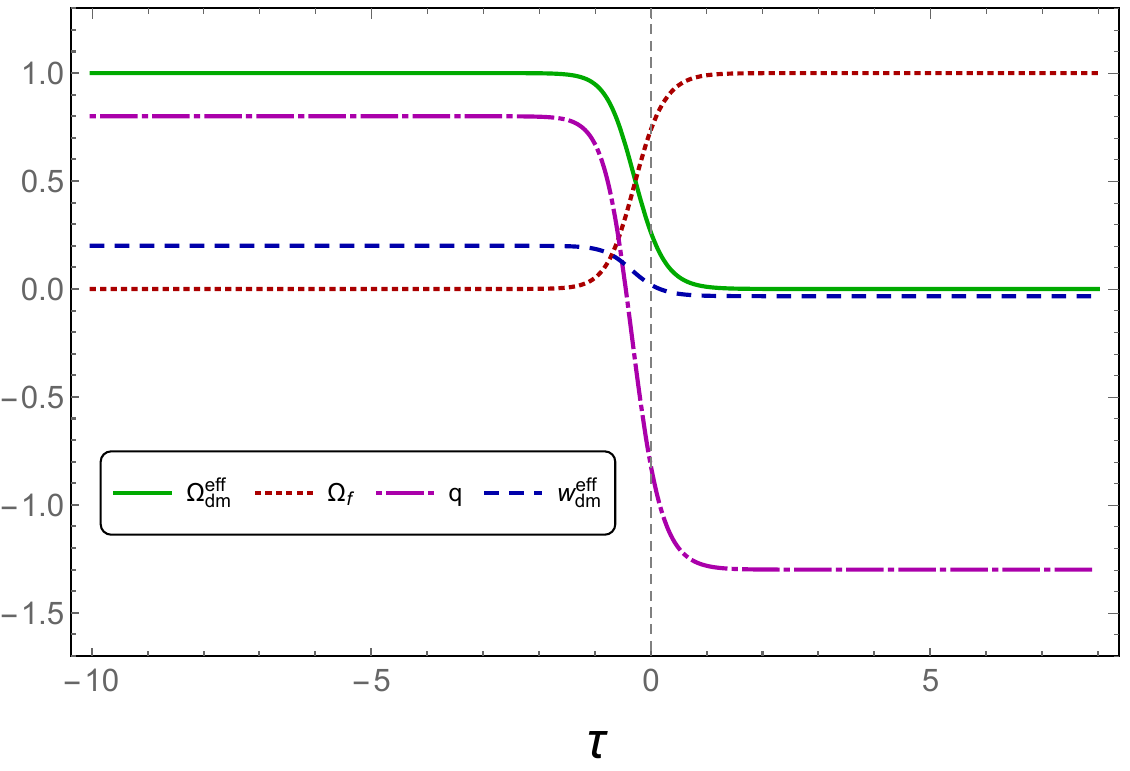} 
 \caption{The figure illustrates the evolution of key cosmological parameters namely, the dark matter density parameter $\Omega_{\rm dm}^{\rm eff}$, the second fluid density parameter $\Omega_f$, the decelerating parameter $q$, and the effective equation of state of DM $w_{\rm dm}^{\rm eff}$ for the dynamical system (\ref{DS-6-new}) with matter creation rate $\Gamma=2\left(1-\frac{1}{\xi}\right)\frac{\dot{H}}{H}$. {\bf Left Plot:} We have drawn this plot using the model parameters $w=0.1$ and $\xi=0.6$. Here, the trajectories end in DM dominated accelerated quintessence era. As no decelerating DM dominated saddle point emerges (see the upper left plot of Fig. \ref{fig6A}), a past DM-dominated phase is not achievable in this case.  {\bf Middle Plot:} This plot is produced adopting the values of the model parameters $w=-0.6$ and $\xi=1.2$ where the late time evolution of the universe is attracted by DE dominated quintessence era connecting through a DM dominated decelerated era. {\bf Right Plot:} For this plot, we take $w=-1.2$ and $\xi=1.2$ where one can see that trajectories are attracted by DE dominated phantom era at late-times joining through a DM dominated decelerated phase.} 
    \label{fig-evo-6}
\end{figure*}
\begin{figure}
    \centering
    \includegraphics[width=0.497\textwidth]{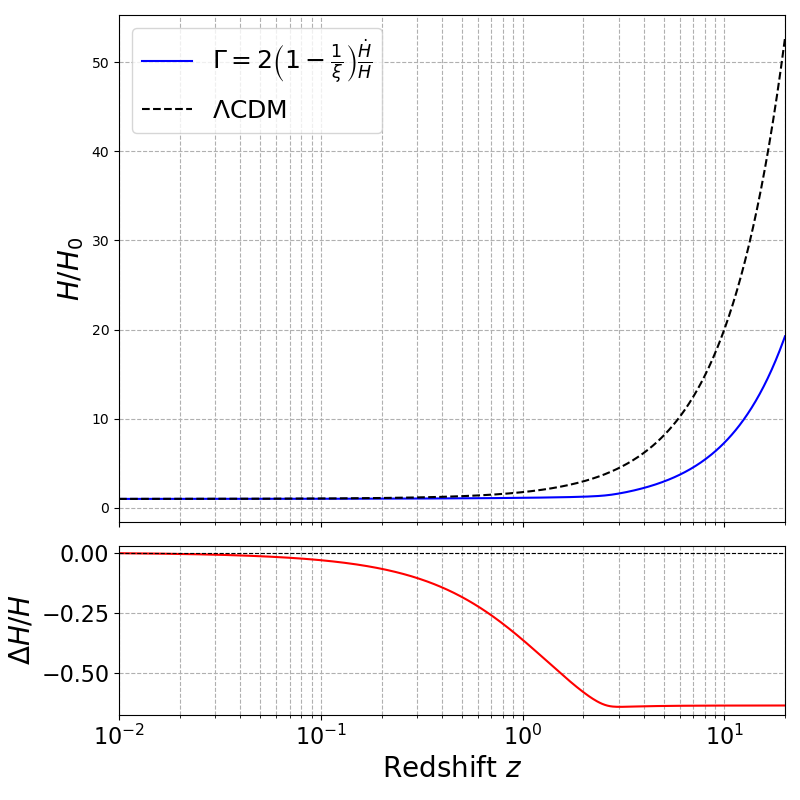}
    \caption{The redshift evolution of $H/H_0$ for the matter creation model $\Gamma= 2\left(1-\frac{1}{\xi}\right)\frac{\dot{H}}{H}$ and the $\Lambda$CDM model (upper panel) and the fractional difference $\Delta H/H = (H_{\rm Model} - H_{\Lambda {\rm CDM}})/H_{\Lambda {\rm CDM}}$ (lower panel) have been shown. For the matter creation model we have taken $w= 0$, $\xi= 0.1$, $\Omega_{f0} =0.04$ while for the $\Lambda$CDM model we set the matter density parameter at present $\Omega_{m0} =0.3$. } 
    \label{fig:exp-hist-M6}
\end{figure}
%%%%%%%   Table 7  %%%%%%%%
\begin{table*}%[t]
\centering
\resizebox{1.0\textwidth}{!}{%
	\begin{tabular}{|c c c c c c c c|}\hline\hline
{\bf Critical point} & {\bf Existence} & {\bf Eigenvalue} & {\bf Stability} & $\mathbf{\Omega_f}$ & $\mathbf{\Omega_{\rm dm}^{\rm eff}}$ & $\mathbf{q}$ & {\bf Acceleration} \\ \hline
%   &&&&&     \\

$S_{1}(0,0)$  & $4(1-2z)+\left(5-10z+9z^2\right)x>0$ & $\left(6(1+w),3(5+9w)\right)$  & {\bf Stable} if $ w < -1 $;  &  1  &  0  &  $\frac{1}{2}(1+3w)$  &  $w<-\frac{1}{3}$  \\  
&   &    & Saddle if $-1<w<-\frac{5}{9}$;   &    &    &    &  \\ 
&   &    & Unstable if $w>-\frac{5}{9}$   &    &    &    & \\ \hline
 
$S_{2}(1,0)$  & $4(1-2z)+\left(5-10z+9z^2\right)x>0$ & $\left(6,-3(5+9w)\right)$   &  Saddle if $w>-\frac{5}{9}$;   & 0   &  1  &  $-\frac{1}{3}$  & Yes \\
 &&& Unstable if $ w < -\frac{5}{9} $ &&&&  \\  \hline
%   &&&&&  \\
$S_{3}\left(1,\frac{1}{2}\right)$  & $4(1-2z)+\left(5-10z+9z^2\right)x>0$  & $\left(-3,-\frac{27}{4}(1+w)\right)$   & {\bf Stable} if $w>-1$;   & 0  &  1  & $-1$   & Yes  \\
 &&& Saddle if $w<-1$ &&&&  \\    \hline
%   &&&&&  \\
$S_{4}(0,1)$  & $4(1-2z)+\left(5-10z+9z^2\right)x<0$  & $\left(-12,-6(1+w)\right)$  & {\bf Stable} if $w>-1$;  &  1  & 0   &  $\frac{1}{2}(1+3w)$  & $w<-\frac{1}{3}$ \\ 
&   &    & Saddle if $w<-1$   &    &    &    & \\ \hline
%   &&&&&  \\ 
$S_{5}\left(0,z_c\right)$  &  $w=-1$, $z_c\neq\frac{1}{2}$  & $(\lambda_\pm,0)$  &  {\bf Stable} &  1  &  0  &  $-1$  &  Yes     \\ \hline
%   &&&&& \\
 $S_6\left(x_c,\frac{1}{2}\right)$  &  $4(1-2z)+\left(5-10z+9z^2\right)x>0$, & $\left(-3x,0\right)$  & {\bf Stable} &  $1-x_c$  &  $x_c$  &  $-1$  &  Yes     \\ 
  & $w=-1$, $x_c\neq 0$  &  &   &&& &    \\ \hline 

 $S_7\left(x_c,0\right)$  &  $4(1-2z)+\left(5-10z+9z^2\right)x>0$, & $\left(\frac{2}{3}(4+5x),0\right)$  & Unstable  &  $1-x_c$  &  $x_c$  &  $-\frac{1}{3}$  &  Yes     \\
 &  $w=-\frac{5}{9}$ &   &      &    &    &    &       \\ \hline \hline
\end{tabular}%
 }
\caption{Summary of the critical points, their existence, stability and the values of the cosmological parameters at those points for the dynamical systems (\ref{RG1-DS-7}) and (\ref{RG2-DS-7}) with the matter creation rate $\Gamma=-\frac{4H_0^2+5H^2}{2H\left(H^2-H_0^2\right)}\dot{H}$. The eigenvalues $\lambda_\pm$ are given by $\lambda_\pm=\pm 12\left(2z_c-1\right)$. In this context, $\lambda_+$ is associated with the interval $0\leq z_c<\frac{1}{2}$, whereas $\lambda_-$ applies when $\frac{1}{2}<z_c\leq 1$.}
	\label{seventh-table}
\end{table*}
\begin{figure*}
    \centering
    \includegraphics[width=0.33\textwidth]{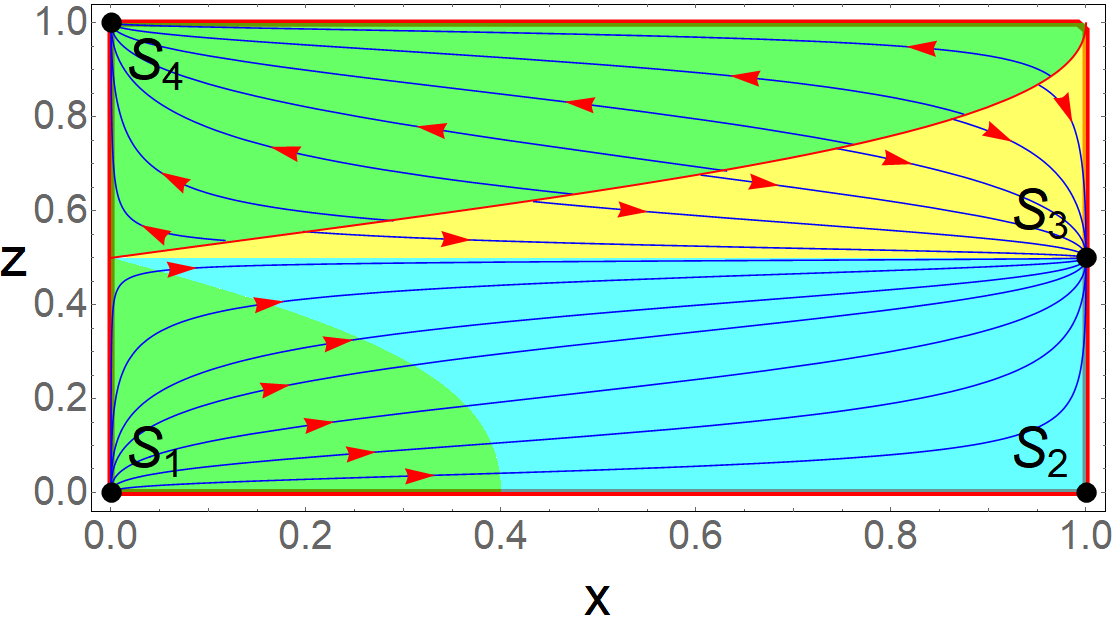}
    \includegraphics[width=0.33\textwidth]{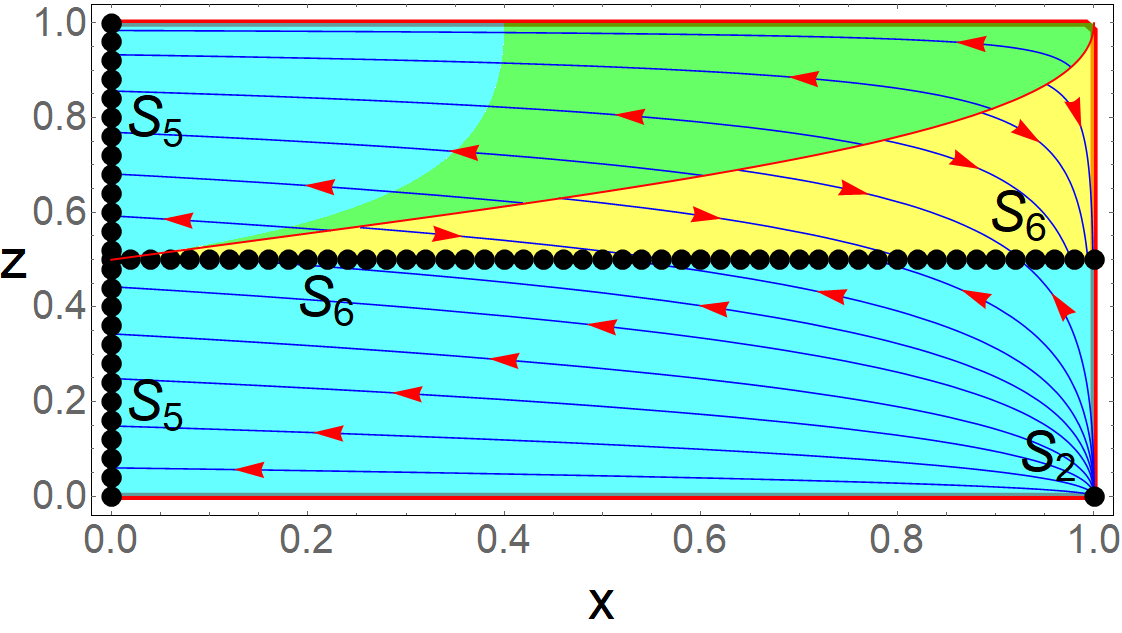}
    \includegraphics[width=0.32\textwidth]{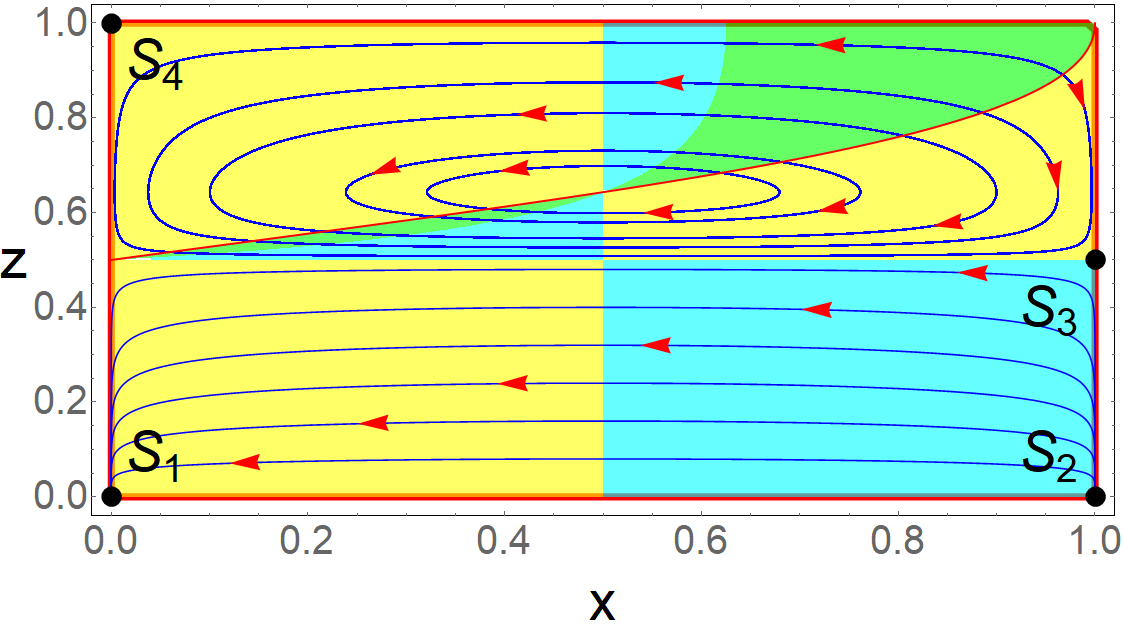}
    \caption{Description of the phase space controlled by the matter creation rate $\Gamma=-\frac{4H_0^2+5H^2}{2H\left(H^2-H_0^2\right)}\dot{H}$. {\bf Left Plot:} The phase plot of the system (\ref{RG1-DS-7}) and (\ref{RG2-DS-7}) when we have assumed $w=0.1$. {\bf Middle Plot:} The phase space of the system (\ref{RG1-DS-7}) and (\ref{RG2-DS-7}) when the EoS $w$ takes the value $-1$. {\bf Right Plot:} The phase plot of the system (\ref{RG1-DS-7}) and (\ref{RG2-DS-7}) considering $w=-2$. Here the green region corresponds to the decelerating phase ($q >0$), the cyan region represents the accelerating phase with $-1<q<0$ and the yellow region corresponds to the super accelerating phase (i.e. $q <-1$). The red solid curve indicating the singular curve $4(1-2z)+\left(5-10z+9z^2\right)x=0$ divides the domain ${\bf R}=[0,1]^2$ into two parts.} 
    \label{fig7A}
\end{figure*}
\begin{figure}
   \centering   
\includegraphics[width=0.495\textwidth]{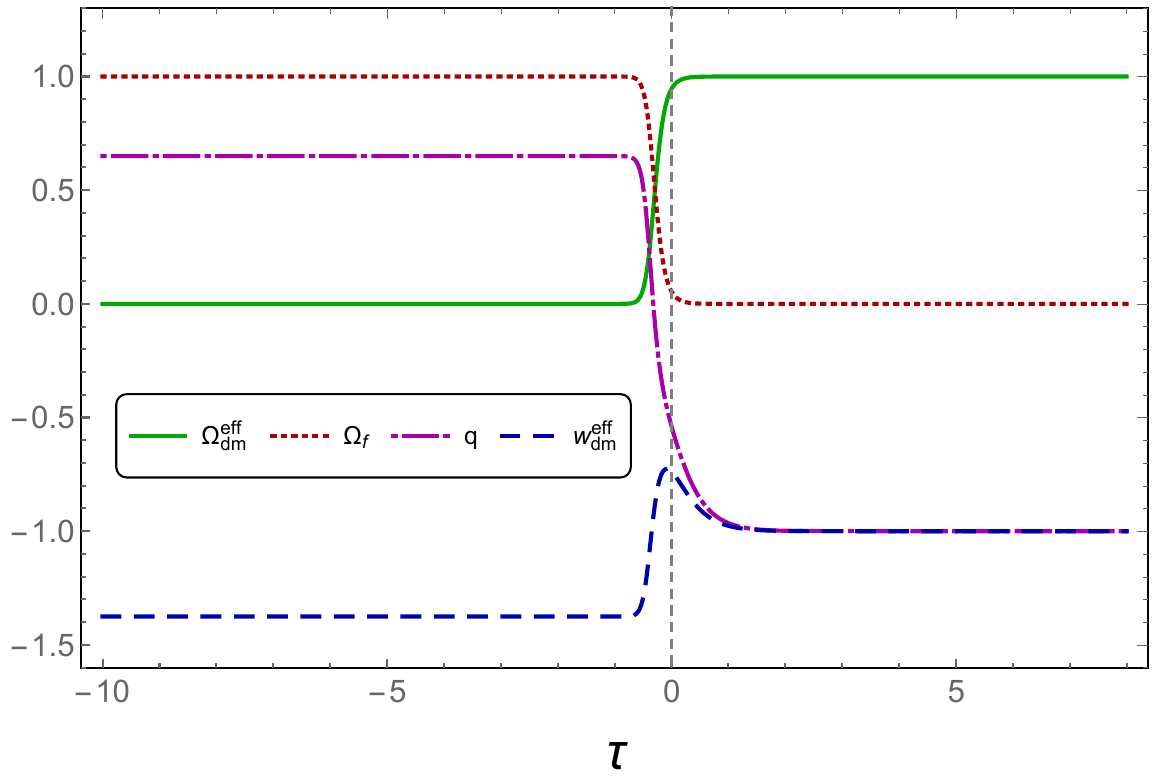}  
 \caption{The figure represents the evolution of the different cosmological parameters like the DM density parameter $\Omega_{\rm dm}^{\rm eff}$, the second fluid density parameter $\Omega_f$, the decelerating parameter $q$, and the effective equation of state of DM $w_{\rm dm}^{\rm eff}$ of the dynamical system (\ref{RG1-DS-7}) for the matter creation rate $\Gamma=-\frac{4H_0^2+5H^2}{2H\left(H^2-H_0^2\right)}\dot{H}$ with the EoS $w=0.1$. Numerical simulation reveals that the final fate of the universe is attracted by DM dominated accelerated cosmological constant era. The absence of a decelerating DM dominated saddle point, as seen in the left plot of Fig. \ref{fig7A}, rules out the possibility of a past DM dominated phase.} 
    \label{fig-evo-7}
\end{figure}
\begin{figure}
    \centering
    \includegraphics[width=0.497\textwidth]{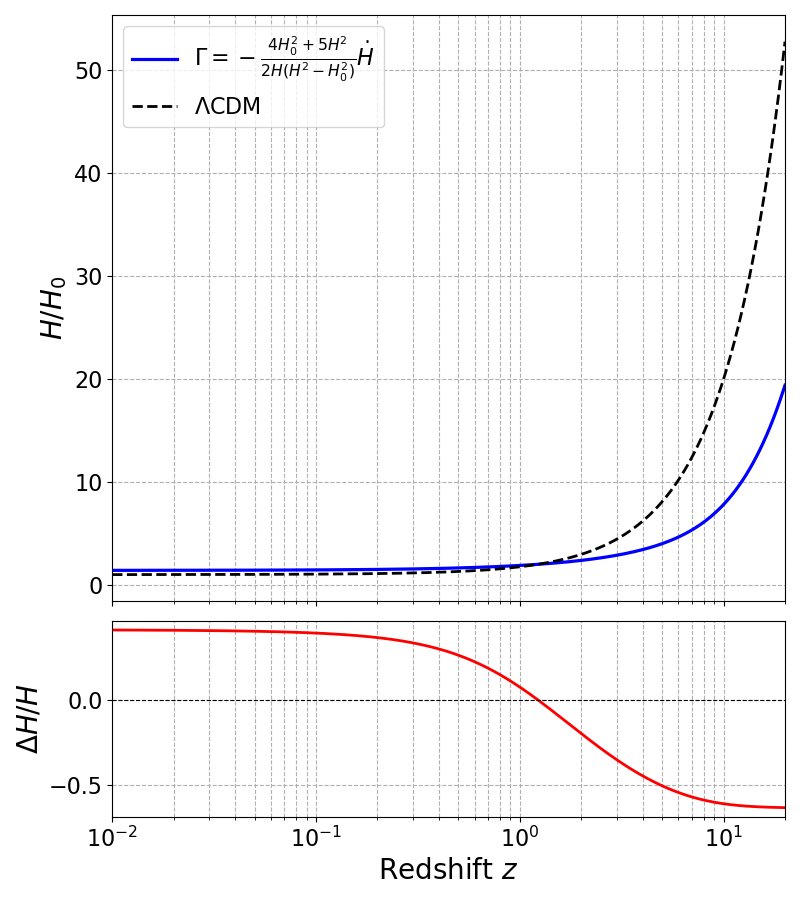}
    \caption{The redshift evolution of $H/H_0$ for the matter creation model $\Gamma=-\frac{4H_0^2+5H^2}{2H(H^2-H_0^2)}\dot{H}$ and the $\Lambda$CDM model (upper panel) and the fractional difference $\Delta H/H = (H_{\rm Model} - H_{\Lambda {\rm CDM}})/H_{\Lambda {\rm CDM}}$ (lower panel) have been shown. For the matter creation model we have taken $w= 0$, $\Omega_{f0} =0.04$ while for the $\Lambda$CDM model we set the matter density parameter at present $\Omega_{m0} = 0.3$.}
    \label{fig:exp-hist-M7}
\end{figure}

{\bf When $\xi>0$}, the dynamical system (\ref{DS-6-new}) yields four isolated critical points: $N_1(0,0)$, $N_2(1,0)$, $N_3(0,1)$ and $N_4(1,1)$, all of which are independent of the model parameters. Additionally, the system exhibits three critical lines: $N_5(0,z_c)$, $N_6(x_c,0)$ and $N_7(x_c,1)$. The conditions for existence, eigenvalues, stability characteristics, density parameters, and decelerating parameter corresponding to these critical points and lines are presented in the Table. \ref{six-table}.

The critical point $N_1$ is second fluid dominated point and it is stable for $\xi>1+w$ with $w<-1$ (see the lower left plot in Fig. \ref{fig6A}), saddle for $\xi>1+w$ with $w>-1$ and unstable for $\xi<1+w$. This point gives accelerated solution if $w<-\frac{1}{3}$ and decelerated solution if $w>-\frac{1}{3}$.

The critical point $N_2$ shows DM dominated accelerated solution for $\xi<\frac{2}{3}$ and decelerated solution if $\xi>\frac{2}{3}$. This point is saddle for $\xi<1+w$ and unstable for $\xi>1+w$.

Again, the point $N_3$ corresponds to DM dominated and is accelerated (decelerated) for $w<-\frac{1}{3}$ ($w>-\frac{1}{3}$). The point is stable if the model parameter satisfies the relation $\xi>1+w$ with $w>-1$ (shown in the upper middle plot in Fig. \ref{fig6A}), otherwise it is saddle by nature.

Similarly to the critical point $N_2$, the point $N_4$ is dominated by the second fluid and has the same accelerated and decelerated region. This point gives stable behavior in $\xi<1+w$ (in the upper left plot in Fig. \ref{fig6A}) and saddle nature in $\xi>1+w$.  

The critical line $N_5$ appears only for $w=-1$. This critical line consists of non-hyperbolic critical points and serves as an attractor (see the upper right plot in Fig. \ref{fig6A}). The critical line $N_5$ represents DE dominated accelerated solution.

The critical lines $N_5$ and $N_6$ exist for $\xi=1+w$. As $\xi>0$, the relation implies $w>-1$. Thus, these two lines indicate accelerated (decelerated) solution for $w<-\frac{1}{3}$ $\left(w>-\frac{1}{3}\right)$. It also follows that at these critical lines DM and DE coexist. The critical line $N_6$ is unstable in nature, while $N_7$ shows stable characteristics. 

The evolution of key cosmological parameters: the DM density parameter $\Omega_{\rm dm}^{\rm eff}$, the second fluid density parameter $\Omega_f$, the decelerating parameter $q$, and the effective EoS of DM $w_{\rm dm}^{\rm eff}$ are shown in Fig. \ref{fig-evo-6} as a function of the time variable $\tau$ for various values of the model parameters $\xi$ and $w$. The left plot in Fig. \ref{fig-evo-6} illustrates the universe’s evolution from a second-fluid-dominated decelerating epoch (a past DM dominated epoch is absent) to a DM dominated phase characterized by quintessence-type accelerated expansion. Here, $w_{\rm dm}^{\rm eff}$ maintains a negative value, acting effectively as dark energy. The numerical solution shown in the middle plot of Fig. \ref{fig-evo-6} depicts the cosmic evolution from a DM dominated decelerated phase at early times to a DE dominated era exhibiting late-time accelerated expansion consistent with quintessence, and $w_{\rm dm}^{\rm eff}$ stays positive during the entire evolution. The right plot of Fig. \ref{fig-evo-6} presents the evolution of the universe from an early stage dominated by DM to a phase of accelerated expansion at late times, driven by an effective DE component exhibiting phantom-like behavior. From the plot, $w_{\rm dm}^{\rm eff}$ is observed to be slightly negative during the late phase of evolution.

{\bf When $\xi<0$}, the system of equations (\ref{DS-6-new-x})-(\ref{DS-6-new-z}) exhibits a singularity along the line $x=\frac{\xi}{\xi-1}$. To address this, we partition the domain ${\bf R}$ into two subdomains: the first for $0\leq x<\frac{\xi}{\xi-1}$, and the second for $\frac{\xi}{\xi-1}<x\leq 1$. We then perform a phase space stability analysis separately within each subdomain. In the first region, the system is regularized by introducing a new time variable $\tau$, defined as $dN=\left(\frac{\xi}{\xi-1}-x\right)d\tau$, while in the second region, we use $dN=\left(x-\frac{\xi}{\xi-1}\right)d\tau$.

In the subdomain corresponding to $0\leq x<\frac{\xi}{\xi-1}$, the original dynamical system (\ref{DS-6-new}) can be expressed in a regularized form as: 
\begin{subequations} \label{RG1-DS-6-new}
\begin{align}
    \frac{dx}{d\tau}=& -\frac{3x(1-x)(1+w-\xi)}{1-\xi},   \label{RG1-DS-6-new-x} \\
    \frac{dz}{d\tau}=& -\frac{3\xi z(1-z)(1+w(1-x))}{2(1-\xi)}. \label{RG1-DS-6-new-z} 
\end{align} 
\end{subequations} 
We now intend to characterize the qualitative behavior of the system \eqref{RG1-DS-6-new} in the subdomain $0\leq x<\frac{\xi}{\xi-1}$. The system \eqref{RG1-DS-6-new} admits critical points $P_1(0,0)$, $P_3(0,1)$ and critical lines $P_5\left(0,z_c\right)$, $P_6(x_c,0)$ and $P_7(x_c,1)$ with $x_c<\frac{\xi}{\xi-1}$. These are detailed in Table. \ref{six-table}, which includes eigenvalues, stability information, and relevant cosmological parameters.  

Both the critical points $P_1$, $P_3$ are DM dominated points, representing acceleration for $w<-\frac{1}{3}$ and deceleration for $w>-\frac{1}{3}$. The point $P_1$ is stable (shown in the lower right plot in Fig. \ref{fig6A}) if the condition $\xi<1+w$ along with $w<-1$ is satisfied, whereas,  $P_3$ is stable (in the lower middle plot in Fig. \ref{fig6A}) if $\xi<1+w$, $w>-1$.

The critical line $P_5$ exists when the second fluid's EoS coincides with the cosmological constant. It is always stable and indicates DE dominated accelerated solution.

The critical lines $P_6$ and $P_7$ appear for $\xi=1+w$. As $\xi<0$, the relation indicates $w<-1$. Thus, these two lines represent an accelerated (decelerated) solution for $w<-\frac{1}{3}$ $\left(w>-\frac{1}{3}\right)$. It is also noted that at these critical lines DM and DE coexist. The critical line $P_6$ is stable, while $P_7$ shows unstable behavior.

For the subdomain $\frac{\xi}{\xi-1}<x\leq1$ within ${\bf R}$, we write the dynamical system (\ref{DS-6-new}) in a regularized form as follows:
\begin{subequations} \label{RG2-DS-6-new}
\begin{align}
    \frac{dx}{d\tau}=& \frac{3x(1-x)(1+w-\xi)}{1-\xi},   \label{RG2-DS-6-new-x} \\
   \frac{dz}{d\tau}=& \frac{3\xi z(1-z)(1+w(1-x))}{2(1-\xi)}. \label{RG2-DS-6-new-z} 
\end{align} 
\end{subequations}
In the subdomain $\frac{\xi}{\xi-1}<x\leq1$, the autonomous system \eqref{RG2-DS-6-new} reveals two isolated critical points, $P_2(1,0)$ and $P_3(1,1)$,  as well as two critical lines, $P_6(x_c,0)$ and $P_7(x_c,1)$ with $x_c>\frac{\xi}{\xi-1}$. Detailed characteristics of these solutions are listed in Table. \ref{six-table}.

Both the critical points $P_2$, $P_4$ are the second fluid dominated accelerating solutions. The point $P_2$ is stable (see the lower middle and right plots in Fig. \ref{fig6A}) if $\xi<1+w$ where $P_4$ is saddle along with $w<-1$ is satisfied. Again, $P_4$ is unstable in the parameter space $\xi<1+w$ where $P_2$ is saddle. It is also noted that we have already discussed about the qualitative features of the critical lines $P_6$ and $P_7$ in the previous section.

We close this section with Fig. \ref{fig:exp-hist-M6} showing the redshift evolution of $H/H_0$ for the matter creation model  $\Gamma=2\left(1-\frac{1}{\xi}\right)\frac{\dot{H}}{H}$ with respect to the $\Lambda{\rm CDM}$ model (upper panel) alongside the fractional difference $\Delta H/H$ (lower panel). 
We find that the model aligns closely with $\Lambda{\rm CDM}$ at low redshifts, while deviations grow at higher redshifts due to particle creation, highlighting potential observational signatures.

\subsection{Model: $\Gamma=-\frac{4H_0^2+5H^2}{2H\left(H^2-H_0^2\right)}\dot{H}$}\label{model-7}
Finally, we consider a model governed by the inhomogeneous matter creation rate $\Gamma=-\frac{4H_0^2+5H^2}{2H\left(H^2-H_0^2\right)}\dot{H}$, which is unique among the series of the present matter creation models as it does not involve any free parameter.

Based on this matter creation rate, the corresponding autonomous system can be formulated in terms of the dimensionless variables $x$ and $z$ as 
\begin{subequations} \label{DS-7}
\begin{align}
    x'=& \frac{3x(1-x)\left(5+9w(1-z)^2-10z+9z^2\right)}{4(1-2z)+\left(5-10z+9z^2\right)x},   \label{DS-7-x} \\
    z'=& \frac{6z(1-z)(1-2z)(1+w(1-x))}{4(1-2z)+\left(5-10z+9z^2\right)x}. \label{DS-7-z} 
\end{align} 
\end{subequations}
In this final model, the phase space region ${\bf R}$ contains a singularity for the above system along the curve $4(1-2z)+\left(5-10z+9z^2\right)x=0$. Accordingly, we partition the domain into two regions: one in which this expression is positive and another in which it is negative. The corresponding deceleration parameter is given by $q=\frac{2(1-2z)-\left(5-10z+9z^2\right)x+6w(1-x)(1-2z)}{4(1-2z)+\left(5-10z+9z^2\right)x}$. The phase space stability analysis is conducted independently in each region.

{\bf When} the condition $4(1-2z)+\left(5-10z+9z^2\right)x>0$ is satisfied, we adopt a new time variable $\tau$, defined through $dN=\left(4(1-2z)+\left(5-10z+9z^2\right)x\right)d\tau$ to obtain the following regularization of the system (\ref{DS-7}) as: 
\begin{subequations} \label{RG1-DS-7}
\begin{align}
    \frac{dx}{d\tau}=& 3x(1-x)\left(5+9w(1-z)^2-10z+9z^2\right),   \label{RG-DS-7-x} \\
    \frac{dz}{d\tau}=& 6z(1-z)(1-2z)(1+w(1-x)). \label{RG-DS-7-z} 
\end{align} 
\end{subequations}
The regularized system \eqref{RG1-DS-7} is topologically equivalent to the system (\ref{DS-7}) in the subdomain $4(1-2z)+\left(5-10z+9z^2\right)x>0$ of ${\bf R}$. In the subdomain $4(1-2z)+\left(5-10z+9z^2\right)x>0$ of ${\bf R}$, the system \eqref{RG1-DS-7} possesses three isolated critical points $-$ $S_1(0,0)$, $S_2(1,0)$, and $S_3\left(1,\frac{1}{2}\right)$ $-$ alongside three critical lines: $S_5(0,z_c)$, $S_6\left(x_c,\frac{1}{2}\right)$, and $S_7(x_c,0)$. Their corresponding properties are listed in Table. \ref{seventh-table}.

The point $S_1$ is second fluid dominated accelerating and decelerating solution for $w<-\frac{1}{3}$ and $w>-\frac{1}{3}$, respectively. Examining of the eigenvalues, one can conclude that this point is stable if $w<-1$ (see the right plot in Fig. \ref{fig7A}), saddle if $-1<w<-\frac{5}{9}$ and unstable if $w>-\frac{5}{9}$.

The critical point $S_2$ corresponds to DM dominated accelerating solution. It is either saddle for $w>-\frac{5}{9}$ or unstable for $w<-\frac{5}{9}$.

The point $S_3$ also indicates DM dominated accelerated solution. Investigating the eigenvalues, one can say that this point is stable for $w>-1$ (shown in the left plot in Fig. \ref{fig7A}) and saddle for $w<-1$.

The critical line $S_5$ exists if conditions $w=-1$ and $z_c\neq\frac{1}{2}$ are satisfied and consists of non-hyperbolic critical points. For $z_c<\frac{1}{2}$, it indicates that this critical line always behaves like an attractor (depicted in the middle plot in Fig. \ref{fig7A}) and represents DE dominated accelerated solution.

The critical line $S_6$ where both the fluid coexist, appears also for $w=-1$ and $x_c\neq 0$. The non-hyperbolic critical points lie on this line are always accelerated and the critical line acts as an attractor (in the middle plot in Fig. \ref{fig7A}).

Again, the line $S_7$ where DM and second fluid coexist, arises for $w=-\frac{5}{9}$. It is accelerating and unstable.

The evolution of the DM and the second fluid density parameters, $\Omega_{\rm dm}^{\rm eff}$ and $\Omega_f$, along with the decelerating parameter $q$ and the effective EoS of DM $w_{\rm dm}^{\rm eff}$, is depicted in Fig. \ref{fig-evo-7} as a function of the time variable $\tau$, for the EoS parameter $w=0.1$. Numerical simulation describes a smooth transition from a decelerating universe dominated by the second fluid to an accelerated expansion regime, where DM dominance leads to dynamics analogous to those of a cosmological constant.  It is also observed that a sufficiently long DM dominated era is absent, and $w_{\rm dm}^{\rm eff}$ remains negative through out the evolution, behaving like dark energy. 

{\bf When $4(1-2z)+\left(5-10z+9z^2\right)x<0$}, the regularization of the system (\ref{DS-7}) is carried out by introducing a new time variable via $dN=-\left(4(1-2z)+\left(5-10z+9z^2\right)x\right)d\tau$. This results in the following regularized system which is topologically equivalent with the system (\ref{DS-7}) in the subdomain $4(1-2z)+\left(5-10z+9z^2\right)x<0$ of ${\bf R}$.
\begin{subequations} \label{RG2-DS-7}
\begin{align}
    \frac{dx}{d\tau}=&- 3x(1-x)\left(5+9w(1-z)^2-10z+9z^2\right),   \label{RG2-DS-7-x} \\
    \frac{dz}{d\tau}=& -6z(1-z)(1-2z)(1+w(1-x)). \label{RG2-DS-7-z} 
\end{align} 
\end{subequations}
In the subdomain $4(1-2z)+\left(5-10z+9z^2\right)x<0$ of ${\bf R}$, the system \eqref{RG2-DS-7} admits one isolated critical point $S_4(0,1)$, and a critical line $S_5(0,z_c)$ as presented in Table \ref{seventh-table}.

The critical point $S_4$ represents second fluid dominated solution, indicating the accelerating phase for $w<-\frac{1}{3}$ and decelerating phase for $w>-\frac{1}{3}$. Examining the eigenvalues, it is clear that $S_4$ is stable for $w>-1$ (highlighted in the left plot in Fig. \ref{fig7A}) and saddle for $w<-1$.

In this region, the critical line $S_5$ appears for $w=-1$ and $z_c\neq\frac{1}{2}$ and for $\frac{1}{2}<z_c\leq 1$, characterizing itself as an attractor, representing DE dominated accelerating solutions.

In Fig. \ref{fig:exp-hist-M7} we present $H/H_0$ for the matter creation model $\Gamma=-\frac{4H_0^2+5H^2}{2H\left(H^2-H_0^2\right)}\dot{H}$ versus $\Lambda{\rm CDM}$ (upper panel), with $\Delta H/H$ in the lower panel. We see that the evolution matches $\Lambda{\rm CDM}$ at low redshifts, but noticeable deviations appear at higher redshifts.

Finally, we performed the stability analysis of other matter creation rates that are more general compared to the one parameter matter creation models discussed above. The models and their analyses are presented in  {\bf Appendix-A}. However, according to the results, these new models do not add anything significantly new compared to what we have observed in the one parameter matter creation models.

\section{Summary and Conclusions}
\label{sec-summary}

Matter creation cosmology was proposed as an alternative to the DE and MG theories and because of this elegant nature, this came to the limelight of modern cosmology. In this article we have studied matter creation cosmologies by applying the powerful technique of dynamical analysis aiming to closely examine whether such cosmological scenarios can predict various phases of our universe's evolution. We have considered a two-fluid cosmological scenarios consisting of a pressure-less DM responsible for the matter creation  and a perfect fluid with constant EoS $w$. 

 We have assumed various matter creation rates and performed their stability analysis. The results are summarized in Tables~\ref{first-table} --  \ref{seventh-table} and the  graphical presentations are shown in Figs. \ref{fig1A} -- \ref{fig-evo-7}.

The main ingredient in this context is the EoS of the second fluid, $w$, which plays a pivotal role in the dynamics of the models. Since the matter creation models were designed mainly for avoiding the need of DE characterized by negative pressure, equivalently, $w< 0$, therefore, scenarios with $w \geq 0$ are appealing in this context. Whilst in order to be complete in this direction, we have also considered $w < 0$. According to our results, for both  $w\geq 0$ and $w < 0$, we have obtained a variety of interesting critical points.   

\subsection{$w \geq 0$} 

Considering the case with $w\geq 0$, we find that all the one parameter matter creation models correctly predict a transition of our universe from its past decelerating phase to the present accelerating universe, but the final state of the universe differs for different models. For $\Gamma = \Gamma_0$, $\Gamma_0 H^{-1}$ and $\Gamma_0 H^{-2}$, we see the following pattern of our universe: {\it past decelerating phase (dominated by the DM)} $\longrightarrow$ {\it present accelerating universe with  $-1 \leq  q < 0$} $\longrightarrow$  {\it a DM dominated accelerating phase with $q =-1$ }. For $\Gamma = \Gamma_0 H$, $2\left(1-\frac{1}{\beta}\right)\frac{\dot{H}}{H}$ and $-\frac{4H_0^2+5H^2}{2H\left(H^2-H_0^2\right)}\dot{H}$, we only notice that our universe transits from its {\it past decelerating phase dominated by the second fluid, without the occurrence of any DM dominated epoch to the present accelerating phase with $-1 \leq q < 0$ and continues accelerating dominated only by DM}. The model $\Gamma = \Gamma_0 H^2$ gives some interesting results related to the final stage of the universe in the future. In this case the pattern of the universe is as follows: {\it past decelerating phase} $\longrightarrow$ {\it present accelerating phase with $-1 < q < 0$} $\longrightarrow$ {\it DM dominated decelerating phase (i.e. $q >0$)}/ {\it scaling decelerating phase} ($(\Omega_{\rm dm}^{\rm eff}, \Omega_{f}) \neq (0, 0)$)/ {\it DM dominated super accelerating phase (i.e. $q < -1$)}. 

A similar trend is noticed when we consider the remaining matter creation models with more than one parameter. That means there is nothing new in the dynamics when we extend the one parameter models considered in this article (see again Appendix-A).

\subsection{$w < 0$} 

It should be noted that $w<0$ can be divided into three parts as follows: $-1< w< 0$, $w =-1$ (cosmological constant) and $w< -1$ (phantom regime). 

When $ -1<w < 0$, we find that for the models $\Gamma = \Gamma_0$, $\Gamma_0 H^{-1}$ and $\Gamma_0 H^{-2}$, 
our universe transits from the past DM dominated decelerating phase $\longrightarrow$ present accelerating phase and finally it continues with an accelerating phase with $q = -1$. 
On the other hand, for the same region of $w$ (i.e. $-1 < w < 0$), for $\Gamma = \Gamma_0 H$, our universe transits from its past decelerating phase dominated by the second fluid to the DM dominated accelerating universe. In the sub region $-1<w<-\frac{1}{3}$, corresponding to the model $\Gamma=2\left(1-\frac{1}{\beta}\right)\frac{\dot{H}}{H}$, the universe undergoes a transition from a decelerating phase dominated by DM to a late-time accelerated expansion driven by DE, consistent with a quintessence-like behavior.

When $w =-1$, we notice that for the models $\Gamma_0$, $\Gamma_0 H^{-1}$, $\Gamma_0 H^{-2}$, we have found two types of universe's evolution. The first one is that the universe leaves its past DM dominated decelerating phase and ends in a completely DE dominated accelerating epoch (the same dynamics is also observed for the models with $\Gamma  = \Gamma_0 H$ and $2\left(1-\frac{1}{\beta}\right)\frac{\dot{H}}{H}$). 
In the second scenario, the universe exits its past DM dominated decelerating phase and finishes in a state where DM and DE both exist, i.e.  $(\Omega_{\rm dm}^{\rm eff},\Omega_f)\neq(0,0)$, concluding that the coincidence problem can be reduced without invoking the interaction in the dark sector~\cite{Amendola:1999er}.

When $w < -1$, we find that for the models $\Gamma 
 = \Gamma_0$, $\Gamma_0 H$, $\Gamma_0 H^{-1}$, $\Gamma_0 H^{-2}$, $2\left(1-\frac{1}{\beta}\right)\frac{\dot{H}}{H}$, our universe exits its past DM dominated decelerating phase and enters into the present accelerating phase with $-1 < q < 0$ and finally it ends in a DE dominated super accelerating phase (i.e. $q< -1$).

Concerning the models with more than one parameters, we notice that except the accelerating scaling attractors obtained either for $-1 < w <0$, or $w < -1$,  their dynamics is almost analogous to the one parameter models.

In addition, from Figs. \ref{fig:exp-hist-M1}, \ref{fig:exp-hist-M2}, \ref{fig:exp-hist-M3}, \ref{fig:exp-hist-M4}, \ref{fig:exp-hist-M5}, \ref{fig:exp-hist-M6}, \ref{fig:exp-hist-M7}, one can observe that the redshift evolution of the normalized Hubble parameter $H/H_0$ remains consistent with the $\Lambda{\rm CDM}$ behavior at low redshifts, indicating the viability of the model in describing the present cosmic expansion. However, noticeable deviations appear at higher redshifts, where the effect of particle creation becomes dynamically significant. These deviations suggest that these matter creation models can leave distinct signatures on the expansion history, providing a possible means to observationally discriminate them from the standard $\Lambda{\rm CDM}$ cosmology.

In summary, based on the results of the article, it is fairly clear that matter creation cosmologies  are phenomenologically very rich and appealing. As far as we are concerned with the existing literature, phase space structures of the matter creation cosmologies in terms of a wide variety of the critical points, observed in this article, are very rich and undoubtedly, matter creation scenarios can be considered to be compelling alternatives to both DE and MG for understanding the evolution of the universe.

%-----------------------------------------------------------
\section{Acknowledgments}
SH acknowledges the financial support from the University Grants Commission (UGC), Govt. of India (NTA Ref. No: 201610019097). JdH is supported by the Spanish grants PID2021-123903NB-I00 and RED2022-134784-T
funded by MCIN/AEI/10.13039/501100011033 and by ERDF ``A way of making Europe''. SP and TS acknowledge the financial support from the Department of Science and Technology (DST), Govt. of India under the Scheme  ``Fund for Improvement of S\&T Infrastructure (FIST)'' (File No. SR/FST/MS-I/2019/41).

\section{Appendix-A}
\label{sec-appendix}

Here we consider various matter creation models which are more general versions of the one parameter matter creation models. As we shall discuss here, we do not obtain anything significantly new results in these general models.  

\subsection{Model: $\Gamma=\Gamma_0+\Gamma_1 H$}

We start with the matter creation model $\Gamma = \Gamma_0 + \Gamma_1 H$ which has two free parameters, namely, $\Gamma_0$ and $\Gamma_1$, and this model represents the linear combination of the matter creation models, namely, $\Gamma = \Gamma_0$ and $\Gamma \propto H$. For this model, the dynamics within the two-fluid system is described by  
\begin{subequations} \label{DS-8}
\begin{align}
    x'=& x(1-x) \left[3w+\frac{\alpha z}{1-z}+\beta\right],   \label{DS-8-x} \\
    z'=& \frac{3}{2}z(1-z)\left[1+w(1-x)-\frac{x}{3} \left(\frac{\alpha z}{1-z}+\beta \right) \right], \label{DS-8-z} 
\end{align} 
\end{subequations}
in which $\alpha~(=\Gamma_0/H_0)$ and $\beta~(=\Gamma_1)$ are the dimensionless positive parameters and after regularization, eqn. (\ref{DS-8}) becomes 
\begin{subequations} \label{RG-DS-8}
\begin{align}
    \frac{dx}{d\tau}=& x(1-x) \left[3w (1-z)+\alpha z+\beta (1-z)\right],   \label{RG-DS-8-x} \\
    \frac{dz}{d\tau}=& \frac{3}{2}z(1-z)\Big[\left(1+w(1-x)\right)(1-z)\nonumber \\ &-\frac{x}{3}\left(\alpha z+\beta (1-z)\right) \Big], \label{RG-DS-8-z} 
\end{align}
\end{subequations}
where $\tau$ is defined as $dN=(1-z)d\tau$. In Table~\ref{tab:gen-2} we present the critical points for this model and their nature. The critical line $F_7$ where both the fluid share energy density, giving stable qualitative nature in the phantom region $(w<-1)$, represents acceleration if we choose $\beta>1$. Thus, there is a possibility to reduce the coincidence problem in the phantom region otherwise this model can offer the results which we have noticed for the Model \ref{model-1} ($\Gamma = \Gamma_0$) and \ref{model-2} ($\Gamma = \Gamma_0 H$).

\subsection{Model: $\Gamma=\Gamma_0+\Gamma_3 H^{-1}$}

We consider the second model in this series as $\Gamma=\Gamma_0+\Gamma_3 H^{-1}  $ having two parameters $\Gamma_0$ and $\Gamma_3$ and this model presents the   linear combination of the models $\Gamma = \Gamma_0$ and $\Gamma \propto H^{-1}$.  
For the two-fluid system, the autonomous system reads 
\begin{subequations} \label{DS-9}
\begin{align}
    x'=& x(1-x) \left[3w+\frac{\alpha z(1-z)+\mu z^2}{(1-z)^2}\right],   \label{DS-9-x} \\
    z'=& \frac{3}{2}z(1-z)\Big[1+w(1-x) \nonumber \\ &-\frac{x}{3} \left(\frac{\alpha z(1-z)+\mu z^2}{(1-z)^2} \right) \Big], \label{DS-9-z} 
\end{align} 
\end{subequations}
where $\alpha~(=\Gamma_0/H_0),~\mu~(=\Gamma_3/H_0^2)$ are the dimensionless  positive free parameters. Similarly, this system admits  a singularity at $z =1$. Thus, we regularize this system and after regularization, the system becomes, 
\begin{subequations} \label{RG-DS-9}
\begin{align}
    \frac{dx}{d\tau}=& x(1-x) \left[3w(1-z)^2+\alpha z(1-z)+\mu z^2\right],   \label{RG-DS-9-x} \\
    \frac{dz}{d\tau}=& \frac{3}{2}z(1-z)\Big[\left(1+w(1-x)\right)(1-z)^2 \nonumber \\ &-\frac{x}{3} \left(\alpha z(1-z)+\mu z^2 \right) \Big], \label{RG-DS-9-z} 
\end{align} 
\end{subequations}
where $\tau$ is defined through $dN=(1-z)^2d\tau$. In Table~\ref{tab:gen-2} we present the critical points and their corresponding analysis. Here, also cosmological implications are similar to the Models \ref{model-1} ($\Gamma = \Gamma_0$), \ref{model-4} ($\Gamma = \Gamma_0 H^{-1}$) and \ref{model-5} ($\Gamma = \Gamma_0 H^{-2}$).

\subsection{Model: $\Gamma=\Gamma_0+\Gamma_2 H^2$}

We now consider the following model $\Gamma=\Gamma_0+\Gamma_2 H^2$ with $\Gamma_0$ and $\Gamma_2$ as two free parameters. For the two-fluid system, we get the following autonomous system
\begin{subequations} \label{DS-10}
\begin{align}
    x'=& x(1-x) \left[3w+\frac{\alpha z^2+\gamma (1-z)^2}{z(1-z)}\right],   \label{DS-10-x} \\
    z'=& \frac{3}{2}z(1-z)\Big[1+w(1-x)-\frac{x}{3} \left(\frac{\alpha z^2+\gamma (1-z)^2}{z(1-z)} \right) \Big], \label{DS-10-z} 
\end{align} 
\end{subequations}
where $\alpha~(=\Gamma_0/H_0)$ and $\gamma~(=\Gamma_2 H_0)$ are the dimensionless  positive parameters. Notice that the system (\ref{DS-10}) has singularities at $z =0$ and $z= 1$. We regularize the system (\ref{DS-10}) and this now reads 
\begin{subequations} \label{RG-DS-10}
\begin{align}
    \frac{dx}{d\tau}=& x(1-x) \left[3w z(1-z)+\alpha z^2+\gamma (1-z)^2\right],   \label{RG-DS-10-x} \\
    \frac{dz}{d\tau}=& \frac{3}{2}z(1-z)\Big[\left(1+w(1-x)\right)z(1-z) \nonumber \\ &-\frac{x}{3} \left(\alpha z^2+\gamma (1-z)^2 \right) \Big], \label{RG-DS-10-z} 
\end{align} 
\end{subequations}
where $\tau$ is introduced as $dN=z(1-z)d\tau$. In Table~\ref{tab:gen-2} we present the results. The cosmological features are previously found in the Models \ref{model-1} ($\Gamma = \Gamma_0$) and \ref{model-3} ($\Gamma = \Gamma_0 H^2$).

\subsection{Model: $\Gamma=\Gamma_1 H+\Gamma_2 H^2$}

We now consider the linear combination of the models $\Gamma \propto H$ and $\Gamma \propto H^2$, that means, $\Gamma = \Gamma_1 H+\Gamma_2 H^2$ where $\Gamma_1$ and $\Gamma_2$ are constants. For the two-fluid system, the dynamics of the model is described by the following equation 
\begin{subequations} \label{DS-11}
\begin{align}
    x'=& x(1-x) \left[3w+\frac{\beta z+\gamma (1-z)}{z}\right],   \label{DS-11-x} \\
    z'=& \frac{3}{2}z(1-z)\Big[1+w(1-x)-\frac{x}{3} \left(\frac{\beta z+\gamma (1-z)}{z} \right) \Big], \label{DS-11-z} 
\end{align} 
\end{subequations}
where $\beta=\Gamma_1$, $\gamma=\Gamma_2 H_0$ are the dimensionless  positive constants, and this system admits a singularity at $z =0$. After regularizing the system (\ref{DS-11}) one arrives at 
\begin{subequations} \label{RG-DS-11}
\begin{align}
    \frac{dx}{d\tau}=& x(1-x) \left[3w z+\beta z+\gamma (1-z)\right],   \label{RG-DS-11-x} \\
    \frac{dz}{d\tau}=& \frac{3}{2}z(1-z)\Big[\left(1+w(1-x)\right)z-\frac{x}{3} \left(\beta z+\gamma (1-z) \right) \Big]. \label{RG-DS-11-z} 
\end{align} 
\end{subequations}
where $\tau$ is introduced via $dN=zd\tau$. We perform the phase space analysis for this system and in Table~\ref{tab:gen-3} we present the results. One can notice that the critical line $J_7$ alleviates cosmic coincidence problem for $-1<w<-\frac{1}{3}$ and the other consequences of this creation rate can be recovered from the Models \ref{model-2} ($\Gamma = \Gamma_0 H$) and \ref{model-3} ($\Gamma = \Gamma_0 H^2$).

\subsection{Model: $\Gamma=\Gamma_1 H+\Gamma_3 H^{-1}$}

Considering the linear combinations of the models, namely,  $\Gamma \propto H$ and $\Gamma \propto H^{-1}$, we consider the following model $\Gamma = \Gamma_1 H+\Gamma_3 H^{-1}$ in which $\Gamma_1$ and $\Gamma_3$ are constants. 
The dynamics for the two-fluid system is described by the following equation 
\begin{subequations} \label{DS-12}
\begin{align}
    x'=& x(1-x) \left[3w+\beta+\frac{\mu z^2}{(1-z)^2}\right],   \label{DS-12-x} \\
    z'=& \frac{3}{2}z(1-z)\Big[1+w(1-x)-\frac{x}{3} \left(\beta+\frac{\mu z^2}{(1-z)^2} \right) \Big], \label{DS-12-z} 
\end{align} 
\end{subequations}
where $\beta=\Gamma_1$ and $\mu=\Gamma_3/H_0^2$ are the dimensionless  positive parameters.  Notice that the system (\ref{DS-12}) has a singularity at $z =1$. We regularize the system (\ref{DS-12}) which finally reduces to 
\begin{subequations} \label{RG-DS-12}
\begin{align}
    \frac{dx}{d\tau}=& x(1-x) \left[(3w+\beta)(1-z)^2+\mu z^2\right],   \label{RG-DS-12-x} \\
    \frac{dz}{d\tau}=& \frac{3}{2}z(1-z)\Big[\left(1+w(1-x)\right)(1-z)^2 \nonumber \\ &-\frac{x}{3} \left(\beta(1-z)^2+\mu z^2 \right) \Big], \label{RG-DS-12-z} 
\end{align} 
\end{subequations}
where $dN=(1-z)^2d\tau$. In Table~\ref{tab:gen-3} we present the critical points and their stability. Here, in the phantom region $(w<-1)$, the critical point $L_7$ is a candidate of solving cosmic coincidence problem, otherwise the results which this model can offer, are already presented in the Models \ref{model-2} ($\Gamma = \Gamma_0 H$) and \ref{model-4} ($\Gamma = \Gamma_0 H^{-1}$).    

\subsection{Model: $\Gamma=\Gamma_0+\Gamma_1 H+\Gamma_2 H^2$}

We now consider a three parameter matter creation model of the form $\Gamma=\Gamma_0+\Gamma_1 H+\Gamma_2 H^2$ where $\Gamma_0$, $\Gamma_1$ and $\Gamma_2$ are constants. 
For the two-fluid system, the dynamics of the model can be described by the following two dimensional system  
\begin{subequations} \label{DS-13}
\begin{align}
    x'=& x(1-x) \left[3w+\frac{\alpha z^2+\beta z (1-z)+\gamma (1-z)^2}{z(1-z)}\right],   \label{DS-13-x} \\
    z'=& \frac{3}{2}z(1-z)\Big[1+w(1-x) \nonumber \\ &-\frac{x}{3} \left(\frac{\alpha z^2+\beta z(1-z)+\gamma (1-z)^2}{z(1-z)} \right) \Big], \label{DS-13-z} 
\end{align} 
\end{subequations}
where $\alpha~(=\Gamma_0/H_0),~\beta~(=\Gamma_1)$, $\gamma=\Gamma_2 H_0$, are the dimensionless  positive constants, however, the above system has two singularities, one at $z=0$ and the other at $z =1$. We regularize the system (\ref{DS-13}) and the final system reads 
\begin{subequations} \label{RG-DS-13}
\begin{align}
    \frac{dx}{d\tau}=& x(1-x) \Big[3w z(1-z)+\alpha z^2 \nonumber \\ &+\beta z (1-z)+\gamma (1-z)^2\Big],   \label{RG-DS-13-x} \\
    \frac{dz}{d\tau}=& \frac{3}{2}z(1-z)\Big[(1+w(1-x))z(1-z) \nonumber \\ &-\frac{x}{3} \left(\alpha z^2+\beta z(1-z)+\gamma (1-z)^2 \right) \Big], \label{RG-DS-13-z} 
\end{align} 
\end{subequations}
where $dN=z(1-z)d\tau$. We perform the phase space analysis and in Table~\ref{tab:gen-3} we present the corresponding results. The cosmological results are not new in this context. Meanwhile, we have obtained this results for the Models \ref{model-1} ($\Gamma = \Gamma_0$), \ref{model-2} ($\Gamma = \Gamma_0 H$) and \ref{model-3} ($\Gamma = \Gamma_0 H^2$). 

%%%%%%%   Table 8  %%%%%%%%
\begin{table*}[t]
\centering
\resizebox{1.0\textwidth}{!}{%
	\begin{tabular}{|c c c c c c|}\hline\hline
{\bf Creation rate} & {\bf Critical point} & {\bf Existence} & {\bf Stability} & $\mathbf{\Omega_{dm}^{\rm eff}}$ & {\bf Acceleration}  \\ \hline
%   &&&&&     \\

$\Gamma_0+\Gamma_1 H$  & $F_0(0,0)$ & Always  & {\bf Stable} if $3w+\beta<0,~w<-1$;  & $0$ & Yes if $w<-1$  \\ 
 & $F_1(1,0)$ & Always & {\bf Stable} if $ 3w+\beta>0,~\beta\geq 3 $;  & $1$ & Yes if $\beta>1$  \\
 & $F_2(0,1)$ & Always & Unstable & $0$  & Undetermined  \\
 & $F_3(1,1)$ & Always & Always Saddle & $1$  & Always yes  \\
 & $F_4\left(1,\frac{3-\beta}{3-\beta+\alpha}\right)$ & $\beta\leq 3$ & {\bf Stable} if $w>-1$ & $1$  & Always yes  \\
 & $F_5\left(0,z_c\right)$ & $w=-1$ & {\bf Stable} if $\beta-3+(3-\beta+\alpha)z_c>0$ & $0$  & Always yes  \\
 & $F_6\left(x_c,\frac{3-\beta}{3-\beta+\alpha}\right)$ & $w=-1,~\beta\leq 3 $ & Always {\bf Stable} & $x_c$  & Always yes  \\
 & $F_7\left(x_c,0\right)$ & $3w+\beta=0$ & {\bf Stable} if $w<-1$ & $x_c$  & Yes if $\beta>1$  \\ \hline
 
$\Gamma_0+\Gamma_3 H^{-1}$ & $G_0(0,0)$ & Always & {\bf Stable} if $w<-1$; & $0$  & Yes if $w<-\frac{1}{3}$  \\
 & $G_1(1,0)$ & Always & Saddle if $w>0$, Unstable if $w<0$; & $1$  & No  \\
 & $G_2(0,1)$ & Always & Unstable/Saddle & $0$  & Undetermined   \\
 & $G_3(1,1)$ & Always & Saddle & $1$  & Always yes  \\ 
 & $G_4\left(1,\frac{6}{6+\alpha+\sqrt{\alpha^2+12\mu}}\right)$ & Always & Always {\bf Stable} & $1$  & Always yes  \\ 
 & $G_5\left(x_c,0\right)$ & $w=0$ & Unstable & $x_c$  & No  \\ 
 & $G_6\left(x_c,\frac{6}{6+\alpha+\sqrt{\alpha^2+12\mu}}\right)$ & $w=-1$ & Always {\bf Stable} & $x_c$  & Always yes  \\ 
 & $G_7\left(0,z_c\right)$ & $w=-1$ & {\bf Stable} if $z_c>(<)\frac{6}{6+\alpha+\sqrt{\alpha^2+12\mu}}$, $3+\alpha-\mu>(<)0$; & $0$  & Always yes  \\ 
 &     &     & Unstable if $z_c<(>)\frac{6}{6+\alpha+\sqrt{\alpha^2+12\mu}}$, $3+\alpha-\mu>(<)0$ &   & 
  \\ \hline

 $\Gamma_0+\Gamma_2 H^2$ & $I_0(0,0)$ & Always & Unstable/Saddle & $0$  & Undetermined  \\
 & $I_1(1,0)$ & Always & Always {\bf Stable} & $1$  & Always yes  \\
 & $I_2(0,1)$ & Always & Unstable/Saddle & $0$  & Undetermined   \\
 & $I_3(1,1)$ & Always & Saddle & $1$  & Always yes  \\ 
 & $I_{4\pm}\left(1,\frac{3+2\gamma\pm\sqrt{9-4\alpha\gamma}}{2(3+\alpha+\gamma)}\right)$ & $4\alpha\gamma\leq 9$ & $I_{4+}:$ {\bf Stable} if $w>-1$, otherwise Saddle & $1$  & Always yes  \\ 
 &    &    & $I_{4-}:$ Unstable/Saddle    &      &  \\
 & $I_{5\pm}\left(x_c,\frac{3+2\gamma\pm\sqrt{9-4\alpha\gamma}}{2(3+\alpha+\gamma)}\right)$ & $4\alpha\gamma\leq 9,~w=-1$ & $I_{5+}:$ {\bf Stable}; $I_{5-}:$ Unstable & $x_c$  & Always yes  \\ 
 & $I_6\left(0,z_c\right)$ & $w=-1$ & {\bf Stable} if $\frac{3+2\gamma-\sqrt{9-4\alpha\gamma}}{2(3+\alpha+\gamma)}<z_c<\frac{3+2\gamma+\sqrt{9-4\alpha\gamma}}{2(3+\alpha+\gamma)}$ & $0$  & Always yes  \\ 
 &     &     & Unstable if $z_c<\frac{3+2\gamma-\sqrt{9-4\alpha\gamma}}{2(3+\alpha+\gamma)}$ or $z_c>\frac{3+2\gamma-\sqrt{9-4\alpha\gamma}}{2(3+\alpha+\gamma)}$ &   & 
  \\ \hline 
 \hline
\end{tabular}%
 }
\caption{Properties of the critical points of the dynamical system (\ref{RG-DS-8}), (\ref{RG-DS-9}) and (\ref{RG-DS-10}) for the matter creation rate $\Gamma=\Gamma_0 + \Gamma_1 H$, $\Gamma_0+\Gamma_3 H^{-1}$ and $\Gamma_0+\Gamma_2 H^2 $ respectively. 
	   }
	\label{tab:gen-2}
\end{table*}

%%%%%%%   Table 9  %%%%%%%%
\begin{table*}[t]
\centering
\resizebox{1.0\textwidth}{!}{%
	\begin{tabular}{|c c c c c c|}\hline\hline
{\bf Creation rate} & {\bf Critical point} & {\bf Existence} & {\bf Stability} & $\mathbf{\Omega_{dm}^{\rm eff}}$ & {\bf Acceleration}  \\ \hline
%   &&&&&     \\

 $\Gamma_1 H+\Gamma_2 H^2$ & $J_0(0,0)$ & Always & Unstable/Saddle & $0$  & Undetermined \\
 & $J_1(1,0)$ & Always & Always {\bf Stable} & $1$  & Always yes  \\
 & $J_2(0,1)$ & Always & {\bf Stable} if $w\leq-\frac{\beta}{3}$, $w>-1$ & $0$  & Yes if $w<-\frac{1}{3}$   \\
 & $J_3(1,1)$ & Always & {\bf Stable} if $w>-\frac{\beta}{3}$, $\beta<3$ & $1$  & Yes if $\beta>1$  \\ 
 & $J_4\left(1,\frac{\gamma}{\gamma+3-\beta}\right)$ & $\beta\leq 3$ & Saddle if $w>-1$; Unstable if $w<-1$ & $1$  & Always yes  \\ 
 & $J_5\left(x_c,\frac{\gamma}{\gamma+3-\beta}\right)$ & $w=-1,~\beta\leq 3$ & Unstable & $x_c$  & Always yes  \\ 
 & $J_6\left(0,z_c\right)$ & $w=-1$ & {\bf Stable} if $\gamma-(\gamma+3-\beta)z_c<0$ & $0$  & Always yes  \\ 
 & $J_7\left(x_c,1\right)$ & $\beta+3w=0$ & {\bf Stable} if $\beta<3$ & $x_c$  & Yes if $\beta>1$  \\ \hline
 
  $\Gamma_1 H+\Gamma_3 H^{-1}$ & $L_0(0,0)$ & Always & {\bf Stable} if $w\leq-\frac{\beta}{3},~w<-1$ & $0$  & Yes if $w<-\frac{1}{3}$ \\
 & $L_1(1,0)$ & Always & {\bf Stable} if $w>-\frac{\beta}{3},~\beta\geq 3$ & $1$  & Yes if $\beta>1$  \\
 & $L_2(0,1)$ & Always & Unstable/Saddle & $0$  & Undetermined   \\
 & $L_3(1,1)$ & Always & Saddle & $1$  & Always yes \\ 
 & $L_4\left(1,\frac{3-\beta-\sqrt{(3-\beta)\mu}}{3-\beta-\mu}\right)$ & $\beta\leq 3$ & {\bf Stable} if $w>-1,~\beta+\mu\neq 3$ & $1$  & Always yes  \\ 
 & $L_5\left(x_c,\frac{3-\beta-\sqrt{(3-\beta)\mu}}{3-\beta-\mu}\right)$ & $w=-1,~\beta\leq 3$ & {\bf Stable} if $\beta<3,~\beta+\mu\neq 3$ & $x_c$  & Always yes  \\ 
 & $L_6\left(0,z_c\right)$ & $w=-1$ & {\bf Stable} if $z_c<\frac{3-\beta-\sqrt{(3-\beta)\mu}}{3-\beta-\mu}$; Unstable if $z_c>\frac{3-\beta-\sqrt{(3-\beta)\mu}}{3-\beta-\mu}$ & $0$  & Always yes  \\ 
 & $L_7\left(x_c,0\right)$ & $\beta+3w=0$ & {\bf Stable} if $\beta>3$ & $x_c$  & Yes if $\beta>1$  \\ \hline

 $\Gamma_0+\Gamma_1 H+\Gamma_2 H^2$ & $M_0(0,0)$ & Always & Unstable/Saddle & $0$  & Undetermined  \\
 & $M_1(1,0)$ & Always & Always {\bf Stable} & $1$  & Always yes  \\
 & $M_2(0,1)$ & Always & Unstable/Saddle & $0$  & Undetermined   \\
 & $M_3(1,1)$ & Always & Saddle & $1$  & Always yes  \\ 
 & $M_{4\pm}\left(1,\frac{3-\beta+2\gamma\pm\sqrt{(3-\beta)^2-4\alpha\gamma}}{2(3+\alpha-\beta+\gamma)}\right)$ & $(3-\beta)^2-4\alpha\gamma\geq 0$, & $M_{4+}:$ {\bf Stable} if $w>-1$, Saddle if $w<-1$ & $1$  & Always yes  \\ 
 &    & $3+2\alpha>\beta,~3+2\gamma>\beta$;   & $M_{4-}:$ Unstable if $w<-1$, Saddle if $w>-1$     &      &  \\
 & $M_{5\pm}\left(x_c,\frac{3-\beta+2\gamma\pm\sqrt{(3-\beta)^2-4\alpha\gamma}}{2(3+\alpha-\beta+\gamma)}\right)$ & $(3-\beta)^2-4\alpha\gamma\geq 0,~w=-1$, & $M_{5+}:$ {\bf Stable}; $M_{5-}:$ Unstable & $x_c$  & Always yes  \\ 
 && $3+2\alpha>\beta,~3+2\gamma>\beta$; &&&   \\
 & $M_6\left(0,z_c\right)$ & $w=-1$ & {\bf Stable} if $\frac{3-\beta+2\gamma-\sqrt{(3-\beta)^2-4\alpha\gamma}}{2(3+\alpha-\beta+\gamma)}<z_c<\frac{3-\beta+2\gamma+\sqrt{(3-\beta)^2-4\alpha\gamma}}{2(3+\alpha-\beta+\gamma)}$, & $0$  & Always yes  \\ 
 &&& $(3-\beta)^2-4\alpha\gamma\geq 0,~3+2\alpha>\beta,~3+2\gamma>\beta$; &&  \\
 &     &     & Unstable if $z_c<\frac{3-\beta+2\gamma-\sqrt{(3-\beta)^2-4\alpha\gamma}}{2(3+\alpha-\beta+\gamma)}$ or $z_c>\frac{3-\beta+2\gamma+\sqrt{(3-\beta)^2-4\alpha\gamma}}{2(3+\alpha-\beta+\gamma)}$, &   &  
  \\
  &&& $(3-\beta)^2-4\alpha\gamma\geq 0,~3+2\alpha>\beta,~3+2\gamma>\beta$; &&     \\   \hline\hline

\end{tabular}%
 }
\caption{Properties of the critical points of the dynamical system (\ref{RG-DS-11}), (\ref{RG-DS-12}) and (\ref{RG-DS-13}) for the matter creation rate $\Gamma=\Gamma_1 H+\Gamma_2 H^2$, $\Gamma_1 H+\Gamma_3 H^{-1}$ and $\Gamma_0 + \Gamma_1 H + \Gamma_2 H^2 $ respectively. 
	   }
	\label{tab:gen-3}
\end{table*}

%-----------------------------------------------------------
\bibliography{biblio}
%-------------------------------------------------------------
\end{document}